\documentclass[times,11pt]{article}
\usepackage{amsmath,mathptmx,amssymb}
\usepackage{tikz}
\newtheorem{assumption}{Assumption}[section]
\newtheorem{problem}{Problem}[section]
\newtheorem{lemma}{Lemma}[section]
\newtheorem{proposition}{Proposition}[section]
\newtheorem{definition}{Definition}[section]
\newtheorem{theorem}{Theorem}[section]
\newtheorem{example}{Example}[section]
\newtheorem{remark}{Remark}[section]
\usepackage{graphicx} 
\usepackage {epstopdf}
\usepackage{caption}
\usepackage{subcaption}
\usepackage{xcolor}
\graphicspath{{Figures/}}

\begin{document}
	
\title{Observer-based switched-linear system identification}
	
	\author{Fethi Bencherki \thanks{Department of Electrical and Electronics Engineering, Eski\c{s}ehir Technical 
			University, 26555 Eski\c{s}ehir, Turkey. E-mail: fethi bencherki@eskisehir.edu.tr} \and Semiha T\"urkay 
		    \thanks{Department of Electrical and Electronics Engineering, Eski\c{s}ehir Technical University, 26555 
		    Eski\c{s}ehir, Turkey. E-mail: semihaturkay@eskisehir.edu.tr} \and H\"useyin~Ak\c{c}ay
		    \thanks{{\em Corresponding author}. Department of Electrical and Electronics Engineering, Eski\c{s}ehir Technical 
		    	University, 26555 Eski\c{s}ehir, Turkey. E-mail: huakcay@eskisehir.edu.tr}}
	
	
	\maketitle

	\begin{abstract}
		In this paper, we present a methodology to identify discrete-time state-space switched-linear systems (SLSs)
		from input-output measurements. Continuous-state is not assumed to be measured. The key step is a 
		deadbeat observer based transformation to a switched auto-regressive with exogenous input (SARX) model. 
		This transformation reduces the state-space identification problem to a SARX model estimation problem. 
		Overfitting issues are tackled. The switch and parameter identifiability and the persistence of excitation 
		conditions on the inputs are discussed in detail. The discrete-states are identified in the observer domain by 
		solving a non-convex sparse optimization problem. A clustering algorithm reveals the discrete-states under mild
		assumptions on the system structure and the dwell times. The switching sequence is estimated from the input-output 
		data by the multi-variable output error state space (MOESP) algorithm and a variant modified from it. A convex relaxation 
		of the sparse optimization problem yields the block basis pursuit denoising (BBPDN) algorithm. Theoretical findings 
		are supported by means of a detailed numerical example. In this example, the proposed methodology is also compared to
		another identification scheme in hybrid systems literature.\\
		
		{\bf AMS:} 93B30, 93B15, 93C30, 93C05 93C95.
	\end{abstract}


\section{Introduction}
Linear time-varying (LTV) systems are frequently used to model systems which have non-stationary 
properties and undergo small amplitude vibrations. Control design,  realization theory, and 
identification of the LTV systems have received increasing attention in the past years 
\cite{Toth:2010,Mohammadpour&Scherer:2012,Petreczky&Toth&Mercere:2016}. Linear parameter varying (LPV)
systems form a particular type of time-varying systems where the variation depends explicitly
on a time-varying signal referred as the scheduling sequence. In state-space realizations, this 
results in the system matrices changed according to this scheduling sequence. 

The state-space models are preferred over the input-output models since in the former multiple inputs and 
outputs are efficiently handled. Besides, advanced control synthesis methods are readily applied to 
the LPV state-space descriptions via the linear fractional transformation \cite{Scherer:2001}. Recent 
studies \cite{Barker&Balas:2000,Giarre&Bauso&Falugi&Bamieh:2006} have shown potential of the LPV system theory for 
industrial applications in which systems depend on a known scheduling vector. A subspace method to 
identify multi-input/multi-output (MIMO) LPV state-space systems with affine parameter dependence was 
proposed in \cite{Verdult&Verhaegen:2002}. A major problem is large dimensions of data 
matrices when the scheduling sequence varies arbitrarily. A numerically efficient implementation was 
presented in \cite{VanWingerden&Verhaegen:2009} using the kernel method \cite{Verdult&Verhaegen:2005}.
Subspace identification of the MIMO--LPV systems using periodic scheduling sequence was studied in 
\cite{Felici&Wingerden&Verhaegen:2007}.

A special class of the LPV systems is the class of piece-wise affine (PWA) models of discrete-time nonlinear 
and hybrid systems. A PWA model is obtained by partitioning the state and the input set into a finite number 
of polyhedral regions. In each region,  linear or affine submodels share the same continuous state. The PWA 
models are hybrid models with dynamical behavior switching among the submodels according to some discrete-event 
space. They have universal approximation properties, that is, any nonlinear phenomenon can be approximated 
by a PWA model. Equivalence between the PWA systems and several classes of hybrid systems was established 
in \cite{Heemels&DeSchutter&Bemporad:2001}. Results on analysis, computation, stability, and control of hybrid systems 
have appeared \cite{Paoletti&Juloski&Ferrari-Trecate&Vidal:2007,Paoletti&Roll&Garulli&Vicino:2007}. 

Identification of a PWA model is performed in three stages: estimation of the submodel parameters, 
estimation of the hyperplanes defining partitioning of the state, and estimation of the input set. For 
models in the regression form, inputs are the regressors. This is a classification problem, that is,
each datum is to be associated with a most suitable submodel. It is a very hard problem unless 
partitioning of the state is fixed {\em a priori}. In  \cite{Ferrari-Trecate&Muselli&Liberati&Morari:2003}, 
piece-wise affine auto-regressive with exogenous input (PWARX) models were considered with clustering, linear 
identification, and pattern recognition techniques to identify both the submodels and the polyhedral partitioning of 
the regressor set. In \cite{Vidal&Soatto&Ma&Sastry:2003}, an algebro-geometric approach to piecewise-linear 
(PWL) model identification  was proposed. It exploits the connections between the PWL system identification and 
the polynomial factorization/hyperplane clustering. In \cite{Roll&Bemporad&Ljung:2004}, a hybrid identification 
problem was formulated for the hinging hyperplane ARX and the Wiener PWARX models and solved by mixed-integer linear 
and quadratic programs. When errors are amplitude bounded, a three-stage procedure that uses a modified greedy 
algorithm for data classification and submodel estimation was proposed in \cite{Bemporad&Garulli&Paoletti&Vicino:2005}.

\subsection{Related work}

A switched ARX (SARX) model is a hybrid affine model in which a finite number of submodels change only at 
switches that partition the time interval. The PWA model class is obtained by replacing the regressors with 
the scheduling sequences. Hence, identification algorithms developed in one model domain may be adapted to 
another with little effort. 

The SARX models were studied in \cite{Ohlsson&Ljung&Boyd:2010,Bako:2011,Ozay&Sznaier&Lagoa&Camps:2011,Ohlsson&Ljung:2013}. 
In all of these works, multiple-input/single-output (MISO) model structures were used. The segmentation 
problem, that is, the decomposition of a time-varying system into submodels whose parameters are piece-wise
constant in time was formulated in \cite{Ohlsson&Ljung&Boyd:2010} as a least-squares estimation problem with 
sum-of-norms regularization over the state parameter jumps and solved by a standard convex optimization algorithm. 
A reformulation by a kernel function was introduced in \cite{Ohlsson&Ljung:2013}. 
In \cite{Ozay&Sznaier&Lagoa&Camps:2011}, this identification problem was cast as a sparsity maximization problem when noise is amplitude bounded and solved by a greedy optimization algorithm. When noise is quadratically bounded, 
a convex relaxation was also introduced. The algorithm proposed in \cite{Bako:2011}
maximizes sparsity by assigning maximum number of the data points to a hyperplane generated by a 
submodel. A convex relaxation by the basis pursuit method was also introduced in this work.

The algebro-geometric method proposed in \cite{Vidal&Soatto&Ma&Sastry:2003} for the PWARX models was extended in 
\cite{Huang&Wagner&Ma:2004} to the state-space models by embedding the input-output data in a higher dimensional 
space. The submodels were extracted by the generalized principal component analysis algorithm. This method 
is suitable only for small data batches and high signal-to-noise ratio (SNR).

A subspace algorithm was proposed in \cite{Verdult&Verhaegen:2004} to identify the SLSs. Although the minimum 
dwell time requirement is modest, the scheduling sequence is assumed to be known. The state-space identification 
algorithm proposed in \cite{BakoVanLuongLauerBloch2013} does not restrict the minimum dwell time, yet 
assumes that continuous-state measurements are available. Without a constraint on the minimum dwell time, 
a state space identification algorithm for the SLSs was proposed in \cite{Bako&Mercere&Guillaume&Vidal&Lecoeuche:2009}. 
This algorithm is based on the observability results derived in \cite{Vidal&Chiuso&Soatto:2002} for jump linear systems. 
Though the minimum dwell time is not constrained, the SLS is assumed pathwise observable. A non-convex optimization based 
identification algorithm was proposed in \cite{Sefidmazgi&Kordmahalleh&Homaifar&Karimoddini&Tunstel:2016} assuming that
the switching sequence has a bounded variation. In this algorithm, the switching sequence is randomly initialized 
and the submodels and the initial states are estimated by the Past Outputs Multivariable Output-Error 
State-Space (PO-MOESP) subspace algorithm \cite{Verhaegen:1994}. The switching sequence is updated by solving a
binary integer programming problem. Next, the submodel parameters are updated. Updating of the submodel clusters 
and the state-space parameter matrices by a coordinate descent algorithm  is continued until a local or the global 
minimum is attained.

Detection and estimation of jumps in linear systems has been extensively studied in the literature
\cite{Willsky:1976,Basseville&Nikiforov:1993,Jikuya&Verhaegen:2002,Pekpe&Mourot&Gasso&Ragot:2004,Borges&Verdult&Verhaegen&Botto:2005}.
A deadbeat observer based generalized likelihood ratio (GLR) test was proposed in \cite{Jikuya&Verhaegen:2002} for 
the detection and estimation of jumps in the LTI system states. The deadbeat observer controls window size in the 
GLR test. The GLR test was extended in \cite{Pekpe&Mourot&Gasso&Ragot:2004} to the SLSs in the state-space form. In 
\cite{Pekpe&Mourot&Gasso&Ragot:2004}, first the number of the local models and the switching sequence were estimated from 
the GLR test. Then, the Markov parameters of the local models were estimated. In the last step, similar local 
models were merged. 

\subsection{Motivation for state-space framework}

With  few exceptions the contributions surveyed above deal with the PWARX-MISO models. Many existing control 
analysis and synthesis design methods, on the other hand, rely on the state-space models. The subspace, or more 
generally, the realization algorithms include some of the very popular methods in system identification. The main 
reason for their success is that they rely on the numerically robust QR factorization and the singular value 
decomposition (SVD) for low-rank matrix approximation from input-output data. Models returned by subspace 
methods are also nearly balanced. 

\subsection{Contributions}

A framework is proposed to identify the discrete states and the switching sequences of the SLSs in the state-space 
form from the input-output measurements. This framework followed by the basis construction procedure in 
\cite{Bencherki&Turkay&Akcay:2021} proposed by the authors of this paper delivers final models suitable for predicting time 
responses of the SLSs to prescribed inputs. The proposed identification framework is demonstrated to be consistent 
under some assumptions on the system structure, the dwell times of the discrete states, and noise amplitude in a 
completely deterministic setting. The switch detection schemes proposed in \cite{Jikuya&Verhaegen:2002,Pekpe&Mourot&Gasso&Ragot:2004}, 
on the other hand, rely on the stochastic noise descriptions. The proposed framework is exact: any 
SLS can be recovered in finite time from its noiseless input-output measurements if every discrete state is active 
in some segment and the minimum dwell time is sufficiently large while in some works \cite{Pekpe&Mourot&Gasso&Ragot:2004}
output transients may be detrimental due to the state approximations. 

\subsection{Organization of the paper}

The contents of this paper are as follows. In Section~\ref{SSSLSsec}, we formulate the state space 
identification problem for single-input/single-output (SISO)-SLS models from input-output data. In
Section~\ref{observertrans}, deadbeat observer-based transformation of the state-space SLS models to 
the SARX models is studied. This is a key step in reducing the state-space identification problem to an 
SARX model estimation problem. Back model transformations from the SARX models to the LTV models and from 
the LTV models to the SLS models are studied. The switches of the SARX model and their identifiability 
from the input-output data are also studied. This technical section prepares the stage for a non-convex and sparse 
optimization problem formulation in the next section to estimate the local models. The role played by model 
transformations is to compress infinite strings of the system Markov parameters into finite sets of 
the observer Markov parameters at the expense of more complicated discrete state sets and the switching sequences
in the transformation domain.   

In Section~\ref{discstest}, a local model set is retrieved by a clustering algorithm from the solution of a 
non-convex and sparse optimization problem over long and constant parameter intervals. This set exhausts all 
discrete states if every discrete state is active in at least one sufficiently long segment. The endpoints of 
such intervals are the switches. The rest of the switches are estimated from the input-output data by a 
MOESP type subspace algorithm \cite{Verhaegen&Dewilde:1992a,Verhaegen&Dewilde:1992b} or a discrete optimization 
algorithm modified from the MOESP algorithm in Section~\ref{switchSARX}.

Convex relaxation of the optimization problem leads to the BBPDN method in Section~\ref{bpdnsec}. This is achieved 
by relaxing the nonconvex mixed $\ell_0/\ell_1$ norm with the convex mixed $\ell_2/\ell_1$ norm. Recovery guarantees 
for the BBPDN \cite{Tropp:2006} and the block orthogonal matching pursuit (BOMP) \cite{Tropp:2007} algorithms 
have been put forward in the compressive sensing/approximation literature \cite{Eldar&Mishali:2009,Eldar&Kuppinger&Bolcskei:2010}. 
They are replaced in this paper by the switch identifiability and the persistence of excitation conditions. 
These conditions not only make recovery of the local modes possible, but also guarantee robustness to amplitude-bounded
noise if SNR is large. Theoretical findings are supported by means of a detailed numerical example in Section~\ref{numinsec}. 
In this example, the proposed method is also compared to a competitive algorithm in hybrid systems literature. 
Concluding remarks with a brief sketch of future work are presented in Section~\ref{concsec}.

\section{Problem statement for the SLS identification }\label{SSSLSsec}
In this paper, we consider a special class of the LTV-SISO systems represented by the state-space equations    
\begin{eqnarray}
	x(k+1) &=& A(k) x(k) +b(k) u(k), \label{ssx} \\
	y(k) &=& c^T(k) x(k) + d(k) u(k) \label{ssy}
\end{eqnarray}
where $u(k) \in \mathbb{R}$, $y(k) \in \mathbb{R}$, $x(k) \in \mathbb{R}^{n}$ are respectively the 
input, the output, the state sequences, and $c^T$ denotes the transpose of a given vector (matrix) $c$. The 
state dimension $n$ is assumed to be known and does not change with time.

Let $\mathbb{N}$ denote the set of positive integers and $\varphi$ be a switching sequence, that is, 
a map from $\mathbb{N}$ onto a finite set $\mathbb{S} =\left\{1,\cdots, \sigma \right\}$ for some fixed 
$\sigma \in \mathbb{N}$. Substitute $l=\varphi(k)$ and suppose that $A(k)=A_l$, $b(k)=b_l$, $c(k)=c_l$, 
$d(k)=d_l$. We denote the set of the discrete states (submodels) ${\mathcal P}_l=(A_l,b_l,c^T_l,d_l)$, 
$l=1,\cdots,\sigma$ by ${\mathcal P}$. The SISO model (\ref{ssx})--(\ref{ssy}) with the state-space
matrices changed by $\varphi$ is an SLS. 

A switching sequence $\varphi(k)$ segments a given interval $\left[1 \;\;N\right]$ into 
disjoint intervals $[k_i\;\;k_{i+1})$ such that   
\begin{equation}\label{varphit}
	\varphi(k)  = \varphi(k_i), \qquad k_i \leq  k < k_{i+1} 
\end{equation}
where $k_0=1$ and $k_i < k_{i^*} \leq N$. Given a segmentation $\chi$ of $[1\;\;N]$, let $\delta_i(\chi)=k_{i+1}-k_i$, 
$0\leq i < i^*$. The minimum dwell time is defined by $\delta_*(\chi) = \min_i \delta_i (\chi)$. Thus, $\delta_i(\chi)$ 
is the waiting time of the discrete state active in $[k_i\;\;k_{i+1})$ and the minimum dwell 
time is the smallest waiting time. We state the requirements on the  model structure as follows.

\begin{assumption}\label{sysasmp}
	The SLS model (\ref{ssx})--(\ref{varphit}) has $\sigma$ stable discrete states with MacMillan degree $n$.
\end{assumption} 

The SLS identification problem for the SISO systems we study in this paper is formulated as follows: 

\begin{problem}\label{problem}
	Given input-output data $u(k),y(k)$, $1 \leq k \leq N$ of the SLS 
	model (\ref{ssx})--(\ref{varphit}) satisfying Assumption~\ref{sysasmp}, estimate the 
	discrete states and the switching sequence. 
\end{problem}

In the course of developing a framework that solves the identification problem posed above, we will 
impose further conditions on the inputs, ${\mathcal P}$, and $\varphi$.

\section{Observer-based transformation to SARX model}\label{observertrans}

Let us add and subtract $g(k) y(k)$ to (\ref{ssx}):
\begin{eqnarray}
	x(k+1) &=& A(k) x(k) + b(k) u(k) +g(k) y(k)- g(k)y(k),  \nonumber 
	\\[-1ex] \label{deadbeattf} \\[-1ex]
	&=& A_{\rm o}(k) x(k)+B_{\rm o}(k) \zeta(k) \nonumber
\end{eqnarray}
where we used (\ref{ssy}), $g(k)$ is a time-varying gain sequence, and 
\begin{eqnarray}
	A_{\rm o}(k) &=&  A(k) +g(k) c^T(k) \in \mathbb{R}^{n \times n}, \label{Aopdef} \\ 
	B_{\rm o}(k) &=&  [ b(k) +g(k) d(k)  \; -g(k)] \in \mathbb{R}^{n \times 2} \label{Bobdef} \\
	\zeta(k)  &=& [u(k)  \;\; y(k)]^T \in \mathbb{R}^2. \label{vdef}
\end{eqnarray} 
Thus, we arrive at the so-called {\em observer} equations
\begin{eqnarray}
	x(k+1) &=& A_{\rm o}(k) x(k)+B_{\rm o}(k) \zeta(k), \label{ssxobsv} \\
	y(k) &=& c^T(k) x(k)+d(k) u(k). \label{ssyobsv}
\end{eqnarray}

The observer response is calculated from (\ref{ssxobsv})--(\ref{ssyobsv})
\begin{equation}\label{yrespobsv}
	y(k)=c^T(k) \Phi_{\rm o}(k,i) x(i) + d(k) u(k)+\sum_{j=i}^{k-1} h_{\rm o}(k,j) \zeta(j)
\end{equation}
for $1 \leq i < k$ by introducing the {\em observer Markov parameters} and the 
{\em observer state transition matrix}
\begin{eqnarray}
	h_{\rm o}(k,i) &=&  c^T(k) \Phi_{\rm o}(k,i+1) B_{\rm o}(i),  \label{defMarkovobsv} \\
	\Phi_{\rm o}(k,i) &=& A_{\rm o}(k-1) \, \cdots \, A_{\rm o}(i), \label{defsttranobsv}   
\end{eqnarray}
for $k>i$ and for $k=i$, $h_{\rm o}(k,k)=[d(k)\;\;0]$ and $\Phi_{\rm o}(k,k)=I_n$. 
Suppose there are $\tau,k^\prime,k^{\prime\prime} \in \mathbb{N}$ such that $k^\prime<\tau$ and for all 
$k^{\prime} \leq k \leq k^{\prime\prime}$, $\Phi_{\rm o}(k,k-\tau)=0$ . Then, (\ref{yrespobsv}) simplifies 
to a linear regression of $u(k)$ and $\zeta(k)$ 
\begin{equation}\label{yrespobsv22}
	y(k)= d(k)u(k)+\sum_{j=k-\tau}^{k-1} h_{\rm o}(k,j) \zeta(j), \;\; k^\prime \leq k \leq k^{\prime\prime}
\end{equation}
which is a time-varying ARX model, and in fact based on (\ref{varphit}) an SARX model, described by $2\tau+1$ 
parameters. An observer with this property is called {\em deadbeat observer}.

\begin{definition}\label{deadbeatdef}
	An LTV discrete-time observer is said to be a deadbeat observer on the interval $[k^\prime\;\;
	k^{\prime\prime}]$ if there exists a gain sequence $g(k) \in \mathbb{R}^n$ and a 
	$\tau < k^\prime$ such that
	\begin{equation}\label{deadbeat}
		\Phi_{\rm o}(k,k-\tau)=0, \qquad {\rm for \; all} \;\; k^\prime \leq k \leq k^{\prime\prime}.
	\end{equation}
\end{definition}

Finding deadbeat observers for arbitrary LTV systems, in particular one with a $\tau$ as small as possible 
is not trivial. Suppose for a moment that the system described by (\ref{ssx})--(\ref{ssy}) is 
time-invariant. Thus, we seek a constant gain $g \in \mathbb{R}^n$. In this case, $A(k)=A$, $b(k)=b$, 
$c^T(k)=c^T$, $d(k)=d$ and
\begin{equation}\label{hodeadbeat}
	h_{\rm o}(k,i)= c^T (A+gc^T)^{k-i-1} [b+gd \;-g], \qquad k>i.
\end{equation}
For an LTI system, (\ref{deadbeat}) translates to $(A+gc^T)^\tau=0$ for some $g \in \mathbb{R}^{n}$. If 
$(A,b,c^T,d)$ is minimal, $(A,c^T)$ is observable and $\tau=n$. Alternatively, by choosing $\tau$ large 
and pushing the eigenvalues of $A$ to zero, 
$\Phi_{\rm o}(k,k-\tau)=0$ may be demanded to hold approximately.  Let us 
illustrate some properties of the deadbeat observers by two numerical examples.

\begin{example}\label{ex1}
	Let the observability pair $(A,c^T)$ be given by 
	\[
	A=\left[\begin{array}{cc} 1 & 1 \\ 0 & 0  \end{array}\right], \qquad c^T=[1 \;\; 2].	
	\]
	Then, $A$ has one eigenvalue at $0$ and from 
	\[
	\left[\begin{array}{c} c^T \\ c^TA  \end{array}\right]=
	\left[\begin{array}{cc} 1 & 2 \\ 1 & 1  \end{array}\right],
	\]
	we see that $(A,c^T)$ is observable. With $g=[g_1 \;\;g_2]^T$ the characteristic equation 
	of $A+gc^T$ is given by
	\[
	\lambda^2-\lambda(1+g_1+2g_2)+g_2=0.
	\]
	The observer gain enforcing $\lambda^2=0$ is uniquely calculated as $g=[-1 \;\; 0]^T$. This is 
	expected since $(A,c^T)$ is observable. Since $(A+gc^T)^2=0$, $\min \tau \leq 2$. The pair 
	$(A+gc^T,c^T)$ is also observable. Hence, $(A+gc^T)^k \neq 0$ for all $k<n$.
\end{example}

\begin{example}\label{ex2}
	Let
	\[
	A_1=\left[\begin{array}{cc} 0 & 1 \\ 0 & 0  \end{array}\right], \;\;\; 	
	A_2=\left[\begin{array}{cc} 0 & 0 \\ 1 & 0  \end{array}\right],\;\;\;c^T=[1 \;\; 1].	
	\]
	Both $A_1$ and $A_2$ have two eigenvalues at $0$ and $(A_1,c^T)$ and $(A_2,c^T)$ are 
	observable. Moreover, $g=[0\;\;0]^T$. The matrices $A_1$ and $A_2$ generate by multiplication 
	only two other matrices
	\[
	A_3=\left[\begin{array}{cc} 1 & 0 \\ 0 & 0  \end{array}\right], \;\;\; 	
	A_4=\left[\begin{array}{cc} 0 & 0 \\ 0 & 1  \end{array}\right]	
	\]
	Hence, $\{A_1,A_2,A_3,A_4\}$ is a finite non-commutative group without identity.
\end{example}

The deadbeat observers transform the state-space models to the ARX models which are easier to estimate 
from the input-output data since infinite strings of the system Markov parameters are packed into  
finite numbers of the regression coefficients. The following result provides an upper bound on 
$\tau$ in Definition~\ref{deadbeatdef}. This upper bound does not depend on $\sigma$. 

\begin{lemma}\label{deadbeatlem}
	Let the SLS model (\ref{ssx})--(\ref{varphit}) be as in Assumption~\ref{sysasmp}. Suppose 
	that $\delta_*(\chi) \geq n$ and $\delta_0(\chi) \geq 2n$. Then, there exists a deadbeat 
	observer and a gain sequence $g(k)$ for (\ref{ssx})--(\ref{ssy}) satisfying a best possible bound 
	$\tau \leq 2 n-1$ on $[1\;\;N]$ and $g(k)=g(k_i)$, $\tau=n$ on $[k_i\;\;k_{i+1})$ for $1 \leq i < i^*$. 
\end{lemma} 

{\em Proof.} See Appendix~\ref{appA}.

The minimum dwell time requirement in Lemma~\ref{deadbeatlem} cannot be dropped. An example using $A_1$
and $A_2$ matrices in Example~\ref{ex2} is $\Phi_{\rm o}(k,i)=A_1 A_2 A_1 \;\cdots$ satisfying  
$\Phi_{\rm o}(k,i) \neq 0$ for all $k$ and $i$ with $k>i$. The last statement in the lemma asserts that
there is a discrete state set satisfying Assumption~\ref{sysasmp} such that the first conclusion
does not hold for a $\tau< 2n-1$. From Lemma~\ref{deadbeatlem}, we may write (\ref{yrespobsv22}) as
\begin{equation}\label{yrespobsv222}
	y(k) = d(k)u(k)+\sum_{j=k-2n+1}^{k-1} h_{\rm o}(k,j) \zeta(j), \;\; 2n \leq k \leq N.
\end{equation}

\subsection{Model conversions SARX-to-LTV-to-SLS}\label{ARXLTVSLScov}
In this subsection, we first study recovery of the system Markov parameters
of (\ref{ssx})--(\ref{varphit}) defined for $k>i$ by 
\begin{equation}\label{defMarkov}
	h(k,i) = c^T(k) \Phi(k,i+1) b(i) 
\end{equation}
and for $k=i$ by $h(k,k)=d(k)$ where $\Phi(k,k)=I_n$ and 
\begin{equation}\label{defsttran}
	\Phi(k,i) = A(k-1) \, \cdots \, A(i) 
\end{equation}
with $I_n \in \mathbb{R}^{n \times n}$ denoting the $n$ by $n$ identity matrix and the gain 
sequence from the observer Markov parameters.

Partition $h_{\rm o}(k,i)$ as $h_{\rm o}(k,i)=[h_{\rm o}^{(1)}(k,i) \;\;h_{\rm o}^{(2)}(k,i)]$ and let
\begin{equation}\label{gammaki}
	\gamma(k,i)=h_{\rm o}^{(1)}(k,i)+h_{\rm o}^{(2)}(k,i)h_{\rm o}^{(1)}(i,i), \;\;k>i \geq 1.
\end{equation}
The following recurrence formula 
\begin{equation}\label{markopasa} 
	h(k,i) = \gamma(k,i)+\sum_{j=i+1}^{k-1} h_{\rm o}^{(2)}(k,j) h(j,i),\;\;k>i+1
\end{equation}
initialized with $h(k,k-1)=\gamma(k,k-1)$ was derived in \cite{Majji&Juang&Junkins:2010}. For the
deadbeat observers, the constraints $h_{\rm o}^{(s)}(k,i)=0$ for $s=1,2$ and $k-i >2n-1$ are 
invoked in (\ref{gammaki}) and (\ref{markopasa}).

Up to a topological equivalence,  the quadruples $A(k)$, $b(k)$, $c^T(k)$, $d(k)$  may be recovered from the system 
Markov parameters if (\ref{ssx})--(\ref{varphit}) is {\em uniform} \cite{Shokoohi&Silverman:1987}. Two LTV 
realizations $(A_1(k),b_1(k),c_1^T(k),d_1(k))$ and $(A_2(k),b_2(k),c_2^T(k),d(k))$ have the 
same Markov parameters if they are {\em topologically equivalent}, that is, if there exists a bounded matrix 
$T(k) \in \mathbb{R}^{n \times n}$ with a bounded inverse $T^{-1}(k)$ such that for all $k \in \mathbb{N}$,
\begin{eqnarray*}
	A_2(k) &=& T(k+1) A_1(k) T^{-1}(k), \\
	b_2(k) &=& T(k+1) b_1(k) \\
	c_2^T(k) &=& c^T_1(k)T^{-1}(k), \\
	d_2(k) &=& d_1(k).
\end{eqnarray*}
The transformation $T(k)$ with bounded inverse $T^{-1}(k)$ is called a {\em Lyapunov transformation}. 
As far as the input-output behavior of an LTV system is concerned, it suffices to estimate $h(k,i)$ 
from the input-output measurements. 

Next, we estimate $g(k)$ by a two-step procedure. In the first step, the {\em observer 
	gain Markov parameters} defined by 
\begin{equation}\label{defMarkovgain}
	h_{\rm m}(k,i) = c^T(k) \Phi(k,i+1) g(i), \;\; k > i+1
\end{equation}
and $h_{\rm m}(i+1,i)=c^T(i+1) g(i)$ are estimated from the observer Markov parameters using the 
recurrence formula 
\begin{equation}\label{markopasam}
	h_{\rm m}(k,i) = -h_{\rm o}^{(2)}(k,i)+\sum_{j=i+1}^{k-1} h_{\rm o}^{(2)}(k,j) \, h_{\rm m}(j,i)
\end{equation}
for $k>i+1$ initialized by $h_{\rm m}(k,k-1)=-h_{\rm o}^{(2)}(k,k-1)$ \cite{Majji&Juang&Junkins:2010}. 
The second step consists of estimating $g(k-1)$ from (\ref{defMarkovgain}). Given $q \in \mathbb{N}$, which
will be fixed later as $q=2n$, concatenate the equations in (\ref{defMarkovgain}) and notice that
\begin{eqnarray}
	\left[\begin{array}{c} h_{\rm m}(k,k-1) \\ \vdots \\ h_{\rm m}(k+q-1,k-1) \end{array}\right] \hspace{3cm} {} \nonumber  \\
	= \left[\begin{array}{c} c^T(k) \\ \vdots \\ c^T(k+q-1)\Phi(k+q-1,k) \\ \end{array}\right] g(k-1) \nonumber 
	\\[-1ex]  \label{Kespre} \\[-1ex]
	= {\mathcal O}_q(k) g(k-1) \nonumber
\end{eqnarray}
where ${\mathcal O}_q(k)$ is the extended observability matrix of (\ref{ssx})--(\ref{ssy})
at $k$. Compute $\hat{O}_q(k)$ for $(\hat{A}(k),\hat{c}^T(k))$ from the realization 
algorithm outlined in Section~\ref{realization} and set
\begin{equation}\label{Kes}
	\hat{g}(k-1) =\hat{\mathcal O}_q^\dag(k) \left[\begin{array}{c} h_{\rm m}(k,k-1) \\ \vdots \\ 
		h_{\rm m}(k+q-1,k-1) \end{array}\right].
\end{equation}
Then, $\hat{g}(k-1)=T(k) g(k-1)$. 

The results in this subsection reduce the identification of the state-space models from the input-output data to 
the estimation of the observer Markov parameters from the input-output data. We outline the above derivations in the 
form of an algorithm. Execution of Step~7 requires $q$-times applications of Steps~1--6. We will 
not apply Algorithm~1 to $[2n\;\;N]$, but to its some specific subsets The above results are summarized in the following.

\begin{table}[h!]
	\small
	\begin{center}
		\begin{tabular}{l}
			\hline 
			{\bf Algorithm~1.} SARX to LTV model conversion \\
			\hline
			{\bf Input:}  $h_{\rm o}(k,i)$, $k-2n <  i \leq k$ \\ 
			1:  Initialize $h(k,k)=h_{\rm o}^{(1)}(k,k)$ \\
			2:  Calculate $\gamma(k,i)$ from (\ref{gammaki}) while $k-2n < i < k$  \\
			3:  Set $h(k,k-1)=\gamma(k,k-1)$ \\
			4:  Estimate $h(k,i)$ from (\ref{markopasa}) 	\\
			5:  Set $h_{\rm m}(k,k-1)=-h_{\rm o}^{(2)}(k,k-1)$ \\
			6:  Estimate $h_{\rm m}(k,i)$ from (\ref{markopasam})  while $k-2n < i <k-1$  \\
			7:  Estimate $g(k-1)$ from (\ref{Kes}) \\
			{\bf Outputs:} $h(k,i)$ and $g(k-1)$ for $i\leq k$  \\
			\hline
		\end{tabular}
	\end{center}
\end{table}	

\begin{lemma}\label{dwelprop}	
	Algorithm~1 back transforms the SARX model (\ref{yrespobsv22}) to the LTV model (\ref{ssx})--(\ref{varphit}).
\end{lemma}

The SARX model is over-parameterized to accommodate the discrete state changes at the switches. The switches 
are not known {\em a priori}. Once they are located, fewer parameters may be used. Zero-padding will only 
require richer inputs for parameter identifiability as opposed to the parsimonious models. In the rest of this 
subsection, we show that (\ref{ssx})--(\ref{varphit}) subject to Assumption~\ref{sysasmp} and $\delta_*(\chi)\geq n$ 
is {\em uniform} on the interval $[2n+1\;\;N-2n+1]$. This means that (\ref{ssx})--(\ref{varphit}) is {\em uniformly bounded},
{\em uniformly observable}, and {\em uniformly controllable}. Recall that (\ref{ssx})--(\ref{varphit}) is uniformly 
controllable if there exist $\kappa_{\rm c}, \delta_{\rm c}\in \mathbb{N}$ and $\alpha_0,\alpha_1,\beta_0,\beta_1>0$ 
such that for all $k >  \delta_{\rm c}$, 
\begin{enumerate}
	\item $G_{\rm c}(k,\kappa_{\rm c}) > 0$, 
	\item $\alpha_0 \, I_n \leq G_{\rm c}^{-1} (k,\kappa_{\rm c}) \leq \alpha_1 \, I_n$,
	\item $\beta_0 \, I_n \leq \Phi^T(k,k-\kappa_{\rm c}) G_{\rm c}^{-1} (k,\kappa_{\rm c})
	\Phi(k,k-\kappa_{\rm c}) \leq \beta_1 \, I_n$
\end{enumerate}
where the notation $X \geq 0$ ($X>0$) means that $X$ is a square and positive semi-define (positive definite) 
matrix and  
\[
G_{\rm c}(k,\kappa_{\rm c}) = \sum_{j=k-\kappa_{\rm c}}^{k-1}\Phi(k,j+1) b(j) b^T(j) \Phi^T(k,j+1). 
\]
Likewise, (\ref{ssx})--(\ref{varphit}) is uniformly observable if there exist a $\kappa_{\rm o} \in \mathbb{N}$ 
and  $\alpha_0^\prime,\alpha_1^\prime, \beta_0^\prime,\beta_1^\prime>0$ such that for all $k \leq \delta_{\rm o}$,
\begin{enumerate}
	\item $G_{\rm o}(k,\kappa_{\rm o}) > 0$,
	\item $\alpha_0^\prime \, I_n \leq G_{\rm o}^{-1} (k,\kappa_{\rm o}) \leq \alpha_1^\prime \, I_n$, 
	\item $\beta_0^\prime \, I_n \leq \Phi^T(k+\kappa_{\rm o},k) G_{\rm o}^{-1} (k,\kappa_{\rm o}) 
	\Phi(k+\kappa_{\rm o},k) \leq \beta_1^\prime \, I_n$
\end{enumerate}
where
\[
G_{\rm o}(k,\kappa_{\rm o}) = \sum_{j=k}^{k+\kappa_{\rm o}-1} \Phi^T(j,k) c(j) c^T(j) \Phi(j,k).
\]

\begin{lemma}\label{lemreal}
	Suppose Assumption~\ref{sysasmp} holds and $\delta_*(\chi) \geq n$.	Then, (\ref{ssx})--(\ref{varphit})
	is uniform on $(\delta_{\rm c}\;\;\delta_{\rm o}]$ with $\kappa_{\rm o}=\kappa_{\rm c}=\delta_{\rm c}=2n$, 
	and $\delta_{\rm o}=N-\kappa_{\rm o}+1$. 
\end{lemma}

{\em Proof.} See Appendix~\ref{appB}.

We capture the requirements on $\varphi$ in the following. 

\begin{assumption}\label{kereste}
	The switching sequence $\varphi$ for the SLS model (\ref{ssx})--(\ref{varphit}) satisfies
	$\delta_*(\chi) \geq n$, $\delta_0(\chi) \geq \kappa_{\rm c}$, and $k_{i^*} \leq \delta_{\rm o}$.  
\end{assumption}
The discrete states for $k \leq \delta_{\rm c}$ and $k>\delta_{\rm o}$ are minimal from 
$\delta_0(\chi) \geq \kappa_{\rm c}$ and $k_{i^*} \leq \delta_{\rm o}$ since they are similar to the discrete 
states at $k=\delta_{\rm c}+1$ and $k=k_{i^*}$. From Lemma~\ref{lemreal}, $G_{\rm o}(k,\kappa_{\rm o})>0$ and if 
$q\geq \kappa_{\rm o}$, ${\mathcal O}_q(k)$ will have full rank. Hence, $\hat{\mathcal{O}}_q^\dag(k)$ is well-defined since $\hat{\mathcal{O}}_q(k)=T^{-1}(k)\mathcal{O}_q(k)$. 

\subsubsection{SLS realization from Hankel matrix pairs}\label{realization}

For $\delta_{\rm c} < k \leq \delta_{\rm o}$, we define a nested sequence of the {\em Hankel} matrices built from
the system Markov parameters
\begin{equation}\label{gHankel}
	{\mathcal H}(k) = \left[ \begin{array}{ccc} h(k,k-1) & \cdots & h(k,k-\kappa_{\rm c}) \\  
		& \ddots & \vdots \\  & \cdots & h(k+\kappa_{\rm o}-1,k-\kappa_{\rm c})   \end{array} \right]
\end{equation}
and factorize them as ${\mathcal H}(k) = {\mathcal O}_{\kappa_{\rm o}}(k){\mathcal R}_{\kappa_{\rm c}}(k-1)$ 
where 
\[
{\mathcal R}_{\kappa_{\rm c}}(k-1)=[b(k-1)\;\cdots\;\Phi(k,k-\kappa_{\rm c}+1) b(k-\kappa_{\rm c})]
\]
is the {\em extended controllability matrix} \cite{Shokoohi&Silverman:1987}. We have already seen that 
${\rm rank}({\mathcal H}(k))=n$ for all $k$. Define the shift matrices of row up and down and column 
left and right by $J_{\rm U}=[0\;\;I_{\kappa_{\rm o}-1}]$ and $J_{\rm D} = [I_{\kappa_{\rm o}-1} \;\;0]$ 
and $J_{\rm L}=[0 \;\;I_{\kappa_{\rm c}-1}]^T$ and $J_{\rm R}=[I_{\kappa_{\rm c}-1}\;\;0]^T$. From the 
factorization formula with $k$ and $k+1$ plugged in, we retrieve $A(k)$ from either of the two formulas
\begin{eqnarray*}
	J_{\rm U} {\mathcal O}_{\kappa_{\rm o}}(k) &=& \left(J_{\rm D} {\mathcal O}_{\kappa_{\rm o}}
	(k+1) \right) A(k), \label{adet1} \\
	A(k)\left({\mathcal R}_{\kappa_{\rm c}}(k-1) J_{\rm R} \right) &=& {\mathcal R}_{\kappa_{\rm c}}
	(k) J_{\rm L} \label{adet2}
\end{eqnarray*}
as follows
\begin{eqnarray*}
	A(k) &=& \left(J_{\rm D} {\mathcal O}_{\kappa_{\rm o}}(k+1) \right)^{\dag} J_{\rm U} 
	{\mathcal O}_{\kappa_{\rm o}}(k) \\
	&=& {\mathcal R}_{\kappa_{\rm c}}(k) J_{\rm L}\left({\mathcal R}_{\kappa_{\rm c}}
	(k-1) J_{\rm R} \right)^{\dag}
\end{eqnarray*}
where $X^\dag$ denotes the unique pseudo-inverse of a given full-rank matrix $X$ defined as 
$(X^TX)^{-1}X^T$. To determine ${\mathcal O}_{\kappa_{\rm o}}(k)$ and ${\mathcal O}_{\kappa_{\rm o}}
(k+1)$. We apply SVD to ${\mathcal H}(k)$ and ${\mathcal H}(k+1)$
\begin{eqnarray}
	{\mathcal H}(k) &=& U_{\kappa_{\rm o}}(k) \Sigma(k) V_{\kappa_{\rm c}}^T(k) \nonumber 
	\\[-1ex] \label{svdHqr} \\[-1ex]
	{\mathcal H}(k+1) &=& U_{\kappa_{\rm o}}(k+1) \Sigma(k+1) V_{\kappa_{\rm c}}^T(k+1) \nonumber
\end{eqnarray}
and let
\begin{eqnarray}
	\hat{\mathcal O}_{\kappa_{\rm o}}(k) &=&  U_{\kappa_{\rm o}}(k) \Sigma^{1/2}(k), \nonumber \\
	\hat{\mathcal R}_{\kappa_{\rm c}}(k-1) &=&  \Sigma^{1/2}(k) V_{\kappa_{\rm c}}^T(k);
	\nonumber
	\\[-1ex] \label{Oqr} \\[-1ex]
	\hat{\mathcal O}_{\kappa_{\rm o}}(k+1)&=&  U_{\kappa_{\rm o}}(k+1) \Sigma^{1/2}(k+1), \nonumber \\
	\hat{\mathcal R}_{\kappa_{\rm c}}(k) &=& \Sigma^{1/2}(k+1) V_{\kappa_{\rm c}}^T(k+1). \nonumber
\end{eqnarray}
They provide estimates of ${\mathcal O}_{\kappa_{\rm o}}(k)$ and ${\mathcal O}_{\kappa_{\rm o}}(k+1)$ 
for some transformations $T(k)$ and $T(k+1)$ as ${\mathcal O}_{\kappa_{\rm o}}(k)=
\hat{\mathcal O}_{\kappa_{\rm o}}(k)T(k)$ and ${\mathcal O}_{\kappa_{\rm o}}(k+1) =
\hat{\mathcal O}_{\kappa_{\rm o}}(k+1)T(k+1)$. Let
\begin{equation}\label{Aes}
	\hat{A}(k)=(J_{\rm D} \hat{\mathcal O}_{\kappa_{\rm o}}(k+1))^{\dag} J_{\rm U} 
	\hat{\mathcal O}_{\kappa_{\rm o}}(k)
\end{equation}
so that
\begin{equation}\label{topA}
	\hat{A}(k) = T(k+1) A(k) T^{-1}(k).
\end{equation}
The estimation of $b(k)$ and $c^T(k)$ is in order. Let
\begin{eqnarray} 
	J_{\rm c} &=& [1 \;0 \;\cdots\;0], \nonumber
	\\[-1.5ex]  \label{CBdef}   \\[-1.5ex]
	J_{\rm b} &=& J_{\rm c}^T. \nonumber
\end{eqnarray}
Then,
\begin{eqnarray}
	c^T(k) &=& J_{\rm c} {\mathcal O}_{\kappa_{\rm o}}(k), \nonumber
	\\[-1ex]    \\[-1ex] 
	b(k) &=& {\mathcal R}_{\kappa_{\rm c}}(k) J_{\rm b}. \nonumber
\end{eqnarray}
As the estimates of $c^T(k)$ and $b(k)$, we set
\begin{eqnarray}
	\hat{c}^T(k) &=& J_{\rm c} \hat{\mathcal O}_{\kappa_{\rm o}}(k), \nonumber
	\\[-1.5ex]\label{CBes} \\[-1.5ex] 
	\hat{b}(k) &=& \hat{\mathcal R}_{\kappa_{\rm c}}(k) J_{\rm b}. \nonumber
\end{eqnarray}
Then, from $ \hat{\mathcal R}_{\kappa_{\rm c}}(k)=T(k+1) {\mathcal R}_{\kappa_{\rm c}}(k)$
\begin{eqnarray}
	\hat{c}^T(k) &=& c^T(k) T^{-1}(k), \nonumber
	\\[-1.5ex] \label{topCB} \\[-1.5ex]
	\hat{b}(k) &=& T(k+1) b(k). \nonumber
\end{eqnarray}
Setting $\hat{d}(k)=h(k,k)$, we see that $(\hat{A}(k),\hat{B}(k),\hat{C}(k),\hat{D}(k))$ is 
similar to ${\mathcal P}_{\varphi(k)}$. The steps above are summarized as a (point-wise in time) 
realization algorithm.

\begin{table}[h!]
	\small
	\begin{center}
		\begin{tabular}{l}
			\hline 
			{\bf Algorithm~2. LTV to SLS model conversion}  \\
			\hline
			{\bf Inputs:}  $h(l+s-1,l-t)$, $1 \leq t \leq \kappa_{\rm c}$; $1 \leq s \leq \kappa_{\rm o}$;
			$l=k,k+1$  \\ 
			1:  Calculate ${\mathcal H}(k)$ and ${\mathcal H}(k+1)$ from (\ref{gHankel}) \\
			2:  Apply SVD to ${\mathcal H}(k)$ and ${\mathcal H}(k+1)$ as in (\ref{svdHqr})	\\
			3:  Estimate ${\mathcal O}_{\kappa_{\rm o}}(k)$ and ${\mathcal O}_{\kappa_{\rm o}}(k+1)$ 
			from (\ref{Oqr}) \\
			4:  Estimate $A(k)$ from (\ref{Aes}) \\
			5:  Estimate $c^T(k)$ and $b(k)$ from (\ref{CBes}) \\
			6:  Set $\hat{d}(k)=h(k,k)$ \\
			{\bf Outputs:} $\hat{A}(k),\hat{b}(k),\hat{c}^T(k),\hat{d}(k)$.  \\
			\hline
		\end{tabular}
	\end{center}
\end{table}

\subsection{The switches of the SARX model}\label{switchsubsec}	

Stack the observer Markov parameters and the input-output data into the parameter and the regression vectors
defined as 
\begin{eqnarray}
	\theta(k) &=& (d(k)\; h_{\rm o}(k,k-1)\;\cdots \;h_{\rm o}(k,k-2n+1))^T, \nonumber
	\\[-1ex] \label{ARtheta} \\[-1ex]
	z(k) &=& (u(k)\;\zeta^T(k-1)\;\cdots\;\zeta^T(k-2n+1))^T \nonumber
\end{eqnarray}
and write (\ref{yrespobsv222}) in the form 
\begin{equation}\label{yt}
	y(k)=z^T(k) \theta(k), \qquad \delta_{\rm c} <k\leq \delta_{\rm o}.
\end{equation}
Recall from Lemma~\ref{deadbeatlem} that $\tau < 2n$ and $\tau=n$ on the union of intervals $[k_i+n\;\;k_{i+1})$,
$1\leq i<i^*$. The latter means that 
\begin{equation}\label{thetared}
	\theta(k)=(d(k)\; h_{\rm o}(k,k-1)\;\cdots \;h_{\rm o}(k,k-n)\;\;0\;\cdots\;0)^T
\end{equation}
on this set. The switches of (\ref{yt}) are the instants $k$ satisfying
\begin{equation}\label{firstorder}
	\delta_\theta (k)= \theta(k) -\theta(k-1) \neq 0.
\end{equation}
Let $\varphi_\theta(k)$ and 
$\{s_1,\cdots,s_{j^*}\}$ denote the switching sequence and its range where $s_0=2n$ so that 
$\delta_{\rm c} <s_{j}\leq \delta_{\rm o}$ for all $j \leq j^*$. For a given segmentation of 
$(\delta_{\rm c}\;\;\delta_{\rm o}]$, {\em i.e.}, covering by semi-closed disjoint intervals 
$\chi_\theta$ as in (\ref{varphit}), we allocate a dwell sequence $\delta_j(\chi_\theta)=s_{j+1}-s_j$, 
$0 \leq j <j^*$ and define the discrete state parameter set as ${\mathcal P}^\theta=\{\theta(k): \, 
\delta_{\rm c} < k \leq \delta_{\rm o} \}$.

We will present several lemmas to link the segments in the SLS model (\ref{ssx})--(\ref{varphit}) to the segments 
in the observer model (\ref{yt}) and vice versa via Lemma~\ref{deadbeatlem}. The link we provide is partial, 
that is, it does not carry complete information on $\chi_\theta$ and ${\mathcal P}^\theta$; yet, this partial 
information will be sufficient to retrieve $\chi$ and ${\mathcal P}$ from the input-output data.

\begin{lemma}\label{ARXdwell}
	Consider the SLS model (\ref{ssx})--(\ref{varphit}). Let the observer be as in Lemma~\ref{deadbeatlem}. 
	If  $k_i,k_{i+1} \in \chi$ and $\delta_i(\chi)>n$, then $[k_i+n\;\;k_{i+1}) \subseteq [s_j\;\;s_{j+1})$
	for some $s_j,s_{j+1} \in \chi_\theta$.
\end{lemma} 

{\em Proof.} See Appendix~\ref{appC}. 

A converse statement is also true. But, first we need an auxiliary result that holds
in the more general LTV setting without requiring the observer be as in Lemma~\ref{deadbeatlem}.

\begin{lemma}\label{ARXdwell2}
	Consider the observer (\ref{yt}) for the SLS model (\ref{ssx})--(\ref{varphit}). Then,
	$h$ and $h_m$ are locally shift-invariant, {\em that is}, $h(k,k-v)=h(l,l-v)$ and	
	$h_{\rm m}(k,k-v)=h_{\rm m}(l,l-v)$ for all $k,l$, and $v\geq 0$ 
	with $s_j+v \leq \min\{k,l\}$ and $\max\{k,l\}< s_{j+1}$.
\end{lemma}

{\em Proof.} See Appendix~\ref{appD}.

Let $\mathcal{H}_{\alpha \beta}(k)$ denote the the entry of $\mathcal{H}(k)$ in (\ref{gHankel}) with 
the row and the column indices $\alpha,\beta$. Recalling $\kappa_{\rm c}=\kappa_{\rm o}=2n$, for all 
$1 \leq \alpha \leq \kappa_{\rm o}$ and $1 \leq \beta \leq \kappa_{\rm c}$, we will require 
\begin{eqnarray*}
	\mathcal{H}_{\alpha\beta}(k) &=&h(k+\alpha-1,k-\beta) \\
	&=& h(k+\alpha,k+1-\beta) = \mathcal{H}_{\alpha\beta}(k+1).
\end{eqnarray*}
Set $l=k+\alpha$ and $l-v=k+1-\beta$ from which we derive $v=\alpha+\beta-1$. We get $s_j+4n-1 \leq k$ with 
$\alpha=\beta=2n$ from the first constraint and $k<s_{j+1}-2n$ with $\alpha=2n$ from the second constraint 
$k+\alpha <s_{j+1}$. It follows from Lemma~\ref{ARXdwell2} that the LTV system is locally shift-invariant on
the interval $[\alpha_j\;\;\tilde{\beta}_j)$ if $\delta_j(\chi_\theta) \geq 6n$ where
\begin{equation}
	\alpha_j=s_j+4n-1 \;\;\; \mbox{and} \;\;\; \tilde{\beta}_j=s_{j+1}-2n.
\end{equation}
Hence, if $k \in [\alpha_j\;\;\tilde{\beta}_j)$, then $\mathcal{H}(k)=\mathcal{H}(k+1)$. Moreover,  $d(k)=d(k+1)$
from Lemma~\ref{ARXdwell2}. Algorithm~2 estimates a realization $(A(k),b(k),c^T(k),d(k))$ of 
(\ref{ssx})--(\ref{varphit}) up to a topological equivalence from the pair $\mathcal{H}(k)$ and $\mathcal{H}(k+1)$. 
Since $T(k)=T(k+1)$, $\varphi(k)=\varphi(\alpha_j)$ for all $k \in [\alpha_j,\tilde{\beta}_j)$. Let 
\[
k_i= \max_{k \in \chi}\{k \leq \alpha_j\}, \;\;\; k_\ell= \min_{k \in \chi}
\{k \geq \tilde{\beta}_j\}.
\]
Suppose $l>i+1$. From the definitions of $k_\ell$ and $k_i$, we get 
$k_i \leq \alpha_j < k_{i+1} < \tilde{\beta}_j \leq k_\ell$. Recall that $\varphi(k)$ is 
constant on $[\alpha_j\;\;\tilde{\beta}_j)$. Since $k_{i+1}$ is an interior point of 
$[\alpha_j\;\;\tilde{\beta}_j)$, $\varphi(k_{i+1})=\varphi(\alpha_j)$. But, 
$\alpha_j \in [k_i\;\;k_{i+1})$. Then, from the equalities $\varphi(k_i)=\varphi(\alpha_j)$ 
and $\varphi(k_i)=\varphi(k_{i+1})$ we reach a contradiction since $k_{i+1} \in \chi$. 
Therefore, $l \leq i+1$. Suppose $l \leq i$. From the chain of the inequalities 
$\tilde{\beta}_j \leq k_l \leq k_i \leq \alpha_j$, we get $\delta_j(\chi_\theta)<6n$ 
reaching a contradiction. Hence, $l=i+1$. The definitions of $k_i$ and $k_l$ yield 
$\tilde{\beta}_j \leq k_{i+1}$ and $k_i \leq \alpha_j$ with $l=i+1$. Reorganize these 
inequalities as
\[
[\alpha_j\;\;\tilde{\beta}_j) \subset [k_i\;\;k_{i+1}) \;\;\;\mbox{and}\;\;\; \delta_j(\chi_\theta)\geq 6n.
\]
The observer gain $g(k-1)$ is estimated from (\ref{Kes}) using  $\hat{c}(\ell)$, $\hat{A}(\ell)$, 
$k \leq \ell \leq k+2n-2$ and  $\hat{c}(k+2n-1)$. Algorithm~2 estimates $c(k+2n-1)$ from the pair $\mathcal{H}(k+2n-1)$ and $\mathcal{H}(k+2n)$. The estimation of the observer gain Markov parameters 
$h_{\rm m}(k+v,k-1)$, $0 \leq v <2n$ does not require additional data since (\ref{markopasam}) uses the present 
and the past samples $h_{\rm o}^{(2)}(k+v,i)$, $i \leq k+v$. Thus, Lemma~\ref{ARXdwell2} will be in use if 
$k+4n-1<s_{j+1}$. Hence, the left-hand side of (\ref{Kespre}) does not depend on $k$ and $\hat{g}(k-1)$ is 
constant if $k \in [\alpha_j\;\;\beta_j]$ where $\beta_j=\tilde{\beta}_j-2n$.

We summarize the results derived above in the following. 
\begin{lemma}\label{cavit}
	Consider the SLS model (\ref{ssx})--(\ref{varphit}) and the observer (\ref{yt}). 
	If  $s_j,s_{j+1} \in \chi_\theta$ and $\delta_j(\chi_\theta) \geq 6n$, then for some $k_i,k_{i+1} \in \chi$, 
	$[\alpha_j\;\;\tilde{\beta}_j) \subseteq [k_i\;\;k_{i+1})$ and if $\delta_j(\chi_\theta) \geq 8n-1$, then 
	$\hat{g}(k-1)$ is constant for all  $k \in [\alpha_j\;\;\beta_j]$. 
\end{lemma}
Similarly to Lemma~\ref{ARXdwell2}, this lemma holds in the LTV setting, without requiring the observer be 
as in Lemma~\ref{deadbeatlem}. The proofs of Lemmas ~\ref{ARXdwell2}--\ref{cavit} are far more difficult than 
that of Lemma~\ref{ARXdwell} due to the possibility $\tau \neq n$.
We state $\chi \subset \chi_\theta$ as a standing assumption and seek conditions which will imply this
assumption. The converse statement, {\em i.e.,} $\chi_\theta \subset \chi$ is not true.

\begin{assumption}\label{main11}
	Every switch of $\chi$ is also a switch of  $\chi_\theta$. 	
\end{assumption}

Using Assumption~\ref{main11}, we improve Lemma~\ref{cavit} as follows.
\begin{lemma}\label{cavit2}
	Consider the SLS model (\ref{ssx})--(\ref{varphit}) and the observer in Lemma~\ref{deadbeatlem}. 
	Suppose that Assumption~\ref{main11} holds. Let $s_j,s_{j+1} \in \chi_\theta$ be with 
	$\delta_j(\chi_\theta) \geq 5n$. Then, there exist $k_i,k_{i+1} \in \chi$  satisfying 
	$0 \leq s_j-k_i < 4n$ and $k_{i+1}=s_{j+1}$.
\end{lemma}

{\em Proof.} See Appendix~\ref{appE}.

Lemma~\ref{cavit2} guarantees a $k_i \in \chi$ satisfying $0\leq s_j-k_i <4n$ when Assumption~\ref{main11} 
holds. Again from Assumption~\ref{main11}, $k_i=s_l\in\chi_\theta$ for some $l \in \mathbb{N}$, but 
we do not assert $l=j-1$. A result that follows from this observation is if $\delta_j(\chi_\theta) \geq 5n$, 
then $\delta_{j-1} < 4n$, {\em that is}, a long constant parameter interval is always preceded by a shorter 
one  when Assumption~\ref{main11} holds. The observer assumption in Lemma~\ref{cavit2} is crucial.

We will state two assumptions implying Assumption~\ref{main11}. To this end, write 
(\ref{yt}) as a sum of the filtered inputs and outputs
\begin{eqnarray}
	y(k) &=&   y^{(1)}(k)+y^{(2)}(k) \nonumber
	\\[-1.5ex]  \label{yt2}   \\[-1.5ex]
	&=&   H_{\rm o}^{(1)}(z;k) u(k) +H_{\rm o}^{(2)}(z;k) y(k) \nonumber
\end{eqnarray}
where with $b_{\rm o}^{(1)}(k)=b(k)+g(k)d(k)$ and $b_{\rm o}^{(2)}(k)=-g(k)$, 
\begin{eqnarray}
	H_{\rm o}^{(1)}(z;k) &=& c^T(k) (zI_n-A_{\rm o}(k))^{-1} b_{\rm o}^{(1)}(k)+ d(k),\label{cukkada1} \\
	H_{\rm o}^{(2)}(z;k) &=& c^T(k) (zI_n-A_{\rm o}(k))^{-1}b_{\rm o}^{(2)}(k). \label{cukkada2}
\end{eqnarray}
From (\ref{cukkada2}), observe that the second term in (\ref{yt2}) is a linear combination of the past 
outputs only. Since $A_{\rm o}(k)$ has all eigenvalues at zero for every $k$, $H_{\rm o}^{(1)}(z;k)$ and
$H_{\rm o}^{(2)}(z;k)$ are time-varying moving-average filters.

\begin{assumption}\label{deadminimal}
	At least one of the following is true:
	\begin{enumerate}
		\item[a.] For every $k \in \chi$, $H_{\rm o}^{(1)}(z;k)$ is minimal, 
		\item[b.] For every $k \in \chi$, $H_{\rm o}^{(2)}(z;k)$ is minimal.  
	\end{enumerate}  
\end{assumption}

Assumption~\ref{deadminimal} holds for almost all discrete state sets and deadbeat observers.
It is a technical assumption required for the consistency of the identification scheme presented 
in the sequel. As $k$ changes in $\chi$, a set of transfer functions of MacMillan degree $n$ is  
generated by  (\ref{cukkada1})--(\ref{cukkada2}) with a cardinality bounded by $2\sigma$. We present 
three numerical examples to illustrate the minimality properties of the transfer functions 
defined in (\ref{cukkada1})--(\ref{cukkada2}). 

\begin{example}
	Let $(A,c^T)$ in Example~\ref{ex1} be the observability pair of a discrete state $(A,b,c^T,d)$. 
	Since $(A,c^T)$ is observable, $(A_{\rm o},c^T)$ is also observable. To check controllability, we
	calculate the observability matrix of $(A,b)$
	\[
	[b \;\;Ab]=\left[\begin{array}{cc} b_1 & b_1+b_2 \\ b_2 & 0 \end{array}\right].
	\]
	Thus, $(A,b)$ is controllable if and only if $b_2 \neq 0$ and $b_1 \neq b_2$. Both conditions
	are necessary by Assumption~\ref{sysasmp}. A simple calculation reveals 
	$H_{\rm o}^1(z) = d+(2b_2+b_1-d)z^{-1}-b_2z^{-2}$ 
	minimal and $H_{\rm o}^2(z)=z^{-2}z$ not. Plug $b_1=-b_2$ in $H_{\rm o}^1(z)$ to see
	it is minimal while $(A,b)$ is uncontrollable. 
\end{example}

\begin{example}
	Let $(A,c^T)$ in Example~\ref{ex2} be the observability pair of $(A,b,c^T,d)$. It is easy to verify that $(A,b)$ 
	is controllable iff $b_2 \neq 0$, $H_{\rm o}^1(z) = d+(b_1+b_2)z^{-1}+b_2 z^{-2}$, and $H_{\rm o}^2(z)=0$. 
	Thus, $H_{\rm o}^1(z)$ is minimal. 
\end{example}

\begin{example} \label{ex3.5}
	Let $c^T=[1 \;\;1]$, $d=1$, and
	\[
	A=\left[\begin{array}{cc}1/2 & 0 \\ 1 & -1/2   \end{array}\right], \qquad b=\left[\begin{array}
		{c} 1 \\ 2 \end{array}\right].
	\]
	It is easy to verify $(A,b,c^T,d)$ minimal, $g=[-1/8 \;\; 1/8]^T$, $H_{\rm o}^1(z) = 1+3z^{-1}+0.25z^{-2}$,
	and $H_{\rm o}^2(z)=-0.25z^{-2}$. Both $H_{\rm o}^1(z)$ and $H_{\rm o}^2(z)$ are minimal. 
\end{example}

\begin{lemma}\label{lemreal2}
	Suppose Assumptions~\ref{sysasmp}, \ref{kereste}, and \ref{deadminimal} hold. Then, on 
	$(\delta_{\rm c},\delta_{\rm o}]$ with 
	$\kappa_{\rm c}=\kappa_{\rm o}=2n$ and $g(k)$ in Lemma~\ref{deadbeatlem}, 
	$H_{\rm o}^{(1)}(q;k):\,u \mapsto y^{(1)}$ or/and $H_{\rm o}^{(2)}(q;k):\,y \mapsto y^{(2)}$ is uniform. 
\end{lemma}

{\em Proof.} This lemma is actually a corollary of Lemma~\ref{lemreal}.

\begin{assumption}\label{distinctas}
	At least one of the following
	\begin{enumerate}
		\item[i.] $\varphi(k) \neq \varphi(l) \Leftrightarrow  c(k) \neq c(l) \;\;\mbox{\rm and/or}\;\; d(k) \neq d(l)$
		\item[ii.] $\varphi(k) \neq \varphi(l) \Leftrightarrow  c(k)  \neq c(l)$
	\end{enumerate}
	holds for all $k,l \in (\delta_{\rm c},\delta_{\rm o}]$.
\end{assumption}

\begin{lemma}~\label{ARXdwell3}
	Consider the SLS model (\ref{ssx})--(\ref{varphit}) with the observer in Lemma~\ref{deadbeatlem}. Suppose 
	$\delta_*(\chi)> 2n$. Then, Assumptions~\ref{deadminimal}a.--\ref{distinctas}i. or 
	\ref{deadminimal}b.--\ref{distinctas}ii. imply Assumption~\ref{main11}.
\end{lemma}

{\em Proof.} See Appendix~\ref{appF}. 

\subsection{Switch identifiability of the SARX model} \label{switchidsec}

The SLS model (\ref{ssx})--(\ref{varphit}) and the deadbeat observer in Lemma~\ref{deadbeatlem} have the 
same input--output map. We also have $\chi \subset \chi_\theta$ by 
Assumptions~\ref{deadminimal}a.--\ref{distinctas}i. or \ref{deadminimal}b.--\ref{distinctas}ii. 
In this subsection, we will study conditions rendering a switch in 
$\chi_\theta$ identifiable from the noiseless input--output data. 

\begin{definition}\cite{Ozay&Sznaier&Lagoa&Camps:2011}\label{switchdef}
	If whenever $\varphi_\theta(s_{j+1}) \neq \varphi_\theta(s_{j+1}-1)$, it is possible to detect the 
	change in the value of $\varphi_\theta$ as soon as $y(s_{j+1})$ is observed, $s_{j+1} \in \chi_\theta$ 
	is causally identifiable. 
\end{definition}
Two discrete states with parameters $\theta_1,\theta_2 \in {\mathcal P}^\theta$ are one-step indistinguishable 
from $z(k)$ if $z^T(k)(\theta_1 -\theta_2) =0$ \cite{Ozay&Sznaier&Lagoa&Camps:2011}. Given 
$s,t \in (\delta_{\rm c}\;\;\delta_{\rm o}]$
with $t\geq s$, we define
\begin{eqnarray*}
	{\mathcal R}(s,t) &=& [z(s) \;\cdots\;z(t)], \nonumber
	\\[-1.5ex]   \\[-1.5ex]
	Y(s,t) &=& [y(s) \;\cdots\;y(t)]^T. \nonumber
\end{eqnarray*}
Denote the linear space spanned by the columns/rows and the rank of a given matrix $X$ by 
${\rm range}(X)$/${\rm span}(X)$ and ${\rm rank}(A)$. The following result \cite{Ozay&Sznaier&Lagoa&Camps:2011}
provides a pair of necessary and sufficient conditions for switch identifiability.

\begin{lemma} \label{lemsparse2}
	A switch $s_{j+1}$ is causally identifiable from the input-output data if and only if
	\begin{eqnarray} 
		0 &\neq&  z^T(s_{j+1})(\theta(s_{j+1})-\theta(s_j)), \nonumber
		\\[-1ex] \label{identlem2} \\[-1ex]
		z(s_{j+1}) &\in&  {\rm range}({\mathcal R}(s_{j},s_{j+1}-1)). \nonumber 
	\end{eqnarray} 
\end{lemma}
In the notation of this paper, sufficiency proof in \cite{Ozay&Sznaier&Lagoa&Camps:2011} shows that there does not exist 
a vector $\mu \in \mathbb{R}^{4n-1}$ explaining the data in $[k_i+n\;\;k_{i+1}]$, {\em that is}, 
$y(k)=z^T(k) \mu$ for all $k_i+n \leq k \leq k_{i+1}$. Here, we disregarded the subinterval $[k_i\;\;k_i+n]$ 
since $\theta(k)$ may not be constant there. In Section~\ref{discstest}, $\mu$ will represent the solution 
of a sparse optimization problem with the feasibility equations $y(k)=z^T(k) \mu$ for all $k_i+n \leq k < k_{i+1}$ 
satisfied also by the deadbeat observer $y(k)=z^T(k) \theta(n)$, cf. (\ref{thetared}). But from 
Lemma~\ref{ARXdwell}, $\theta(k)$ is constant for all $k \in [k_i+n\;\;k_{i+1})$. If the inputs are persistently 
exciting (PE) and $\delta_i(\chi)$ is large, $\theta(k)=\mu$, see Section~\ref{PEsubsec}. Then, 
$\theta(k_i)=\mu=\theta(k_{i+1})$, arriving to a contradiction. We conclude that identifiability of the switches
is not changed in the optimization problems (\ref{optim1}) and (\ref{optim3}).

By replacing $s_{j+1}$ with $s_j$ and decreasing $k$ instead of increasing it Definition~\ref{switchdef}
can be adapted to $s_j$. Likewise, with slight changes, Lemma~\ref{lemsparse2} can be adapted to $s_j$. The 
conditions in the lemma restrict the class of inputs that can be used for identification. If  
(\ref{ssx})--(\ref{varphit}) is strictly proper, then $d(k)=0$ for all $k$ and the first entry of 
$\theta(k)$ may be deleted out. When $\delta_\theta(k)$ is calculated, possible pole-zero cancellations 
in the polynomials $\theta(k)$ and $\theta(k-1)$ must be taken into account before the zero-padding since they are 
defined in terms of the deadbeat observer Markov parameters. 

Since $\tau=n$ in $[k_i+n\;\;k_{i+1})$, ${\mathcal R}(k_i+n,k_{i+1}-1)$ has no more than $2n+1$ nonzero rows 
and by adding the column vectors $z(k_i+n-t)$, $1 \leq t \leq n$ we cannot make ${\mathcal R}(k_i,k_{i+1}-1)$ full-rank. 
Hence, ${\mathcal R}(k_i,k_{i+1}-1)$ is always rank deficient. If there is no switch of $\chi_\theta$ in $(k_i\;\;k_{i+1})$,
to check if $k_i \in \chi_\theta$ is identifiable from the input-output data substitute $s_j=k_i$ and $s_{j+1}=k_{i+1}$ 
in (\ref{identlem2}) which is equivalent to \cite{Ozay&Sznaier&Lagoa&Camps:2011}
\begin{equation}\label{hopdedikbe}
	{\rm rank}(Y(k_i,k_{i+1}) \;\; {\mathcal R}^T(k_i,k_{i+1})) > {\rm rank}({\mathcal R}(k_i,k_{i+1})).
\end{equation}
This inequality is readily checked by counting the number of the nonzero singular values of a given matrix as a 
measure of its rank. This test tolerates noise with small amplitude. In Section~\ref{numinsec}, we will 
illustrate (\ref{identlem2}) or (\ref{hopdedikbe}) by means of a numerical example using multi-sine excitations.
As we mentioned earlier, $\chi_\theta$ and ${\mathcal P}^\theta$ are very complicated sets. There may even be
a continuum of the switches in the interval $(k_i \;\;k_i+n)$. Replace $k_i$ with a $k \in (k_i \;\;k_i+n)$. The 
largest $k$ satisfying (\ref{hopdedikbe}) is the first switch of $\chi_\theta$ to the left of $k_{i+1}$.
As noted there is an ambiguity in locating $k_i$ using the rank test (\ref{hopdedikbe}) by approaching it from $k_{i+1}$.

\section{Estimation of the discrete-states}\label{discstest}

Evaluating and stacking (\ref{yt}) for $\delta_{\rm c} < k \leq \delta_{\rm o}$, we derive the so-called 
measurement equation 
$Y= Z \mu$
\begin{eqnarray*}
	Y &=&(y(N^\prime)\;\cdots\;y(N^{\prime\prime}))^T, \\
	\mu &=& (\theta^T(N^\prime)\;\cdots\;\theta^T(N^{\prime\prime})^T, \\
	Z &=& \left[\begin{array}{ccc} z^T(N^\prime) & & 0 \\ & \ddots & \vdots \\ 
		0 & \cdots & z^T(N^{\prime\prime}) \end{array} \right] \label{Zdef}
\end{eqnarray*}
where $N^\prime=\delta_{\rm c}+1=2n+1$ and $N^{\prime\prime}=\delta_{\rm o}=N-2n+1$. With $\mu(0)=\theta(2n)$, 
write the first-order differences in (\ref{firstorder}) as
\begin{equation}\label{subs}
	\mu = \left[L \;\; 1_{N^{\prime\prime}-2n} \otimes I \right]\left[\begin{array}{c} 
		\eta \\ \mu(0) \end{array}\right]
\end{equation}
where $1_n$ denotes the column vector of ones in $\mathbb{R}^n$ and $A \otimes B$ is the Kronecker product
of two matrices $A$ and $B$, and
\begin{eqnarray*}
	L &=& \left[\begin{array}{ccc} 1  &  &  0 \\ \vdots & \ddots &  \\ 1 &   \cdots & 1 \end{array}\right] \otimes I, \nonumber
	\\[-1ex]   \\[-1ex]
	\eta &=& \left[\begin{array}{c}\delta_\theta(N^\prime) \\ \vdots \\ \delta_\theta(N^{\prime\prime}) 
	\end{array}\right]. \nonumber
\end{eqnarray*}
Let $\hat{N}=N^{\prime\prime}-2n$, $M=Z [L \;\;1_{\hat{N}} \otimes I]$, and $\zeta=[\eta^T \;\;\mu^T(0)]^T$. 
Substituting $M$ and $\zeta$ in $Y= Z \mu$, we derive $Y = M \zeta$.

Now, let $ \mathfrak{I} \left(\left\|\eta(k)\right\|_2 > 0\right)$ denote the indicator function 
defined $1$ when $\left\|\eta(k)\right\|_2 > 0$ and $0$ otherwise. The $\ell_0$ quasi-norm 
of the sequence $\eta(k)$ is defined by 
\begin{equation}\label{loquas}
	\|\eta\|_{0,\mathfrak{I}} =\sum_{k=1}^{\hat{N}} \mathfrak{I} \left( \left\|\eta(k) \right\|_2 >0 \right). 
\end{equation}

Since $k$ is a switch if and only if $\delta_\theta(k) \neq 0$, $\|\eta\|_{0,\mathfrak{I}}$ is the number of the switches 
of the deadbeat observer in (\ref{yt}) in the interval $[N^\prime\;\;N^{\prime\prime}]$. In the noiseless case, we find 
all identifiable switches by solving the following non-convex and non-polynomial (NP)-hard sparse optimization problem
\begin{equation}\label{optim1}
	\min_{\zeta}  \|\eta\|_{0,\mathfrak{I}} \;\;{\rm subject \;to} \;\;  Y= M\zeta.
\end{equation}
Let $\hat{\zeta}=[\hat{\eta}^T\;\;\hat{\mu}^T(0)]^T$ be a solution of (\ref{optim1}) and 
$\tilde{\zeta}=\hat{\zeta}-\zeta$ denote its error. Likewise, $\hat{\mu}$ is the estimate of $\mu$ 
calculated from (\ref{subs}) with $\hat{\zeta}$. Partition
$\tilde{\zeta}$ similarly to $\zeta$. From $M \tilde{\zeta}=0$, we derive 
$ZL \tilde{\eta}+Z 1_{\hat{N}} \otimes \tilde{\mu}(0)=0$. Thus,
\begin{equation}\label{at1}
	\left[\begin{array}{ccc} z^T(N^\prime) &   &  0 \\ \vdots & \ddots & \vdots \\ 
		z^T(N^{\prime\prime}) & \cdots & z^T(N^{\prime\prime}) \end{array}\right] \tilde{\eta}=
	-\left[\begin{array}{c}z^T(N^\prime) \\ \vdots \\ z^T(N^{\prime\prime})\end{array} \right] 
	\tilde{\mu}(0). 
\end{equation}
Denote the rows of a given matrix $X$ by $X_k$. From 
\begin{eqnarray} 
	\tilde{\eta}(k) &=& \hat{\eta}(k)-\eta(k) \nonumber \\
	&=& \hat{\mu}(k)-\hat{\mu}(k-1)-(\mu(k)-\mu(k-1)) \label{naberlan} \\
	&=& \tilde{\mu}(k)-\tilde{\mu}(k-1), \qquad 1 \leq k \leq \hat{N} \nonumber
\end{eqnarray}
and (\ref{at1}), we derive
\begin{eqnarray*}
	z^T(2n+k) \sum_{l=1}^{k} \tilde{\eta}(l) &=& z^T(2n+k)(\tilde{\mu}(k)-\tilde{\mu}(0)) \\
	&=& -z^T(2n+k) \tilde{\mu}(0).
\end{eqnarray*}
Thus, $z^T(2n+k)\tilde{\mu}(k)=0$ for all $1 \leq k \leq \hat{N}$.   

Consider a segment $[k_i\;\;k_{i+1})$ and suppose Assumptions~\ref{deadminimal}a.--\ref{distinctas}i. or 
\ref{deadminimal}b.--\ref{distinctas}ii. hold. Since $k_i,k_{i+1} \in \chi_\theta$, $g(k-1)$ is constant 
on $[k_i+4n-1\;\;k_{i+1}-4n]$ from Lemma~\ref{cavit} if $\delta_i(\chi) \geq 8n-1$ and $\theta(k)$ is constant on 
$[k_i+n\;\;k_{i+1})$ from Lemma~\ref{ARXdwell} if $\delta_i(\chi)>n$. We assume $\delta_i(\chi) \geq 8n-1$ then.
On $[k_i+4n-1\;\;k_{i+1}-4n]$, $\theta(k)$ is sparse as in (\ref{thetared}). Since this interval is a
subset of $[k_i+n\;\;k_{i+1})$, this sparse form extends to the entire $[k_i+n\;\;k_{i+1})$. Thus,
$y(k)=z_n^T(k)\theta_n(k)$ and ${\mathcal R}_n(s,t)=[z_n(s) \;\cdots\;z_n(t)]$ where
\begin{eqnarray}
	z_n(k) &=& (u(k)\;\;\zeta(k-1)\;\cdots \;\zeta(k-n))^T, \nonumber
	\\[-1ex]  \label{ARthetared}   \\[-1ex]
	\theta_n(k) &=& (d(k)\;\;h_{\rm o}(k,k-1)\;\cdots \;h_{\rm o}(k,k-n))^T. \nonumber
\end{eqnarray}
We define $\mu_n(k)$ and $\tilde{\mu}_n(k)$ similarly to $\theta_n(k)$.

Since $\theta(k)=\theta(k_i+n)$, $k_i+n \leq k <k_{i+1}$, $\eta(k-2n)=0$ for all $k_i+n < k <k_{i+1}$. 
But $\hat{\eta}$ is an optimal solution and $\zeta$ is a feasible solution. Then, $\hat{\eta}(k-2n)=0$ and 
hence $\tilde{\eta}(k-2n)=0$ on $(k_i+n\;\;k_{i+1})$ from the first equality in (\ref{naberlan}). Plug $l$ in place 
of $k$ in the last equality in (\ref{naberlan}) and sum over $l \in (k_i+n\;\;k]$ to get
$\tilde{\mu}(k-2n)=\tilde{\mu}(k_i-n)$. From the constraints $z^T(2n+k)\tilde{\mu}(k)=0$ for all $k=1,\cdots,\hat{N}$, 
then 
\[
{\mathcal R}_n^T(k_i+n+1,k_{i+1}-1) \tilde{\mu}_n(k_i-n)=0. 
\]
If 
\begin{equation}\label{PErank}
	{\rm rank}({\mathcal R}_n(k_i+n+1,k_{i+1}-1))=2n+1,
\end{equation}
then $\tilde{\mu}_n(k_i-n)=0$ and hence $\tilde{\mu}(k_i-n)=0$. Thus, $\hat{\mu}(k-2n)$ is sparse as in 
(\ref{thetared}) and satisfies $\hat{\mu}(k-2n)=\mu(k_i-n)$ for all $k_i+n\leq k<k_{i+1}$. We can now apply 
Algorithms~1--2 to $\hat{\theta}_n(k_i+n)$ to extract a state-space realization of $\varphi(k_i)$. The condition 
(\ref{PErank}) is a PE condition.  

\subsection{Persistence of excitation conditions}\label{PEsubsec}

Let us rewrite (\ref{yt2}) by adding noise $e(k)$
\begin{equation}\label{ARX45}
	(1-H_{\rm o}^{(2)} (z;k)) y(k) =H_{\rm o}^{(1)}(z;k) u(k)+e(k). 
\end{equation}
The one-step-ahead predictors of (\ref{yt}) for all $k \in (k_i+n,k_{i+1})$ are  derived from (\ref{ARX45}) as follows
\begin{equation}\label{predARX}
	\hat{y}(k|k-1) = H_{\rm o}^{(1)} (z;k_i) u(k)+ H_{\rm o}^{(2)} (z;k_i) y(k).
\end{equation}
Plug $k=k_i$ in and transform (\ref{ARX45}) into the frequency domain
\begin{equation}\label{sisomod}
	Y(z)  =  (1-H_{\rm o}^{(2)}(z;k_i))^{-1} H_{\rm o}^{(1)}(z;k_i) U(z) +V(z)
\end{equation}
where $V(z)=(1-H_{\rm o}^{(2)}(z;k))^{-1}E(z)$. 

It is a well-known fact \cite{Verhaegen&Verdult:2007} that when $u(k)$ and $y(k)$, with $u(k)$ independent 
from $e(k)$, and with $e(k)$ persistently exciting of any order, the moving average parameters of 
$H_{\rm o}^{(2)} (z;k_i)$ and $H_{\rm o}^{(2)}(z;k_i)$ in (\ref{predARX}) are uniquely determined by minimizing 
the quadratic norm of the prediction errors if $u(k)$ is PE of order $2n+1$. In the noiseless case, 
$z^n H_{\rm o}^{(1)}(z;k_i)$ and $z^n(1-H_{\rm o}^{(2)}(z;k_i))$ must be co-prime and $u(k)$ be PE of order $2n+1$. 
An ergodic or quasi-stationary sequence $u(k)$ is PE of order $\rho$ if its power spectrum is nonzero at least at 
$\rho$ distinct frequencies. The requirements on the SLS model (\ref{ssx})--(\ref{varphit}) and the deadbeat 
observer in (\ref{yt}) are captured in

\begin{assumption}\label{PEsls}
	For every $k_i \in \chi$, the polynomials $z^n H_{\rm o}^{(1)} (z;k_i)$ and $z^n (1-H_{\rm o}^{(2)} (z;k_i))$ are 
	co-prime and the latter is Hurwitz, {\em i.e.,} it has all zeros inside the unit circle. 
\end{assumption}

We illustrate this assumption with a numerical example. Some conditions might be redundant
based on the previous assumptions. We defer a detailed study to future work. The co-prime condition
guarantees $\theta_n(k)$ is irreducible for all $k$.  

\begin{example}
	Let $(A,b,c,d)$ in Example~\ref{ex3.5} be a discrete state at $k_i \in \chi$. We calculate 
	$z^2(1-H_{\rm o}^{(2)}(z;k_i))=z^2+0.25$ verifying it is Hurwitz and $z^2 H_{\rm o}^1(z) = z^2+3z+0.25$. 
	Thus, $z^2(1-H_{\rm o}^{(2)}(z;k_i))$ and $z^2 H_{\rm o}^1(z)$ are co-prime as requested. 
\end{example}

The requirements on the inputs are stated in

\begin{assumption}\label{PEinput}
	The inputs are PE of order at least $2n+1$.  
\end{assumption}

\begin{proposition}\label{propPE}
	Consider the SLS model (\ref{ssx})--(\ref{varphit}) with the observer in Lemma~\ref{deadbeatlem}. Suppose that
	Assumptions~\ref{PEsls}--\ref{PEinput} hold. Then, there exists a $\gamma>0$ such that for every $k_i \in \chi$
	and $s,t \in [k_i\;\;k_{i+1})$ satisfying $t-s \geq \gamma$, 
	\begin{equation}\label{rankPE}
		{\rm rank}({\mathcal R}_n(s,t))=2n+1. 
	\end{equation}
\end{proposition}

{\em Proof.} See Appendix~\ref{appG}.

From Proposition~\ref{propPE}, we see that if Assumptions~\ref{PEsls}--\ref{PEinput} hold and 
$\delta_i(\chi) \geq \gamma+n+2$, then  (\ref{PErank}) is valid. It is not realistic to expect every 
segment to meet this requirement; but, there must be a sufficient number of long segments. 

\subsection{Clustering of the discrete states}\label{clustersec}

The solution of (\ref{optim1}) partitions the interval $[1\;\;\hat{N}]$ where with 
$\hat{\zeta}(0)=\hat{\mu}(0)$ we define its switches by the instants $1 \leq t \leq \hat{N}$ 
\begin{eqnarray*}
	\delta_{\hat{\zeta}}(t) &=& \hat{\zeta}(t)-\hat{\zeta}(t-1) \\
	&=& \hat{\mu}(t)-2\hat{\mu}(t-1)+\hat{\mu}(t-2)\neq 0.
\end{eqnarray*}
Let ${\mathcal T}=\{t_1,\cdots,t_{J^*}\}$ denote the switches of $\hat{\zeta}(t)$, $0\leq t \leq \hat{N}$. 
We set $t_0=0$. Suppose $k_{i+1}$ is identifiable from the input--output data. 
First, from $\eta(k-2n)=0$ for all $k$ in $[k_i+n\;\;k_{i+1})$, note that $\hat{\eta}(k-2n)=0$ on $[k_i+n\;\;k_{i+1})$. 
Thus, there exists a $J \in {\mathcal T}$ with $[k_i+n\;\;k_{i+1}) \subset[t_J\;\;t_{J+1})$. 
Hence, $\hat{\theta}(k)=\hat{\theta}(k_i)$ for all $k_i+n \leq k < k_{i+1}$. Suppose $k_{i+1}$ is not a switch 
of $\hat{\zeta}$. Then, $k_{i+1}$ is in $(t_J,t_{J+1})$ and $\hat{\zeta}(k_{i+1}-2n)=0$. Thus,
$\hat{\theta}(k_{i+1})=\hat{\theta}(k_{i+1}-1)=\hat{\theta}(k_i)$. Since $\hat{\zeta}$ is a feasible solution of 
(\ref{optim1}), $k_{i+1}$ is an identifiable switch. A contradiction. We summarize this result as

\begin{lemma}\label{optim1ident}
	Consider the SLS model (\ref{ssx})--(\ref{varphit}) with the observer in Lemma~\ref{deadbeatlem}. 
	Suppose that $\delta_*(\chi)>2n$ and Assumptions~\ref{deadminimal}a.--\ref{distinctas}i. or 
	\ref{deadminimal}b.--\ref{distinctas}ii. 
	hold. Let $\hat{\zeta}$ be a solution of (\ref{optim1}). Then, every identifiable swithch of $\chi$ is a 
	switch of $\hat{\zeta}$. 
\end{lemma}   

The conditions in Lemma~\ref{optim1ident} guarantees $\chi \subset \chi_\theta$. If $k_i \in \chi$ is 
identifiable, $k_i \in {\mathcal T}$ then. Since $\chi$ is a proper 
subset of $\chi_\theta$, the map $\chi \rightarrow {\mathcal T}$ is not necessarily surjective. We state the 
switch identifiability requirement in the following.

\begin{assumption}\label{switchdetassmp}
	Every switch in $\chi$ is identifiable from the input--output data. 
\end{assumption}

It is worth to emphasize that the switch identifiability is defined with respect to $\chi_\theta$
and ${\mathcal P}^\theta$. 

\begin{remark}
	If $k_{i+1}$ is identifiable from the input--output data, a change in $\theta(k)$ at $k_{i+1}$ is detectable. 
	However, this does not tell anything about $\theta(k_i)$ and $\theta(k_{i+1})$. 
\end{remark}

Suppose Assumption~\ref{switchdetassmp} holds. Let $k_i,k_{i+1} \in \chi_\theta$. Since $\eta(k-2n)=0$ for all 
$k \in [k_i+n\;\;k_{i+1})$, $\hat{\eta}(k-2n)=0$ on $[k_i+n\;\;k_{i+1})$. Then, for some $J \in {\mathcal T}$, 
$[k_i+n\;\;k_{i+1}) \subset [t_J\;\;t_{J+1}]$. Since $k_i\in {\mathcal T}$, $k_i=t_K$ for some $K$ and
$[k_i\;\;k_{i+1}]=[t_K\;\;t_{J+1}]$. Thus, $\chi \subset \chi_{\hat{\zeta}}$ with $\chi_{\hat{\zeta}}$ denoting 
partitioning of $[1\;\;\hat{N}]$ by $\hat{\zeta}$. Suppose Assumptions~\ref{PEsls}--\ref{switchdetassmp} hold 
and $\delta_i(\chi) \geq \gamma+n+2$. Then, 
(\ref{PErank}) holds and, as we discussed earlier, we can apply Algorithms~1--2 to $\hat{\theta}_n(t_J+n)$ 
to extract a realization of $\varphi(k_i)$. Note that $k_i \leq t_J \leq k_i+n$. The following assumption 
ensures that every discrete state in ${\mathcal P}$ visits a long segment so that its parameters can be 
identified from the input-output data.

\begin{assumption}\label{sysasmp3}
	Every discrete state in ${\mathcal P}$ is active in at least one segment with a dwell 
	time of at least $\gamma+2n+2$.
\end{assumption} 

This assumption lets us recover the discrete states by clustering. In clustering,
first a statistics is selected. A statistics based on the realization returned by Algorithm~2
and denoted by the quadruple $\hat{\mathcal P}(k)=(\hat{A}(k),\hat{b}(k),\hat{c}^T(k),\hat{d}(k))$ may be chosen 
as the $\ell_1$-norm of the eigenvalues of $\hat{A}(k)$, {\em that is,} we choose 
${\mathcal M}(\hat{A}(k))=\sum_{i=1}^n|\lambda_i(\hat{A}(k))|$ or 
the time-varying ${\mathcal H}_2$ norm of $\hat{\mathcal P}(k)$. Since ${\mathcal M}(\hat{A}
(\varphi(k))={\mathcal M}(\hat{A}(\varphi(l))$ if $\varphi(k)=\varphi(l)$, in the noiseless 
case, ${\mathcal P}$ may be determined by viewing the graph of 
${\mathcal M}(\hat{A}(\varphi(k)))$. By running a clustering algorithm
\cite{EsterKriegelSanderXu1996,Vassilvitskii&Arthur:2007}, we recover a collection of the discrete states 
in ${\mathcal P}$. This collection exhausts all discrete states in ${\mathcal P}$ by 
Assumption~\ref{sysasmp3}. We summarize these steps as an algorithm.

\begin{table}[h!]
	\small
	\begin{center}
		\begin{tabular}{l}
			\hline 
			{\bf Algorithm~3. Estimation of $\mathcal{P}$}  \\
			\hline
			{\bf Inputs:} $u(k)$ and $y(k)$ for $1 \leq  k \leq N$ \\ 
			1:  Solve (\ref{optim1}) for $\hat{\zeta}$ \\
			2:  Choose segments with $t_{J+1}-t_J\geq \gamma+n+2$ 	\\
			3:  Estimate $h(k,i)$ from Algorithm~1 using $\hat{\theta}_n(t_J+n)$\\
			4:  Estimate the discrete states from Algorithm~2 \\
			5:  Choose a statistics either ${\mathcal M}(\hat{A}(k))$ or 
			$\|\hat{\mathcal P}(k)\|_2$ \\
			6: Estimate $\mathcal{P}$ by running \cite{EsterKriegelSanderXu1996} over the segments in Step~2\\ 
			{\bf Output:} $\hat{\mathcal P}$.  \\
			\hline
		\end{tabular}
	\end{center}
\end{table}

In our numerical study, we will use the density-based clustering algorithm \cite{EsterKriegelSanderXu1996}
implemented by the {\tt dbscan} command in MATLAB. Another popular clustering 
method is the $k$-means clustering algorithm \cite{Vassilvitskii&Arthur:2007} implemented by the 
{\tt kmeans} command in MATLAB. Unlike the $k$-means clustering algorithm, the density-based clustering 
algorithm does not need the number of the clusters to be specified {\em a priori}. We capture the results
derived above in the following.

\begin{theorem}\label{balik}	
	Consider the SLS model (\ref{ssx})--(\ref{varphit}) with noiseless input-output data. 
	If Assumptions~\ref{sysasmp}, \ref{kereste}, \ref{deadminimal}a.--\ref{distinctas}i. or 
	\ref{deadminimal}b.--\ref{distinctas}ii., \ref{PEsls}--\ref{sysasmp3} hold, Algorithm~3 recovers 
	${\mathcal P}$. 
\end{theorem}

\section{Estimation of the switching sequence}\label{switchSARX}

Algorithm~3 delivers not only the discrete state estimates, but also locates the switches at the right endpoints 
of segments with dwell times at least $\gamma+n+2$ under the stated assumptions in Theorem~\ref{balik}. Indeed, 
let $[t_J\;\;t_{J+1})$ be such a segment. Then, $k_i \leq t_J \leq k_i+n$ and $k_{i+1} \leq t_{J+1}$ for some $i$. 
But, $k_{i+1} < t_{J+1}$ is not possible from Assumption~\ref{switchdetassmp}. The switch at $k_i$ may be determined 
up to an uncertainty $n$ by extending $t_J$ along the negative axis and applying (\ref{hopdedikbe}) if it has not 
already been determined by a segment $[t_K\;\;t_{K+1}]$ for some $K\leq J$. More precise estimates may be obtained by using
a GLR test with the discrete state estimates and the input-output data. An option  we have yet to explore
is to use the switching sequence and the discrete state estimates delivered by the algorithms in this paper for
initializing a non-convex optimization algorithm, for example  \cite{Sefidmazgi&Kordmahalleh&Homaifar&Karimoddini&Tunstel:2016}.
Switch detection and estimation is a topic of major interest in hybrid systems literature. 

If $\delta_*(\chi) \geq \gamma+2n+2$, $\varphi(k)$ is completely determined by Algorithm~3 from 
Assumption~\ref{switchdetassmp}. It remains to determine the switching sequence on the shorter segments.
In this section, given a segment $[t_J\;\;t_{J+1})$ satisfying $t_{J+1}-t_J < \gamma+n+2$ we will discuss 
how to determine the discrete state that is active in $[t_J\;\;t_{J+1})$, which is unique by 
Assumption~\ref{switchdetassmp}. This will provide an answer to the value of $\varphi(k)$ on 
$[t_J\;\;t_{J+1})$. 
We will use a subspace identification algorithm to achieve this goal. 

Let $l=\varphi(t_J)$, pick $\vartheta \leq t_{J+1}-t_J$, and form two Hankel matrices from the input--output data
\begin{eqnarray}
	{\mathcal U}_{t_J|\vartheta} &=& \left[\begin{array}{ccc} u(t_J)  &  & u(t_{J+1}-\vartheta) \\   
		& \ddots & \vdots \\ u(t_J+\vartheta-1)  & \cdots & u(t_{J+1}-1) \end{array} \right], \nonumber
	\\[-1ex] \label{Hankelmoesp}\\[-1ex]
	{\mathcal Y}_{t_J|\vartheta} &=& \left[\begin{array}{ccc} y(t_J)  &  & y(t_{J+1}-\vartheta) \\   
		& \ddots & \vdots \\ y(t_J+\vartheta-1)  & \cdots & y(t_{J+1}-1) \end{array} \right], \nonumber
\end{eqnarray}   
and generate the Toeplitz matrices from ${\mathcal P}_l=(A_l,b_l,c_l^T,d_l)$
\[
\Psi_l = \left[\begin{array}{lcc} d_l &  & \\    & \ddots & \vdots \\
	c_l^T A_l^{\vartheta-2} b_l &   \cdots & d_l   \end{array}\right], \;\;\; l \in \mathbb{S}.
\]
The following data  equation
\[
{\mathcal Y}_{t_J|\vartheta} = {\mathcal O}_{\vartheta} {\mathcal X}_{t_J|\vartheta} + \Psi_{\varphi(t_J)} 
{\mathcal U}_{t_J|\vartheta}
\]
can be derived where $\mathcal{X}_{t_J|\vartheta}$ is the compound state matrix defined by
\[
\mathcal{X}_{t_J|\vartheta}=\left[x(t_J)\;\;\cdots\;\;x(t_{J+1}-\vartheta)\right].
\]
Let the LQ decomposition of $[{\mathcal U}_{t_J|\vartheta}^T \;\;{\mathcal Y}_{t_J|\vartheta}^T]^T$ be given by
\begin{equation}\label{LQ}
	\left[\begin{array}{c} {\mathcal U}_{t_J|\vartheta} \\ {\mathcal Y}_{t_J|\vartheta}  \end{array}\right]=
	\left[\begin{array}{cc} L_{11} & 0 \\ L_{21} & L_{22} \end{array}\right] \left[\begin{array}{c} Q_{1}^T \\ 
		Q_2^T  \end{array}\right]
\end{equation}
where $L_{11}$ and $L_{22}$ are lower triangular matrices and $Q_1$ and $Q_2$ are orthogonal matrices with dimensions 
compatible with the row sizes of ${\mathcal U}_{t_J|\vartheta}$ and ${\mathcal Y}_{t_J|\vartheta}$. The MOESP algorithm 
\cite{Verhaegen&Dewilde:1992a,Verhaegen&Dewilde:1992b} estimates the observability range space by performing first the LQ 
decomposition (\ref{LQ}) and then the SVD
\[
L_{22} = \left[U_1 \;\; U_2\right]\left[\begin{array}{cc} \Sigma_1 & 0 \\ 0 & 0 \end{array}\right]
\left[\begin{array}{c} V_1^T \\ V_2^T \end{array}\right] =U_1 \Sigma_1 V_1^T
\]
where $\Sigma_1 \in \mathbb{R}^{n \times n}$. The extended observability matrix estimate is given by
\begin{equation}\label{obsex}
	\hat{\mathcal O}_{\vartheta}=U_1 \Sigma_1^{1/2}.
\end{equation}
Then, we estimate $A_{\varphi(t_J)}$ from (\ref{obsex})
\begin{equation}\label{Aess}
	\hat{A}_{\varphi(t_J)}=(J_{\rm D}\hat{O}_{\vartheta})^\dag J_{\rm U} \hat{\mathcal O}_{\vartheta}
\end{equation}
and calculate ${\mathcal M}(\hat{A}_{\varphi(t_J)})$. We find the active model index by comparing 
${\mathcal M}(\hat{A}_{\varphi(t_J)})$ with the statistics of the discrete states in $\hat{\mathcal P}$. 
The MOESP algorithm retrieves $(A_{\varphi(t_J)},c^T_{\varphi(t_J)})$ up to a similarity transformation if the 
following three conditions
\begin{equation}
	{\rm rank}(\mathcal{X}_{t_J|\vartheta}) = n,\label{ident1}
\end{equation}
\begin{equation}
	{\rm rank}({\mathcal U}_{t_J|\vartheta}) = \vartheta, \label{ident2}
\end{equation}
\begin{equation}
	{\rm span}(\mathcal{X}_{t_J|\vartheta}) \cap {\rm span} (\mathcal{U}_{t_J|\vartheta})=\{0\} \label{ident3}
\end{equation} 
are satisfied.
The first condition is the controllability of the discrete state that is active on the interval $[t_J\;\;t_{J+1})$ while 
the second is satisfied by selecting the PE inputs of order at least $\vartheta$. The last condition requires data be 
collected in an open-loop experiment. A pseudo-code implementing this method is outlined below.

\begin{table}[h!]
	\small
	\begin{center}
		\begin{tabular}{l}
			\hline 
			{\bf Algorithm~4. Estimation of $\varphi(t_J)$ by MOESP}  \\
			\hline
			{\bf Inputs:} $u(k),y(k)$, $1 \leq k \leq N$ and $\hat{\mathcal P}$ \\ 
			{\bf While} $t_{J+1}-t_J<\gamma+n+2$ \\
			1: Compute ${\mathcal U}_{t_J|\vartheta}$ and ${\mathcal Y}_{t_J|\vartheta}$ in (\ref{Hankelmoesp}) \\
			2: Perform the LQ decomposition in (\ref{LQ}) \\
			3: Apply SVD to $L_{22}$ \\
			4: Compute $\hat{A}_{\varphi(t_J
				)}$ from (\ref{obsex})--(\ref{Aess}) \\
			5: Compute $\mathcal{M}(\hat{A}_{\varphi(t_J)})$ \\
			6: If ${\mathcal M}(\hat{A}_s)={\mathcal M}(A_{\varphi(t_J)})$ for some $s\in \mathbb{S}$, 
			set $\varphi(t_J)=s$ \\
			{\bf End} \\
			{\bf Output:} $\varphi(t_J)$.  \\
			\hline
		\end{tabular}
	\end{center}
\end{table}

In Algorithm~4, we substitute $\vartheta=2n+1$. The calculation of $\hat{\mathcal O}_\vartheta$ requires 
${\rm rank}(V_1)=n$. Therefore, $L_{22}$ must have at least $n$ columns. Hence, $t_{J+1}-t_J \geq \vartheta+n=3n+1$. 
The PE condition on the inputs requires in fact a larger gap. In this section, we directly apply 
the PE condition to the SLS model (\ref{ssx})--(\ref{varphit}) instead of the observer model in 
Lemma~\ref{deadbeatlem}. 

Alternatively, let $\Xi(l)={\mathcal Y}_{t_J|\vartheta} - \Psi_{l} {\mathcal U}_{t_J|\vartheta}$, 
$l \in \mathbb{S}$ and if $l=\varphi(t_J)$, $\Xi(l)={\mathcal O}_\vartheta {\mathcal X}_{t_J|\vartheta}$. 
For each $l$, perform an SVD as $\Xi(l)=U_1^{(l)} \Sigma_1^{(l)} V_1^{(l)}+U_2^{(l)} \Sigma_2^{(l)}  V_2^{(l)}$ and 
let $\hat{\mathcal O}_{\vartheta}^{(l)}=U_1^{(l)} [\Sigma_1^{(l)}]^{1/2}$. Define
$\hat{A}_l=(J_{\rm D}\hat{O}_{\vartheta})^\dag J_{\rm U} \hat{\mathcal O}_{\vartheta}$, 
$l \in \mathbb{S}$. If $l$ is the right choice, then ${\mathcal M}(\hat{A}_{l})= {\mathcal M}(\hat{A}_{\varphi(t_J)})$. 
We will call this alternative Algorithm~4$^\prime$. It is easy to implement and works for 
$t_{J+1}-t_J \geq 2n$. For performance guarantees, (\ref{ident1})--(\ref{ident3}) are still needed since this 
algorithm originates from Algorithm~4. 

The main result of this section is stated as follows.

\begin{theorem}\label{main2}
	Consider the SLS model (\ref{ssx})--(\ref{varphit}) with noiseless input--output data. Assume that
	$\chi \subset {\mathcal T}$ and the discrete states in ${\mathcal P}$ are known up to similarity 
	transformations. Suppose that $\delta_*(\chi) \geq \gamma+2n+2$. Then, $\varphi$ is determined on $[1\;\;N]$ 
	by Algorithm~3. If $\delta_*(\chi) < \gamma+2n+2$, then Algorithms~4--4$^\prime$ together recovers 
	$\varphi$ in the segments satisfying (\ref{ident1})--(\ref{ident3}).
\end{theorem}

Since ${\mathcal M}(\hat{A}_l)$ is invariant to similarity transformations, it suffices to know the discrete
states in ${\mathcal P}$ up to similarity transformations without bringing to a common basis. Algorithm~4--4$^\prime$
may falsely detect the discrete states, in particular over the short segments if the inputs are not PE there.
If the SLS model is used to predict outputs, it is necessary to transform the discrete states
to a common basis. One can freely select one similarity transformation only. The rest of the 
transformations are fixed and can be calculated from the input-output data, ${\mathcal P}$, and $\chi$. See 
\cite{Bencherki&Turkay&Akcay:2021}.

\section{Convex relaxation}\label{bpdnsec}

Consider the following robust reformulation of (\ref{optim1})
\begin{equation}\label{optim22}
	\min_\zeta  \|\eta\|_{0,\mathfrak{I}} \;\;{\rm subject\;to} \;\; \|Y -M \zeta\|_2 \leq \varepsilon
\end{equation}
to model inaccuracies in the measurements. Replace $Y$ with $\hat{Y}$ satisfying $\|\hat{Y}-Y\|\leq \varepsilon$ 
and set $\hat{Y}=M\zeta$. We can find an $\varepsilon_0>0$ such that for all $\varepsilon \leq \varepsilon_0$, 
every switch in $\chi$ is identifiable from the input--output data and (\ref{rankPE}) holds by virtue of finiteness 
of $N$. Thus, we see that Algorithm~3 recovers every discrete state in ${\mathcal P}$ within an error that vanishes 
as $\varepsilon$ decreases to $0$. Now, replace ${\mathcal P}$ with $\hat{\mathcal P}$ in Theorem~\ref{main2} and 
note that its conclusion is true for all $\varepsilon \leq \varepsilon_0$, if necessary by reducing $\varepsilon_0$. 
Since $\|\eta\|_{0,\mathfrak{I}}$ has a fixed optimal value for all $\varepsilon \leq \varepsilon_0$ and $\hat{\zeta}$ 
is a feasible solution of (\ref{optim22}), we see that any solution of (\ref{optim22}) is also a sparse vector. 
Conversely starting from any solution of (\ref{optim22}), we reach the same optimal value. We may now replace Step~1 
in Algorithm~3 with (\ref{optim22}).

The BBPDN algorithm is a well-known convex relaxation of (\ref{optim22}). It recovers any block-sparse vector $\zeta$ 
from the measurements $Y=M\zeta$ whenever $M$ satisfies a recovery condition. The recovery conditions for the greedy 
BOMP algorithm happen to coincide with those of the BBDN algorithm. The stability and robustness properties of the
BOMP algorithm are however entirely different than those of the BBPDN algorithm. For example, the BOMP algorithm obeys
a local stability result, {\em that is}, stable recovery is not possible if $\varepsilon > \varepsilon_0$ for some
$\varepsilon_0$ \cite{Donoho&Elad&Temlyakov:2005}. See, \cite{Wen&Zhou&Liu&Lai&Tang:2019} for results of the same kind.
By comparison, stability is global for the BBPDN algorithm, that is, stable recovery is possible for all $\varepsilon>0$
once a recovery condition is satisfied.

In order a recovery condition to hold, it is necessary 
that (\ref{optim1}) has a unique solution, which is not true in general since $u(k)$ may not be a PE of sufficient order 
on all segments. In fact, a deadbeat observer generates constant $\theta(k)$ intervals shorter than $4n$, cf. Lemma~\ref{cavit2} 
and the comment that follows the lemma. Nevertheless, all one needs to recover ${\mathcal P}$, as expressed in 
Theorem~\ref{main2}, is that Assumption~\ref{sysasmp3} holds.

We now formulate an efficient convex optimization problem instead of (\ref{optim1}).  Similarly to the $\ell_0$ 
quasi-norm, we define the mixed $\ell_2/\ell_1$ norm over the interval $(\delta_{\rm c}\;\;\delta_{\rm o}]$ by
\begin{equation}\label{l1quas}
	\|\eta\|_{2,\mathfrak{I}} =\sum_{k=1}^{\hat{N}} \left\|\eta(k) \right\|_2. 
\end{equation}

Suppose that the measurements are corrupted by noise and consider the following convex optimization problem 
\begin{equation}\label{optim3} 
	\min_{\zeta} \, \|\eta\|_{2,\mathfrak{I}} \;\;{\rm subject\;to} \;\;   \|Y-M \zeta\|_2 \leq \varepsilon.
\end{equation}
The feasible parameter sets $Y=M\zeta$ in (\ref{optim1}) $(\epsilon=0)$ and the ellipsoid in (\ref{optim3}) 
$(\epsilon \neq 0)$ are not necessarily bounded. If they were, (\ref{optim3}) would have a unique and finite
solution. Enforcement of the constraint $\|\zeta\|_2 \leq R$ for a large $R$ does not change the optimal value 
of (\ref{optim3}), denoted by ${\zeta}_{2/1}(k)=\theta_{2/1}(k)-\theta_{2/1}(k-1)$, over the long segments as 
will be shown shortly, but ensures a unique and bounded solution everywhere. 

Let $\varepsilon=0$, adjoin $\|\zeta\|_2 \leq R$ to (\ref{optim3}), and consider $[k_i\;\;k_{i+1})$ with 
$k_i,k_{i+1} \in \chi$ identifiable from the input-out data and satisfying $\delta_i(\chi) \geq 8n-1$. 
If $R$ is chosen very large, any solution $\hat{\zeta}(k)$ of (\ref{optim1}) provides a feasible solution 
to (\ref{optim3}). Hence, from ${\zeta}_{2/1}(k)=0$ we get $y(k)=z_n^T(k) \theta_{2/1}(k)$ for all $k \in (k_i,k_{i+1})$.
Then, $\theta_{2/1}(k_i+n)=\hat{\theta}_n(k_i+n)$ whenever (\ref{PErank}) holds, yet for $k=k_i$ or $k=k_{i+1}$, 
$\zeta_{2/1}(k) \neq \hat{\zeta}(k)$ is possible. The clustering results in Subsection~\ref{clustersec} hold 
with $\zeta_{2/1}(k)$ since $k_i,k_{i+1} \in \chi$ are identifiable from the input-output data.

We can express $\|\zeta\|_2 \leq R$ in a suitable form by letting
\begin{equation}\label{blockpara}
	\Gamma_s = \left[\begin{array}{c} h_{\rm o}^T(\delta_{\rm c},\delta_{\rm c}-s) \\ \vdots \\ 
		h_o^T (\delta_{\rm o},\delta_{\rm o}-s) \end{array} \right], \;\;\; 0 \leq s < 2n.
\end{equation}
Recall that for all $k \in [k_i+n\;\;k_{i+1})$,  $\theta(k)$ is a constant and sparse vector having the 
structure (\ref{thetared}) though $k_i$ and $k_{i+1}$ are not known. As a result, we propose penalizing 
all Markov parameters $h_{\rm o}(k,k-l)$, $l>n$. Similarly, we penalize the terms $h_{\rm o}(k,k-l)$, 
$l \leq n$, but with a different weight. 

We propose the following BBPDN problem:
\begin{eqnarray}
	\mathop {\min }\limits_\zeta  {\left\| \eta  \right\|_{2,\mathfrak{I}}} &{}& \quad {\rm subject\;to} 
	\nonumber \\[-1.5ex]  \label{constrain-BPDN}  \\[-1.5ex]
	{\left\| {Y - M\zeta } \right\|_2} \le \varepsilon,&{}&\;\; \sum\limits_{s = n+1}^{2n-1}
	{\left\| \Gamma_s \right\|_2}  \le \rho_1, \;\; \sum\limits_{s = 0}^{n} 
	{\left\| \Gamma_s \right\|_2} \le \rho_2 \nonumber
\end{eqnarray}
which can be written in a regularized form by appending the two constraints to the objective function 
giving rise to
\begin{equation}\label{constrain-BPDN-regul}
	\mathop {\min }\limits_\zeta  {\left\| {Y - M\zeta } \right\|_2} 
	+\lambda {\left\| \eta  \right\|_{2,\mathfrak{I}}} +\sum_{s=0}^{2n-1} \gamma_s \| \Gamma_s\|_2
\end{equation}
where $\gamma_s=\gamma_1$, $s>n$ and $\gamma_s=\gamma_2$, $s\leq n$. We change the hyper-parameters
$\lambda,\gamma_1,\gamma_2 \geq 0$ by gridding to get a family of Pareto-optimal solutions. The summation 
in (\ref{constrain-BPDN-regul}) is over the spatial variables; not over the time as in (\ref{l1quas}). 
The rationale behind this is to limit the rapid fluctuations of the spatial variables while controlling the 
computational load. This is especially true when the solution of (\ref{constrain-BPDN-regul}) is iteratively 
refined. If the second and the third constraints in (\ref{constrain-BPDN}) were replaced with
\[
\sum\limits_{s = 0}^{2n-1} \sum\limits_{k = \delta_{\rm c}}^{\delta_0}    
{\left\| h_{\rm o}(k,k-s) \right\|_2}  \le \rho,
\]
(\ref{constrain-BPDN}) would have a sparse solution, but distributed in time as opposed to the block
sparsity induced by the constraints in (\ref{constrain-BPDN}). Block sparsity is most desirable 
since it does not destroy the switch identifiability while bounding $\zeta$. An identification procedure 
extracting the discrete state parameters one after another was presented in \cite{Bako:2011}. It is based on 
the $\ell_1$ relaxation, but has a different measurement model than ours.

Given $\varepsilon \geq 0$ and $\rho_1,\rho_2 >0$, replace $\hat{\zeta}(k)$ in Step~1 of Algorithm~3 with 
$\zeta_{2/1}(k)$ obtained by solving (\ref{constrain-BPDN}) and save Steps~2--6. The
new algorithm will be called Algorithm~5 henceforth. We summarize the results derived in this section.

\begin{theorem}\label{balikcil}	
	Consider the SLS model (\ref{ssx})--(\ref{varphit}) with Algorithm~5. Suppose that Assumptions~\ref{sysasmp}, 
	\ref{kereste}, \ref{deadminimal}a.--\ref{distinctas}i. or \ref{deadminimal}b.--\ref{distinctas}ii., 
	\ref{PEsls}--\ref{sysasmp3} hold. Then, the following are true:  
	\begin{itemize}
		\item[(a)] If $\varepsilon=0$, there exists a $\rho_0>0$ such that for all $\rho_1,\rho_2\geq \rho_o$,
		Algorithm~5 recovers ${\mathcal P}$ up to similarity transformations. 
		\item[(b)] There exists an $\varepsilon_0>0$ and $\rho_0>0$ such that for all $\varepsilon \leq \varepsilon_0$ 
		and $\rho_1,\rho_2\geq \rho_o$, Algorithm~5 approximately recovers ${\mathcal P}$, {\em that is}, there exists 
		an absolute constant $c>0$ such that
		$$
		\max_{1 \leq t \leq \sigma} \min_{\hat{\mathcal P}_s \in {\mathcal P}} \|\hat{\mathcal P}_s-{\mathcal P}_t\|
		\leq c \varepsilon.
		$$   
		\item[(c)] Assume further that $\chi \subset {\mathcal T}$. In Algorithms~4--4$^\prime$, replace ${\mathcal P}$ 
		with $\hat{\mathcal P}$ estimated by Algorithm~5. Then, there exists an $\varepsilon_0>0$ such that for all 
		$\varepsilon \leq \varepsilon_0$, $\varphi$ is determined on $[1\;\;N]$ by Algorithm~3 if 
		$\delta_*(\chi) \geq \gamma+2n+2$. Furthermore, Algorithms~4--4$^\prime$ togerher
		recovers $\varphi$ in the segments satisfying (\ref{ident1})--(\ref{ident3}) if $\delta_*(\chi) < \gamma+2n+2$. 
	\end{itemize}
\end{theorem}

\subsection{Iterative refinement}
With $\varepsilon=0$ and $\rho_1=\rho_2 =\infty$, (\ref{constrain-BPDN}) is the tightest convex relaxation of 
(\ref{optim1}). A better heuristic, the iterative weighted $\ell_1$ relaxation, was proposed in 
\cite{Candes&Wakin&Boyd:2008}. The iterative refinement procedure when applied to (\ref{constrain-BPDN-regul}) 
assumes the following form

\begin{tabular}{p{7.5cm}}
	\hrule
	\vspace{0.1cm}
	{\bf Algorithm~6. Iteratively reweighted BBPDN} 
	\vspace{0.1cm}
	\hrule
	\vspace{0.1cm}
	\textbf{Given:} $\lambda,\gamma_1,\gamma_2,\alpha \geq 0$ and maximum number of
	iterations $t_{\rm max}$, initialize weights $w_k^{(0)}=1$, $v_s^{(0)}=1$.	
	
	\vspace{0.2cm}	
	Run the following loop:
	\vspace{0.1cm}	
	
	\noindent \textbf{while}  $t \leq t_{\rm max}$
	\begin{enumerate}
		\vspace{0.1cm}
		\item Solve the problem:
		\begin{eqnarray*}\label{constrain-BP3}
			\hspace{-0.5cm} \mathop {\min }\limits_\zeta  &{}& {\left\| {Y - M\zeta } \right\|_2} + 
			\lambda \sum\limits_{k=1}^{\hat{N}}{w_k^{(t)}} {\left\| {\eta(k)} \right\|_2} \\
			&{}& \;\;\; +\sum\limits_{s = 0}^{2n-1} \gamma_s {v_s^{(t)}{\| \Gamma_s \|_2}} 
		\end{eqnarray*}	
		\item Update weights for $1 \leq k \leq \hat{N}$ and $0 \leq s<2n$,
		\begin{eqnarray*}
			w_k^{(t+1)} &=& (\alpha+\|\eta(k)\|_2)^{-1}, \\
			v_s^{(t+1)} &=& (\alpha+\|\Gamma_s\|_2)^{-1}.
		\end{eqnarray*}		
	\end{enumerate}
	\textbf{end while}
	\vspace{0.2cm}
	
	\textbf{Outputs:} $\hat{\theta}(k)$, $\delta_{\rm c}< k \leq \delta_{\rm o}$ and $\hat{\mu}(0)$.
	\vspace{0.1cm}	
	\hrule
\end{tabular}

The solution given by Algorithm~6 may be further refined by running Algorithm~6 with the thresholded weights: 
the weights below a certain value are set to zero and the weights above this value are set to a large number, say 
$10^5$. This will help improve the fit over segments. Many standard convex optimization solvers could 
be used to implement Algorithms~5--6. In this paper, we used the CVX package \cite{Grant&Boyd:2014} which converts 
the optimization problem to a second-order cone program (SOCP) and calls a standard interior-point cone solver. 

\subsection{Summary}

\begin{figure}
	\centering
	\includegraphics[width=0.90\linewidth]{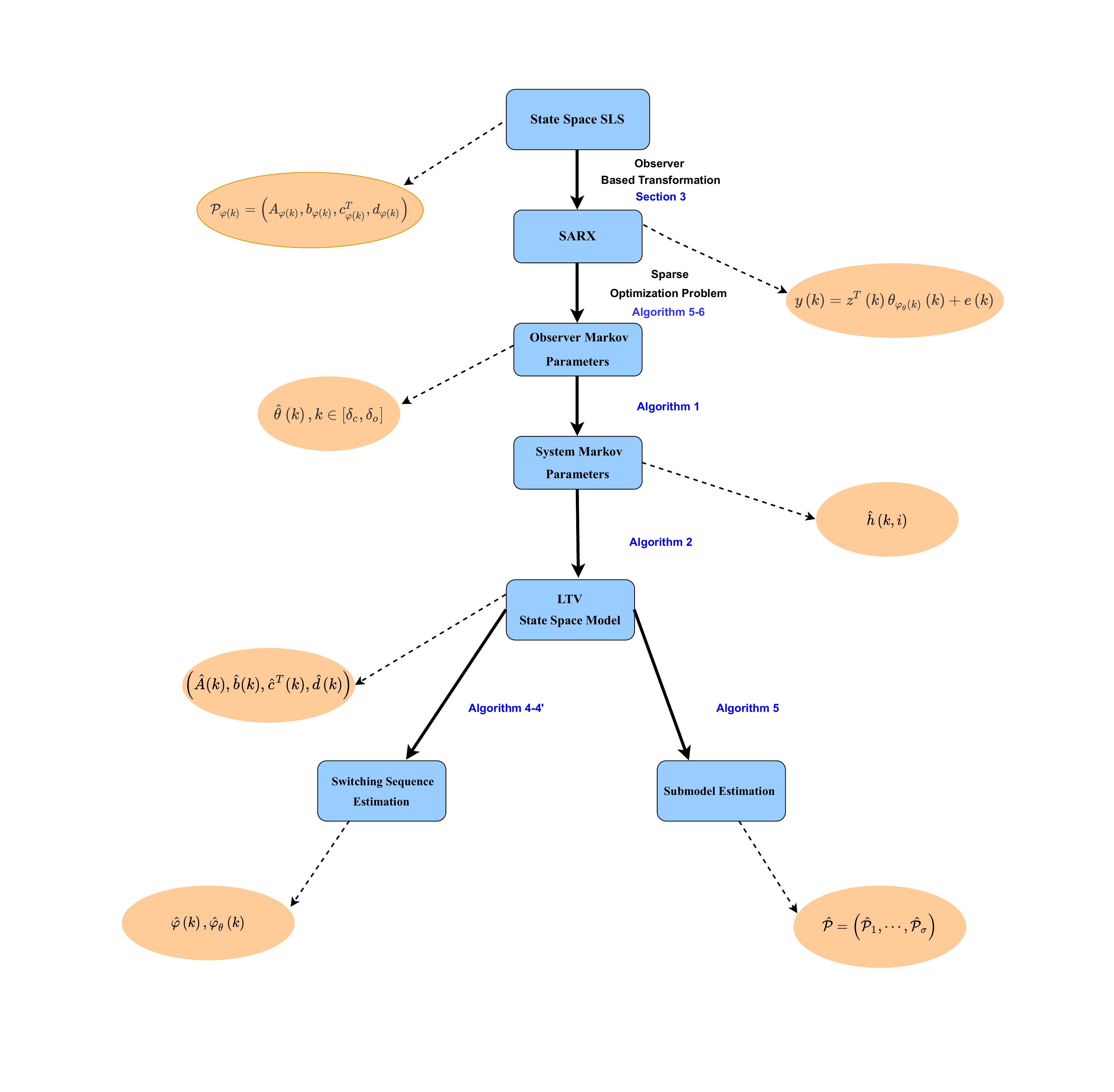}
	\caption{Flowchart of the SLS model identification algorithm from the input-output measurements.}
	\label{figure1001}
\end{figure}

In Figure~\ref{figure1001}, we provide a flowchart to implement the proposed identification scheme.
Except Algorithms~5--6, all algorithms in Figure~\ref{figure1001} are unsupervised learning algorithms
in the terminology of statistical learning theory \cite{Vapnik:1998,Hastie&Tibshirani&Friedman:2001}.
Supervised learning algorithms, in contrast, are prediction algorithms estimating the system structure and
the parameters by minimizing a suitable norm of the prediction errors constructed from the input-output data.
An example is the learning algorithm in \cite{Sefidmazgi&Kordmahalleh&Homaifar&Karimoddini&Tunstel:2016}.
The discrete-state estimates and the switching sequence estimate delivered by the proposed scheme should
further be enhanced by a supervised learning algorithm. 

\section{Numerical example}\label{numinsec}

To illustrate the results derived in this paper, we consider an SLS model with three discrete-states in the
state-space form

\begin{eqnarray*}
	A_1 &=& \left[\begin{array}{lr} 0 & - 1 \\ 0.9 & 0.6 \end{array} \right], \;\;
	b_1 = \left[\begin{array}{r} 0.4 \\ - 1 \end{array}\right], \;\;
	c_1 = \left[\begin{array}{r} -1 \\ -2\end{array}\right], \\
	A_2 &=& \left[\begin{array}{lr} 0.6 & 1\\ - 1 & - 1 \end{array} \right],\;\;
	b_2 = \left[\begin{array}{r} 0.5 \\ 1\end{array} \right],
	c_2 =  \left[\begin{array}{r} - 1 \\ 2\end{array} \right],\\
	A_3 &=& \left[\begin{array}{lr} - 1 & - 2 \\ 1 & 1.5 \end{array} \right],\;\;
	b_3 = \left[\begin{array}{r} 3 \\ 1 \end{array} \right],\;\;
	c_3 = \left[\begin{array}{r}0.9 \\ - 1\end{array}\right],
\end{eqnarray*}
$d_1=0,5$, $d_2= - 1.5$, and $d_3=2.5$.
In this example, we will address several issues in hybrid system identification starting with the 
observer-based  transformation to SARX models.

\subsection{Observer-based transformation to SARX models}\label{deadnumsec}

\begin{figure}
	\centering
	\includegraphics[width=3.0in]{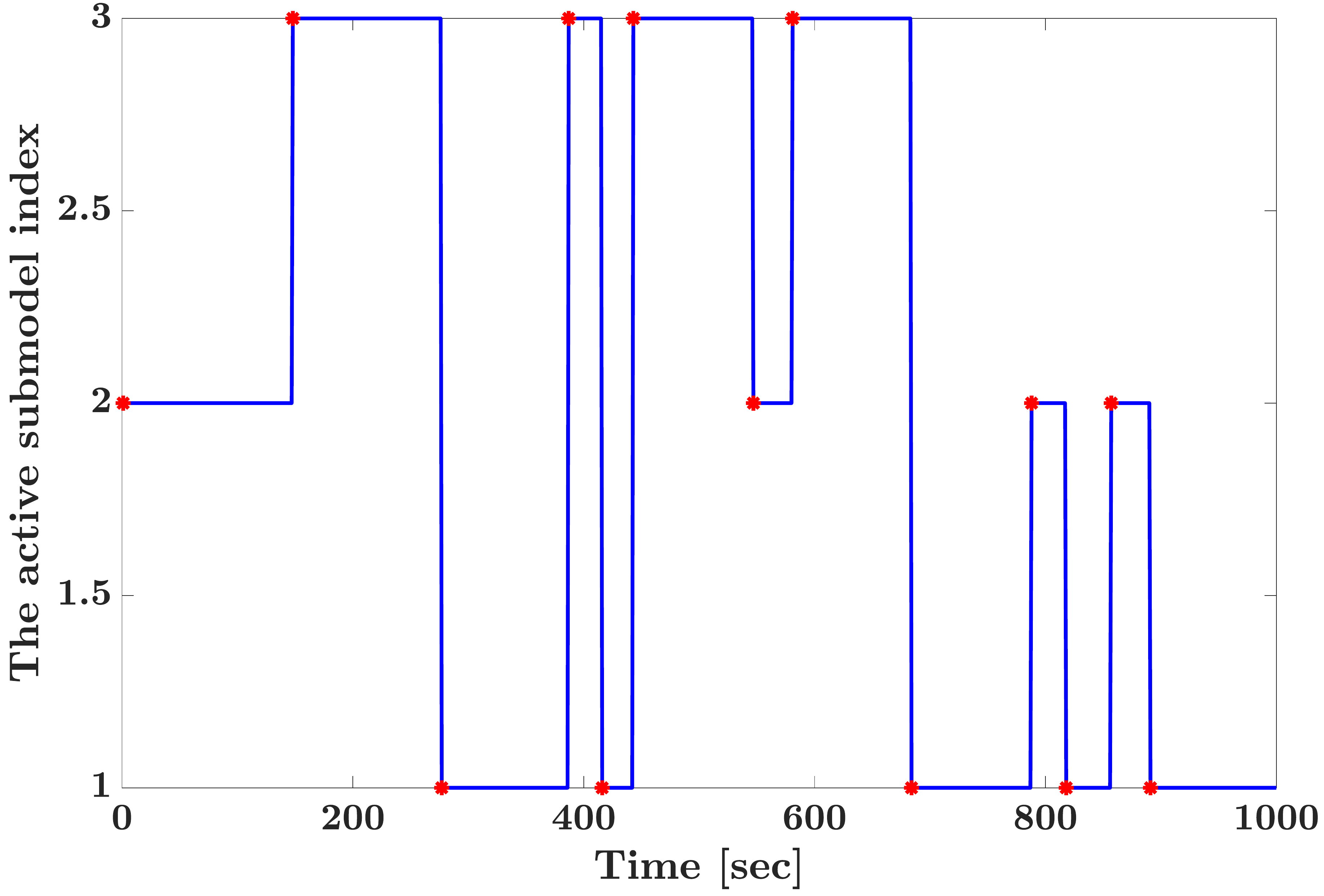}
	\caption{The switching sequence in Subsection~\ref{deadnumsec}.}
	\label{figure1002}
\end{figure}

\begin{figure}
	\centering
	\includegraphics[width=3in]{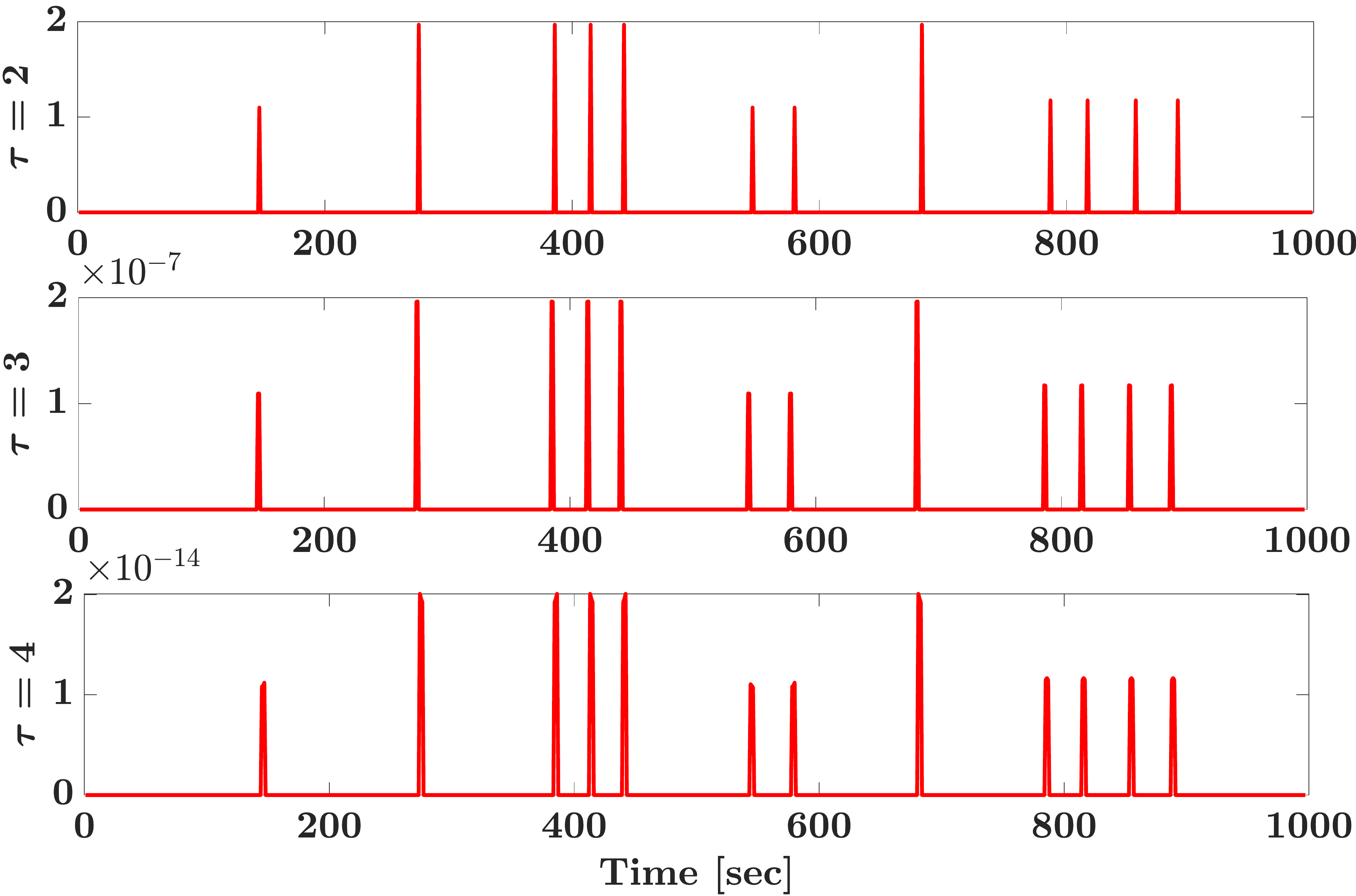}
	\caption{$\|\Phi_{\rm o}(k,k-\tau)\|_F$ as a function of $k$ for $\tau=2,3,4$. }
	\label{figure1003}
\end{figure}

The switching sequence of length $N=1,000$ plotted in Figure~\ref{figure1002} was obtained by sampling from a uniform 
distribution with a minimum dwell time of $\delta _*(\chi)=27$ secs. The SLS-to-SARX model conversion was carried out by the pole 
placement (the {\tt place} command in MATLAB). For $\tau=2,3,4$ and the piece-wise constant gain sequence $g(k)$ in
Lemma~\ref{deadbeatlem}, $\|\Phi_{\rm o}(k,k-\tau)\|_F$ is plotted in Figure~\ref{figure1003} where $\|X\|_F$ denotes the 
Frobenius norm of a given matrix $X$. Definition~\ref{deadbeatdef} holds for all $\tau \geq 3$ and the time-varying ARX 
model (\ref{yrespobsv22}) has order $3$  on $[1\;\;N]$ and $2$ in the subintervals $[ {{k_i} + 2\;\;{k_{i + 1}}})$. 
Since $3 \leq 2n-1$, $\tau=3$ is the best possible from Lemma~\ref{deadbeatlem}.

\subsection{Switch and submodel identifiability }\label{switchident}

\begin{figure}
	\centering
	\includegraphics[width=3in]{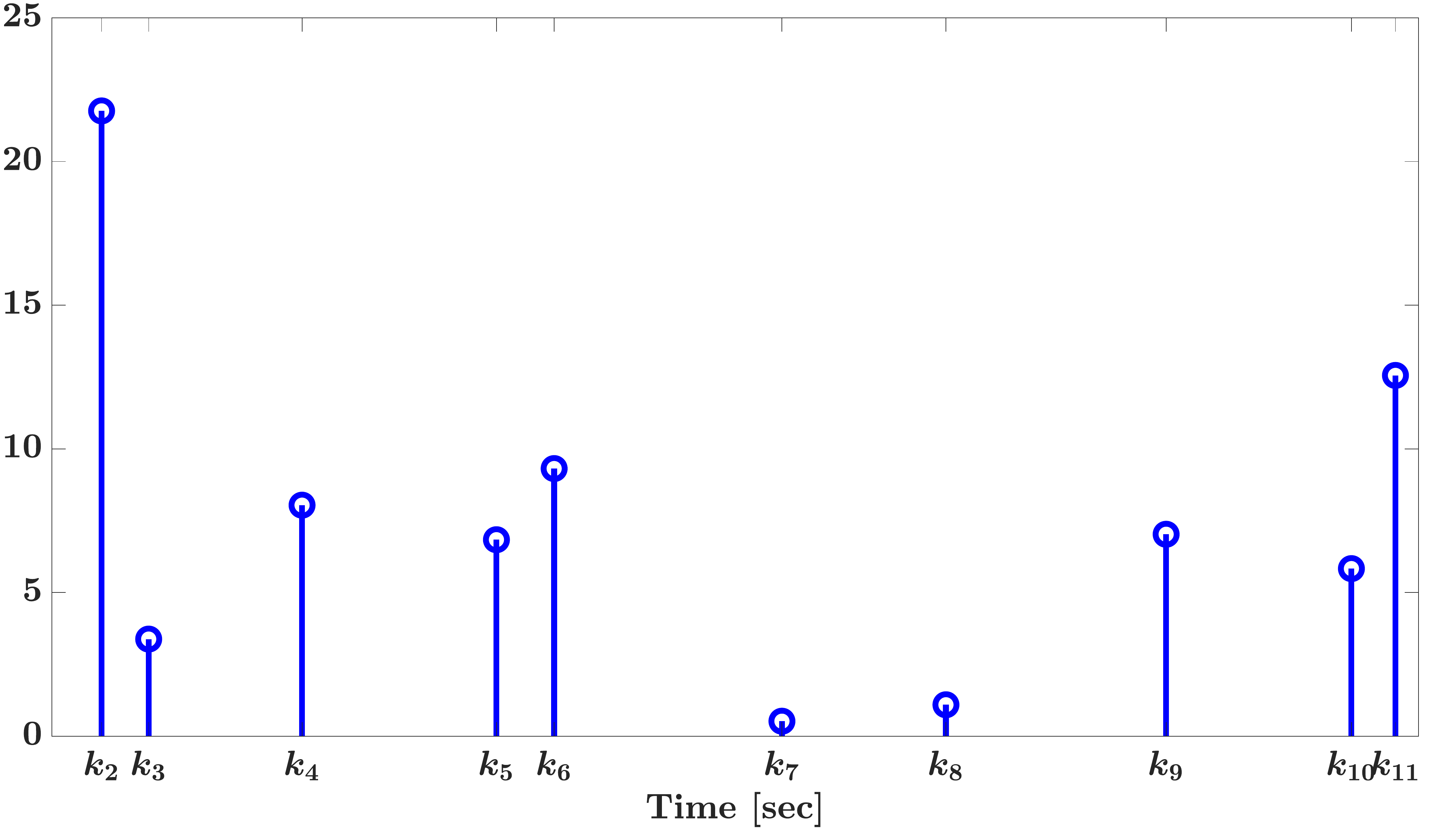}
	\caption{$|z^T(k_{i+1})(\theta(k_{i+1})-\theta(k_i))|$ on $\chi$.}
	\label{figure1004}
\end{figure}

We first check the switch identifiability conditions in Section~\ref{switchidsec} for $\tau=2n=4$ with the gain 
sequence constructed in Lemma~\ref{deadbeatlem}. We generated a switching sequence of length $N=1,000$. The minimum 
dwell time satisfies $\delta_*(\chi)=29$. We excited the SLS model with a multi-sine input consisting of five superposed 
harmonics with frequencies, phases, and amplitudes randomly selected from uniform distributions. We verified the switch
identifiability conditions in a noiseless setup since identifiability of a switch is not lost if perturbations are small 
enough while we checked the identifiability of the discrete-states by injecting noise with a large SNR. The latter approach, 
besides being more realistic, has eliminated delicate parametrization issues in designing the PE inputs for noiseless
identification experiments. Calculating $|z^T(k_{i+1})(\theta(k_{i+1})-\theta(k_i))|$ and plotting it, we see from
Figure~\ref{figure1004} that the first condition in Lemma~\ref{lemsparse2} is satisfied on the switch set $\chi$. 
Recall from Lemma~\ref{ARXdwell3} that $\chi \subset \chi_\theta$. 

\begin{figure}
	\centering
	\includegraphics[width=3in]{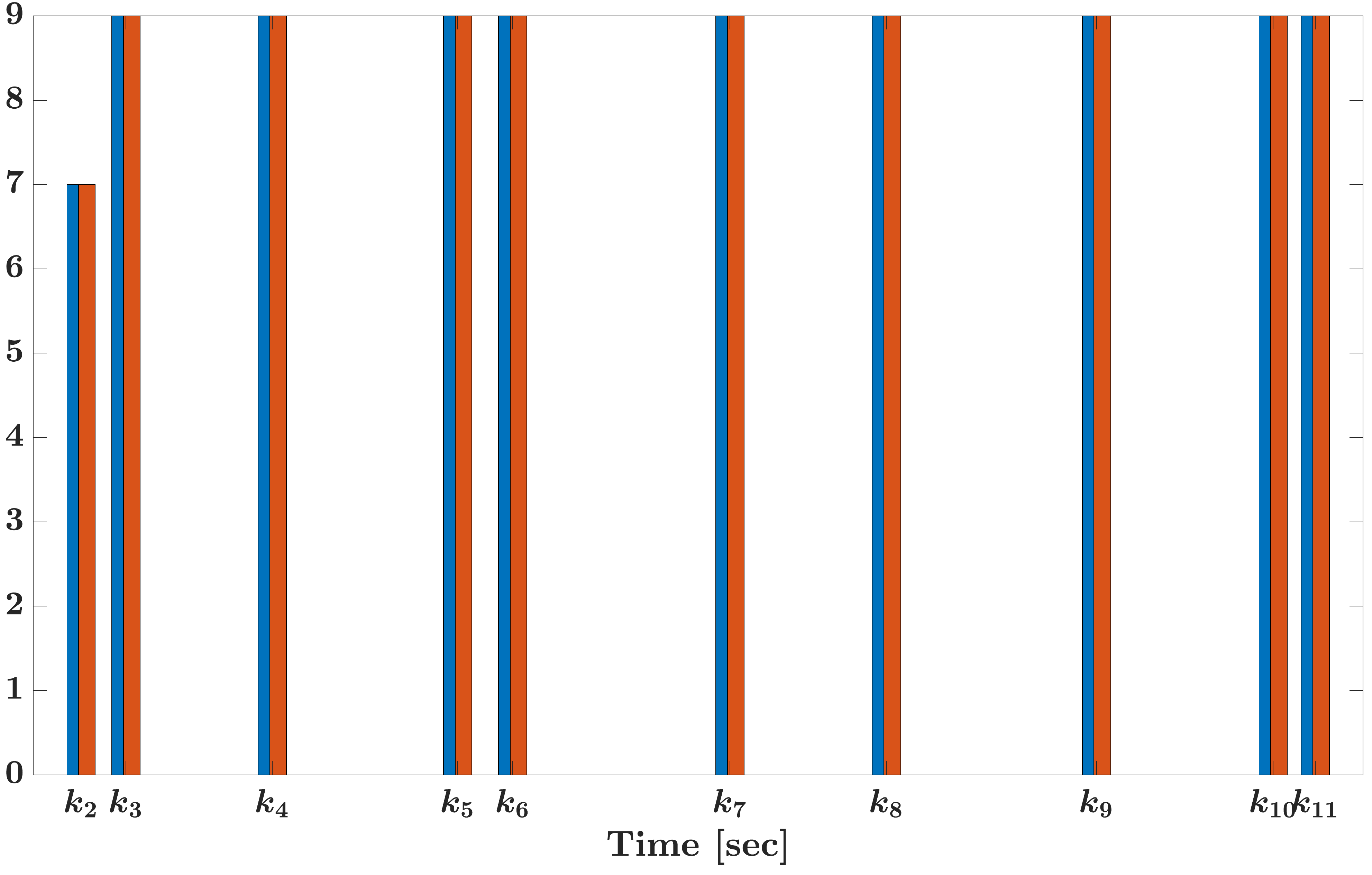}
	\caption{$\|\Sigma_{\mathcal{R}(k_i,k_{i+1}-1})\|_0$ (left bars) and 
		$\|\Sigma_{\mathcal{R}(k_i,k_{i+1}})\|_0$ (right bars). }
	\label{figure1005}
\end{figure}

Next, we verify the second condition in Lemma~\ref{lemsparse2} on $\chi$ by performing two SVDs: one for 
${\mathcal R}(k_i,k_{i+1}-1)$ and one for ${\mathcal R}(k_i,k_{i+1})=[{\mathcal R}(k_i,k_{i+1}-1)\;\;z(k_{i+1})]$. 
The augmentation of $z(k_{i+1})$ into ${\mathcal R}(k_i,k_{i+1}-1)$ does not lead to an increase in the number of 
the significant singular values as shown in Figure~\ref{figure1005} where $\|\Sigma_A\|_0$ denotes the number
of the nonzero singular values of a given matrix $A \in \mathbb{R}^{n \times m}$. We conclude that by using the 
multi-sine excitations $\chi $ is identifiable. 

Lemma~\ref{deadbeatlem} tells us that we could have used $\tau=2n-1$ instead of $\tau=2n$. This choice would not
change the inner product $z^T(k_{i+1})(\theta(k_{i+1})-\theta(k_i))$ because zero-padding in $\theta(k)$
has no effect in this operation. For the second condition in Lemma~\ref{lemsparse2}, permute and split ${\mathcal R}(s,t)$ 
as follows
\[
P_z {\mathcal R}(s,t) = \left[\begin{array}{ccc} u(s) & \cdots & u(t) \\ \vdots & \ddots & \vdots \\ u(s-2n) & \cdots & 
	u(t-2n) \\ y(s-1) & \cdots & y(t-1) \\ \vdots & \ddots & \vdots \\ y(s-2n) & \cdots & y(t-2n) \end{array} \right] = 
\left[\begin{array}{c} {\mathcal R}_u(s,t) \\ {\mathcal R}_y(s,t) \end{array} \right] 
\]
with $P_z$ denoting an appropriate permuatation matrix. If $u(k)$ is PE of a sufficient order, then ${\mathcal R}_u(s,t)$ 
has full rank, that is, ${\rm rank}({\mathcal R}_u(s,t))=2n+1=5$ provided that $t-s$ is large enough. As for the bottom matrix, 
$n \leq{\rm rank}({\mathcal R}_y(s,t)) \leq 2n$. Both limits are attained in Figure~\ref{figure1005}. As $t-s$ diverges,
the lower limit will be more likely to be seen. For intermediate values, $2n-1$ is possible, and the bar graph would include $8$ 
as a possible value. If we had started with the correct parametrization $\theta \in \mathbb{R}^{2\tau+1}$ and $\tau=2n-1$,
then Figure~\ref{figure1005} would have $7$ only possible value, but the identifiability of $\chi$ would not change.    

\begin{figure}
	\centering
	\includegraphics[width=3in]{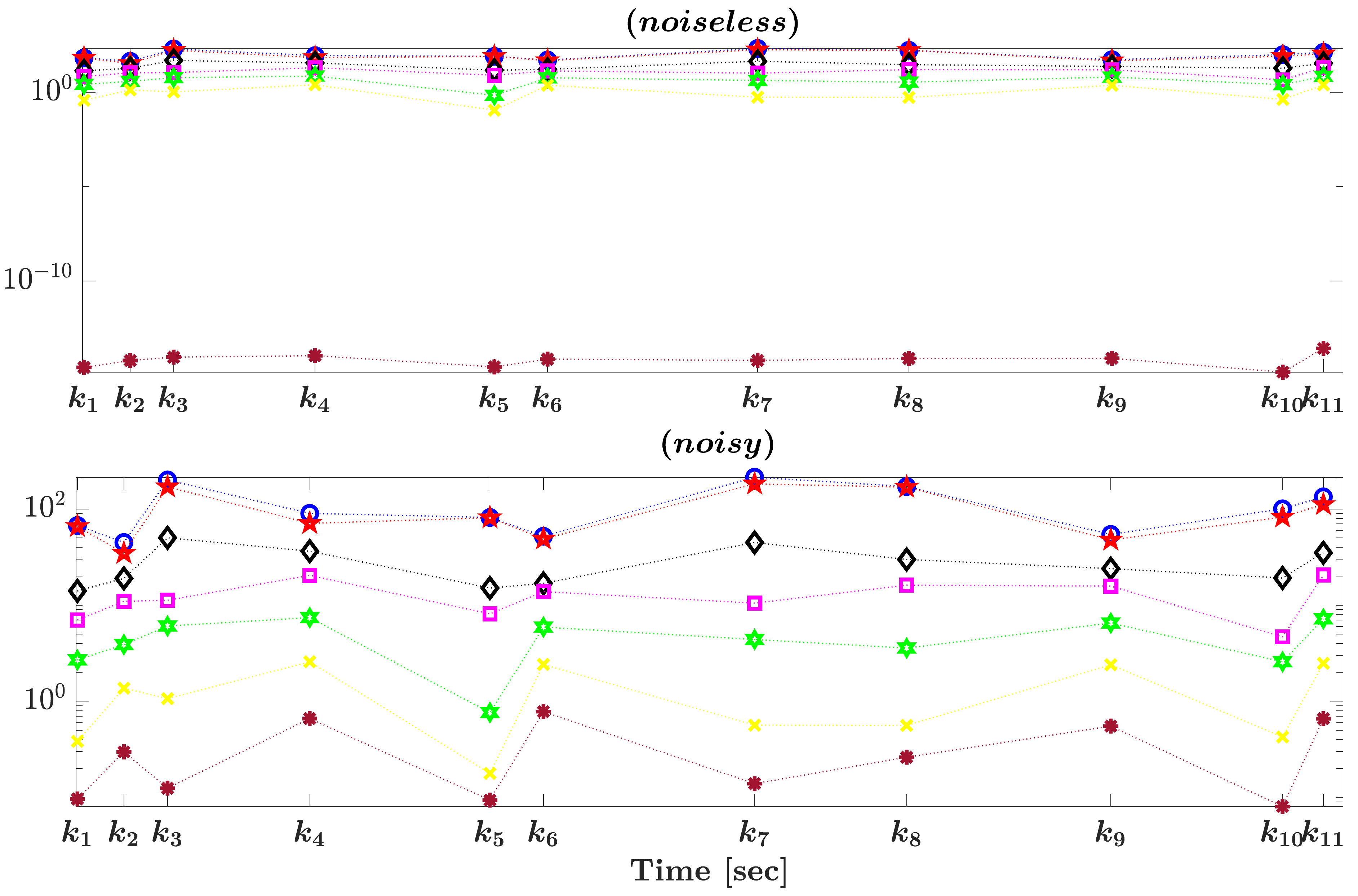}
	\caption{The singular values of ${\mathcal{R}(k_i+2n-1,k_{i+1}-1})$.}
	\label{figure1006}
\end{figure}

Lastly, we check the discrete-state identifiability for the observer order $\tau =2n-1$ on the intervals 
$[k_i+\tau\;\;k_{i+1})$ and the same gain sequence. We corrupted the output measurements by an additive Gaussian 
noise to achieve an SNR of $40$dB. The PE condition is satisfied in both the noiseless and the
noisy cases, see Figure~\ref{figure1006}. On the top subfigure, notice that ${\mathcal R}(k_i+2n-1,k_{i+1}-1)$ has $6$ 
significant singular values exceeding by one the theoretical prediction $2n+1$. Past discrete states in 
$(-\infty\;\;k_i)$ set a nonzero initial state. 

\subsection{Monte Carlo simulations}

We consider the same SLS and start it at $x(0)=[1,0]^T$. We sampled the switching function from a 
uniform distribution by selecting a minimum dwell time $\delta _*(\chi)=26$. The SLS model was 
excited with a multi-sine input consisting of five superposed harmonics generated as in Section 
\ref{switchident}. In the identification experiment, the length of the input-output data was 
$N=2,000$. The output measurements were corrupted by additive Gaussian noise to achieve the
SNRs of 40dB, 30dB, and 20dB. 

\begin{figure}[hbt!]
	\centering
	\includegraphics[width=3in]{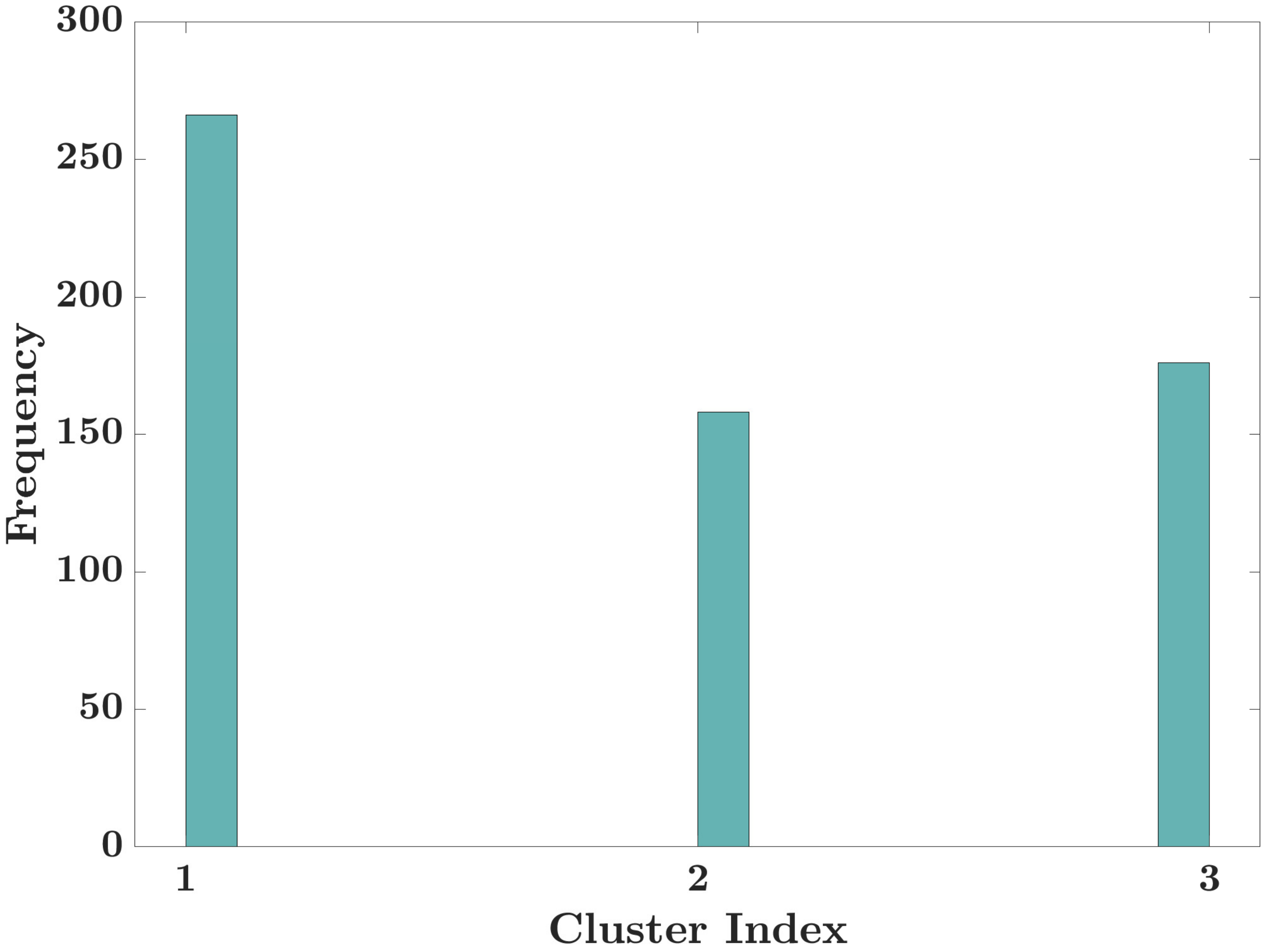}
	\caption{The histogram of the {\tt dbscan} clustering results from Algorithm~5 at 30dB SNR.}
	\label{figure1007}
\end{figure}

We used the CVX package and selected the hyper-parameters of the optimization problem by gridding: at $30$dB $\lambda=2$, 
$\gamma_1=10^5$ to enforce sparsity on the over-parameterized part, but not too large to avoid numerical ill conditioning  
and $\gamma_2=10^{-5}$ to control nonzero components of the parameter vector, cf. (\ref{thetared}), but not too small 
to avoid again numerical ill conditioning. The weights $\gamma_1$ and $\gamma_2$ may be selected more or less the same 
for the three SNRs. However, for best results $\lambda$ should be adjusted according to the SNR. The segments satisfying 
$\delta_j(\chi_\theta) \geq 150$ were chosen to extract discrete-state estimates. Points in these segments accounted about 
$70\%$ of the time interval. Applying Algorithm~5, we estimated the discrete-states. The eigenvalue-based statistics 
${\mathcal M}(\hat{A}_{\varphi(k_i)})$ was applied to them. The {\tt dbscan} clustering algorithm retrieved all discrete 
states. In Figure~\ref{figure1007}, the histogram of the clustering results for a single run of the algorithm at 30dB 
SNR is plotted for the three largest clusters. The number of the discrete states is then correctly estimated as 
$\hat{\sigma}=3$.

\begin{figure}[hbt!]
	\centering
	\includegraphics[width=3in]{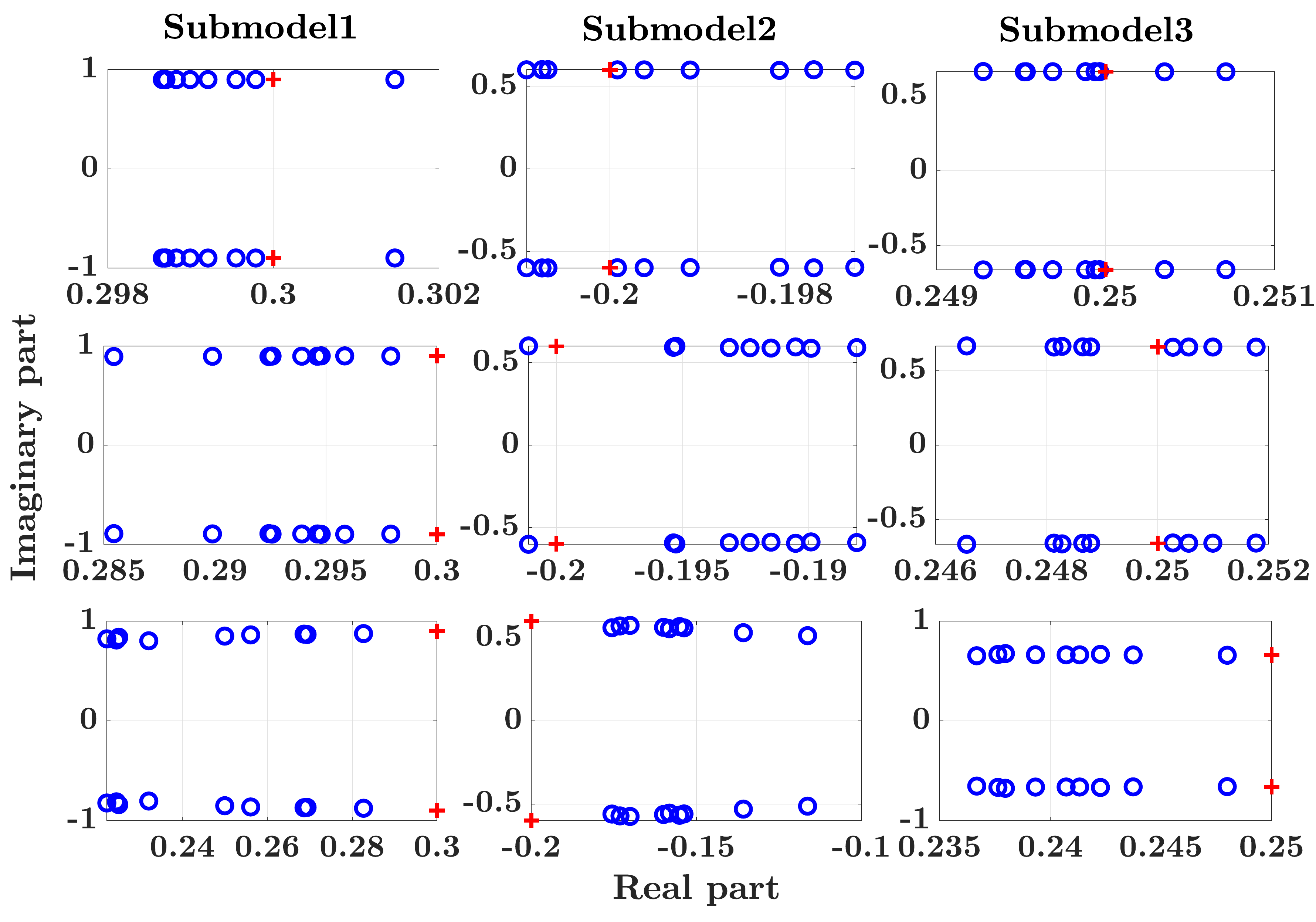}
	\caption{The true and the estimated discrete state eigenvalues: the true '{\bf +}', the estimated '$\circ$' 
		at 40dB, 30dB, and 20dB SNRs, the top to the bottom respectively for different noise realizations. }
	\label{figure1008}
\end{figure}

Figure~\ref{figure1008} shows the true eigenvalues and the eigenvalue estimates of the discrete states at 
three different SNRs. The estimates are very accurate even for the smallest SNR. 

\begin{figure}[hbt!]
	\centering
	\includegraphics[width=3in]{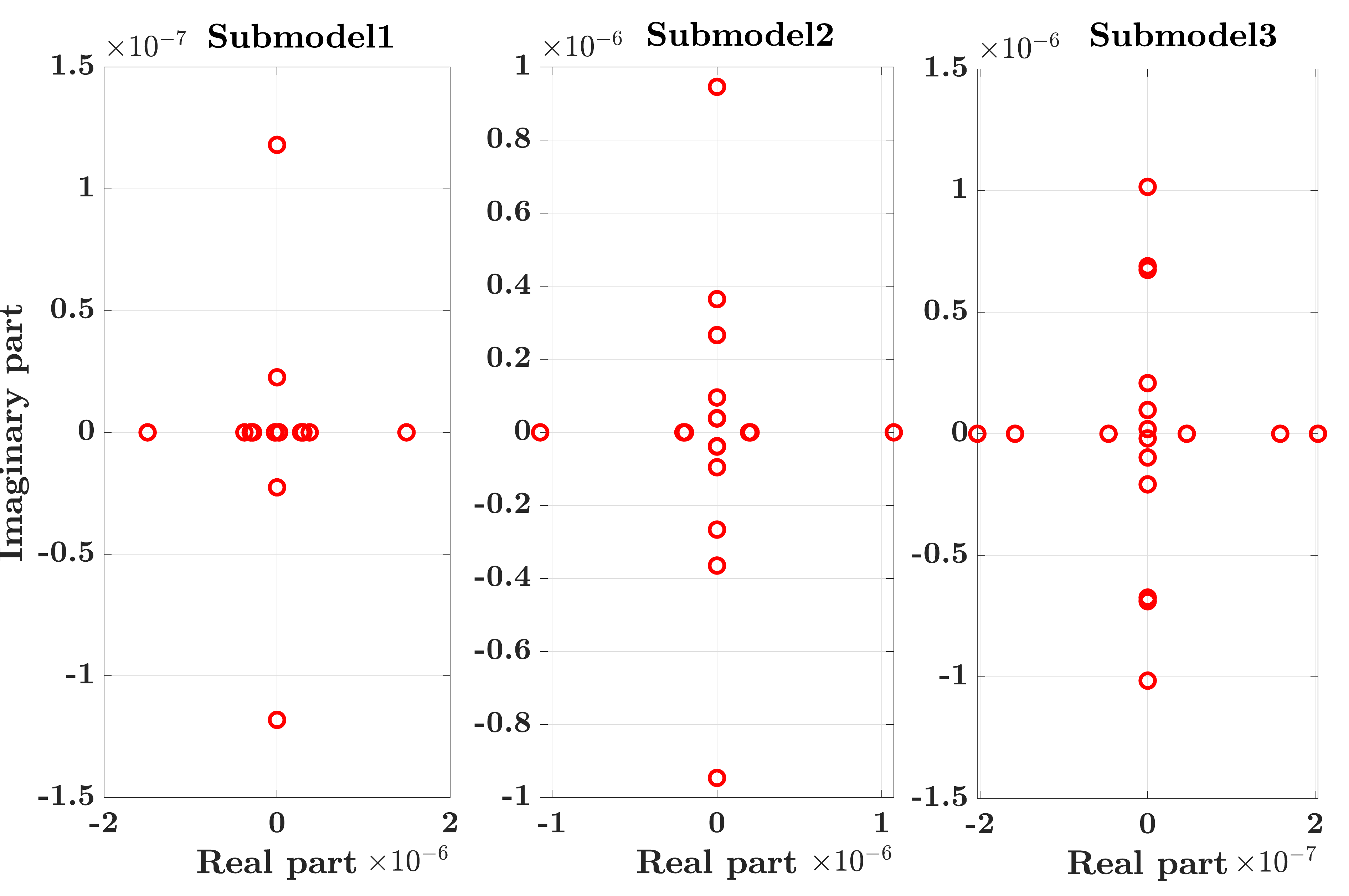}
	\caption{The eigenvalues of $A_{\rm o}(k)$ for the $3$ submodels at 30dB SNR for several runs with different noise realizations.}
	\label{figure1009}
\end{figure}

Now, we verify that the optimal solution of (\ref{optim1}) is indeed a deadbeat observer. Recall that
this condition is implicitly enforced through the feasible parameter set. From (\ref{Kes}), we estimated 
$g(k)$ and calculated the eigenvalues of $A_{\rm o}(k)$ for the $3$ submodels. In Figure~\ref{figure1009}, 
the Monte Carlo simulation results are plotted at 30dB SNR. The submodel eigenvalues require two steps only
to the origin, that is, $\tau= n = 2$.

\begin{figure}[hbt!]
	\centering
	\includegraphics[width=3in]{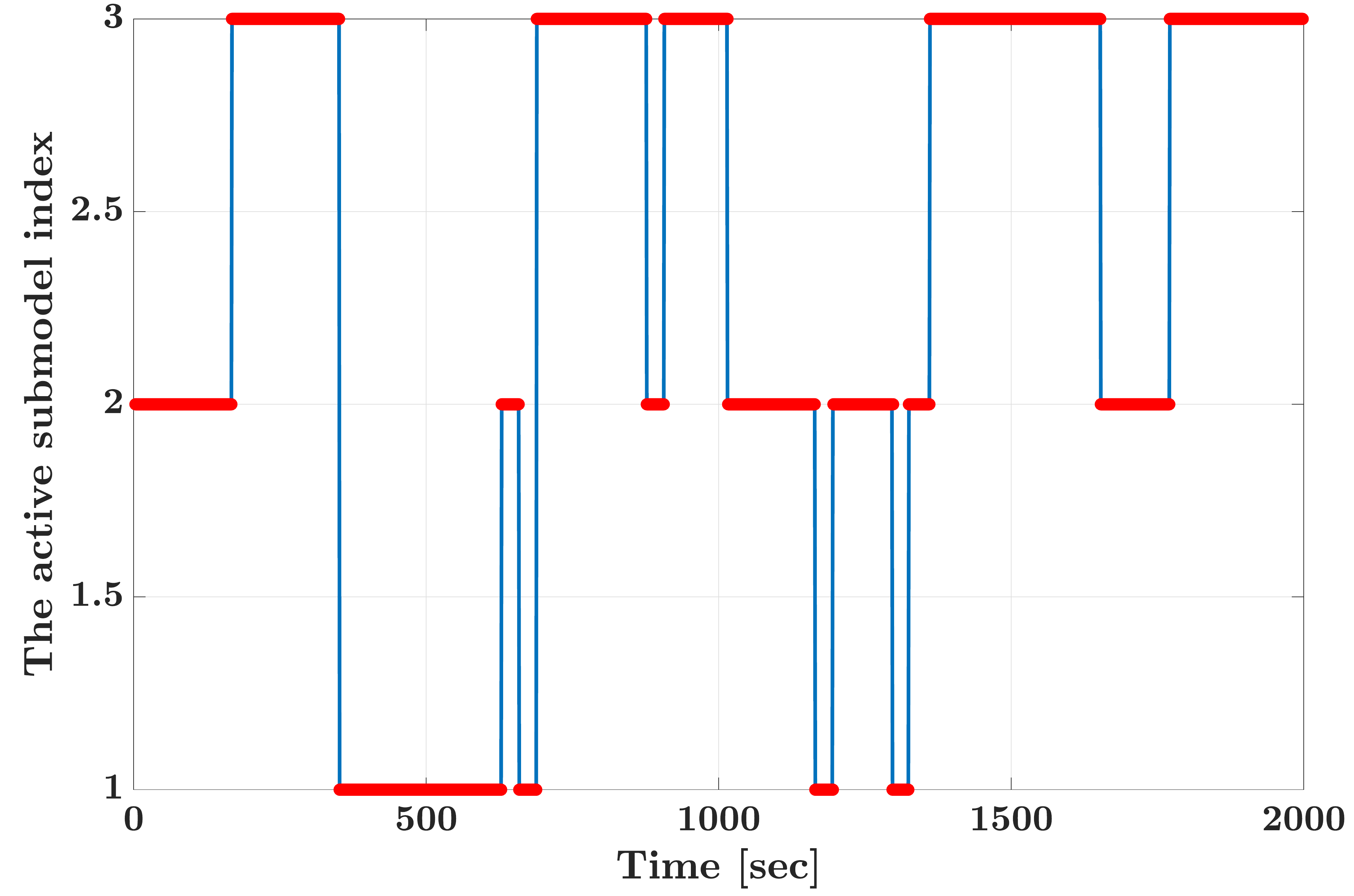}
	\caption{The switching sequence estimate $\hat{\chi}$ (dotted) and the true switching sequence 
		$\chi$ (solid) for the SLS model.}
	\label{figure10010}
\end{figure}

\begin{figure}[hbt!]
	\centering
	\includegraphics[width=3in]{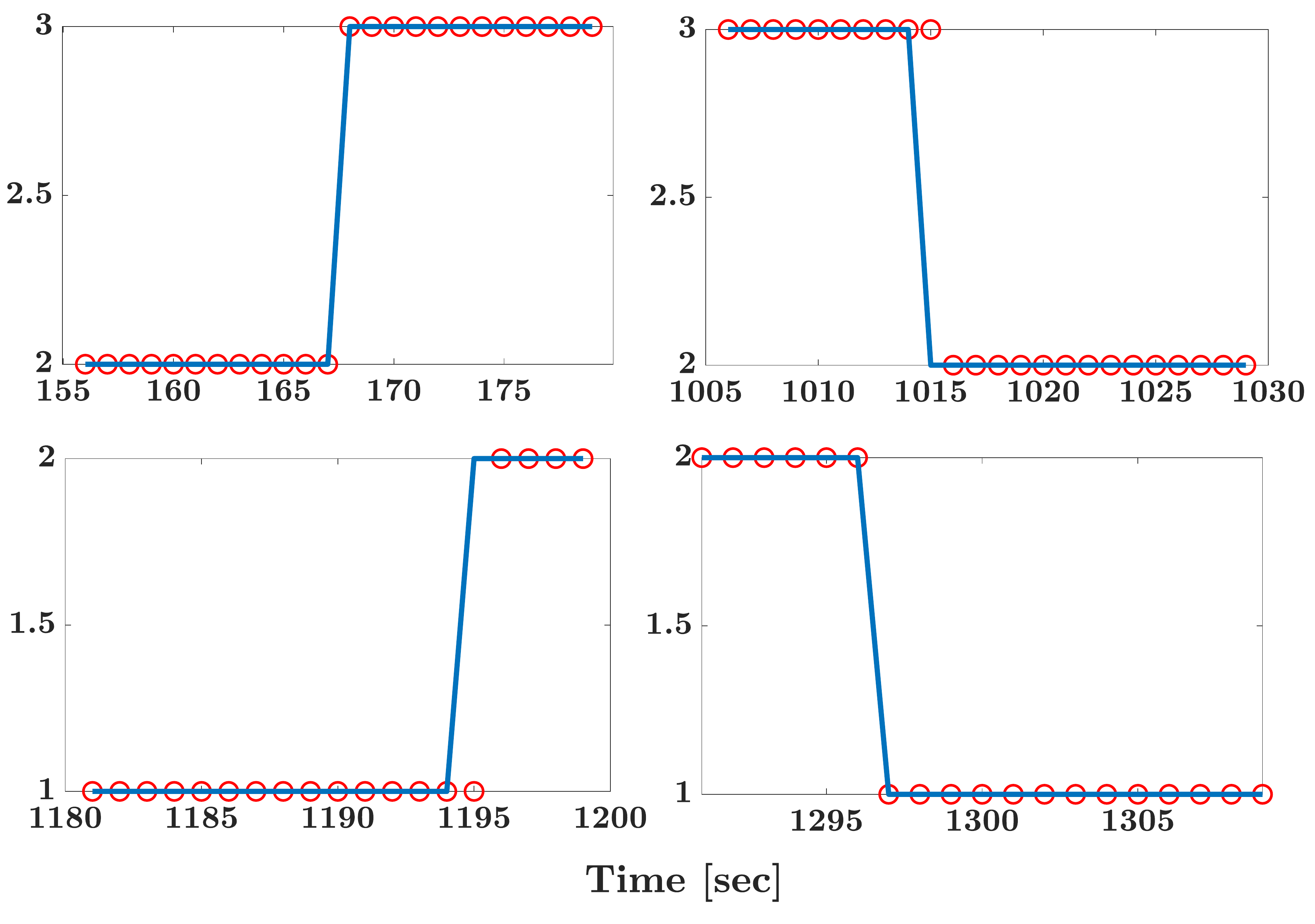}
	\caption{The switching sequence estimate (dotted) and the true switching sequence (solid) obtained by zooming
		from Figure~\ref{figure10010}.}
	\label{figure10011}
\end{figure}

\begin{figure}[hbt!]
	\centering
	\includegraphics[width=3in]{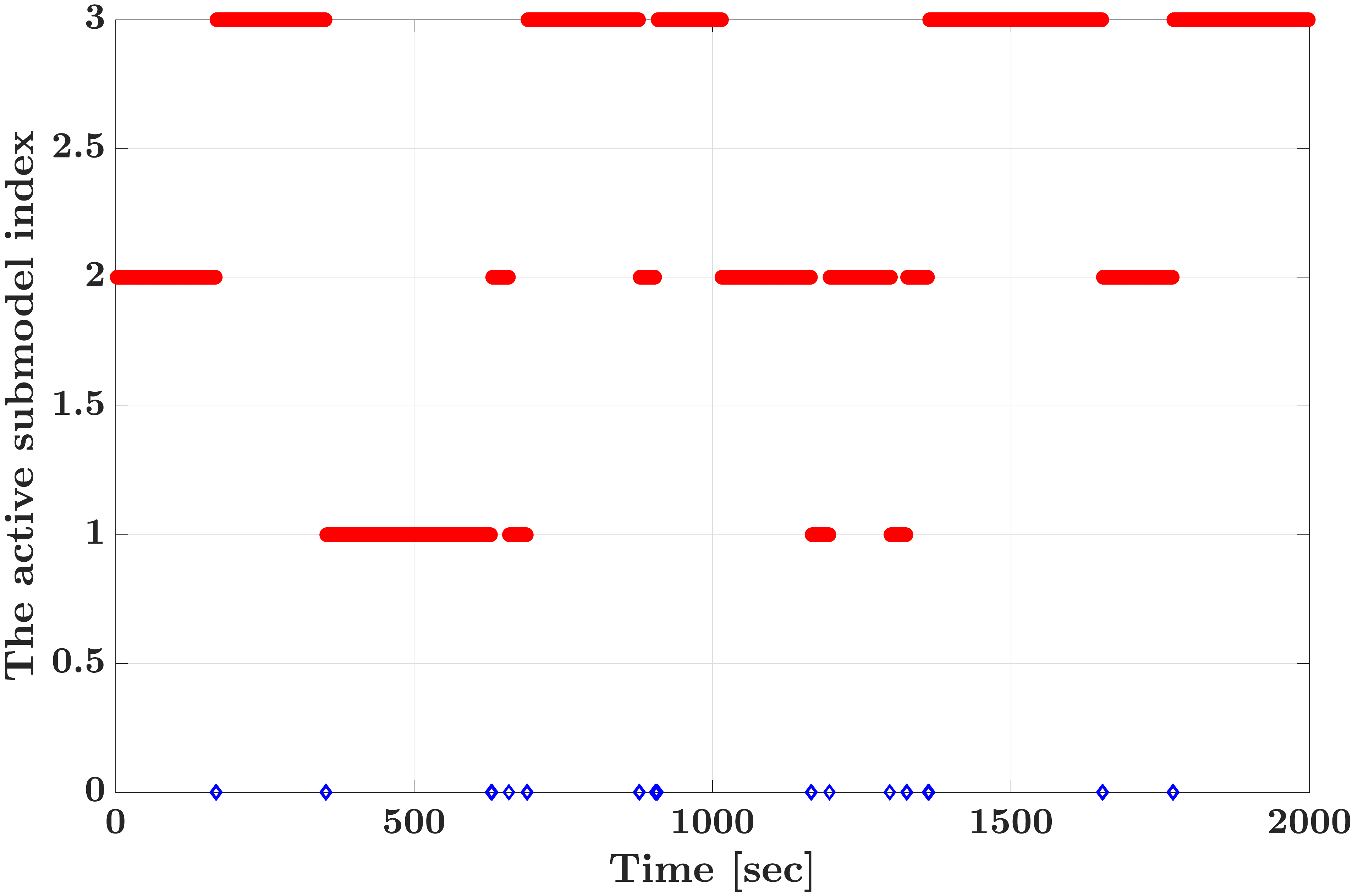}
	\caption{The switching sequence estimate $\hat{\chi}_\theta$ for the SARX model.}
	\label{figure10012}
\end{figure}

Using Algorithm~4$^\prime$ for $2n< k < N-2n$ and excluding points in the long segments, we estimated $\varphi(k)$ at 30dB 
SNR. The estimation results plotted in Figure~\ref{figure10010} show perfect match to the 
true switching sequence except few points zoomed in a separate figure for clarity, see Figure~\ref{figure10011}. In
Figure~\ref{figure10012}, the estimate of $\chi_\theta$ is plotted. Recall that $\chi_\theta$ is derived from $\chi$ under 
an observer transformation. Three groups of points are visible in the figure: points lying in the long segments, points 
lying in the short segments, and points lying in the very short segments. Recall that points in the third group belong 
to intervals with lengths less than $4n$. Points in the first group are determined by Algorithm~5. 

\begin{figure}[hbt!]
	\centering
	\includegraphics[width=3in]{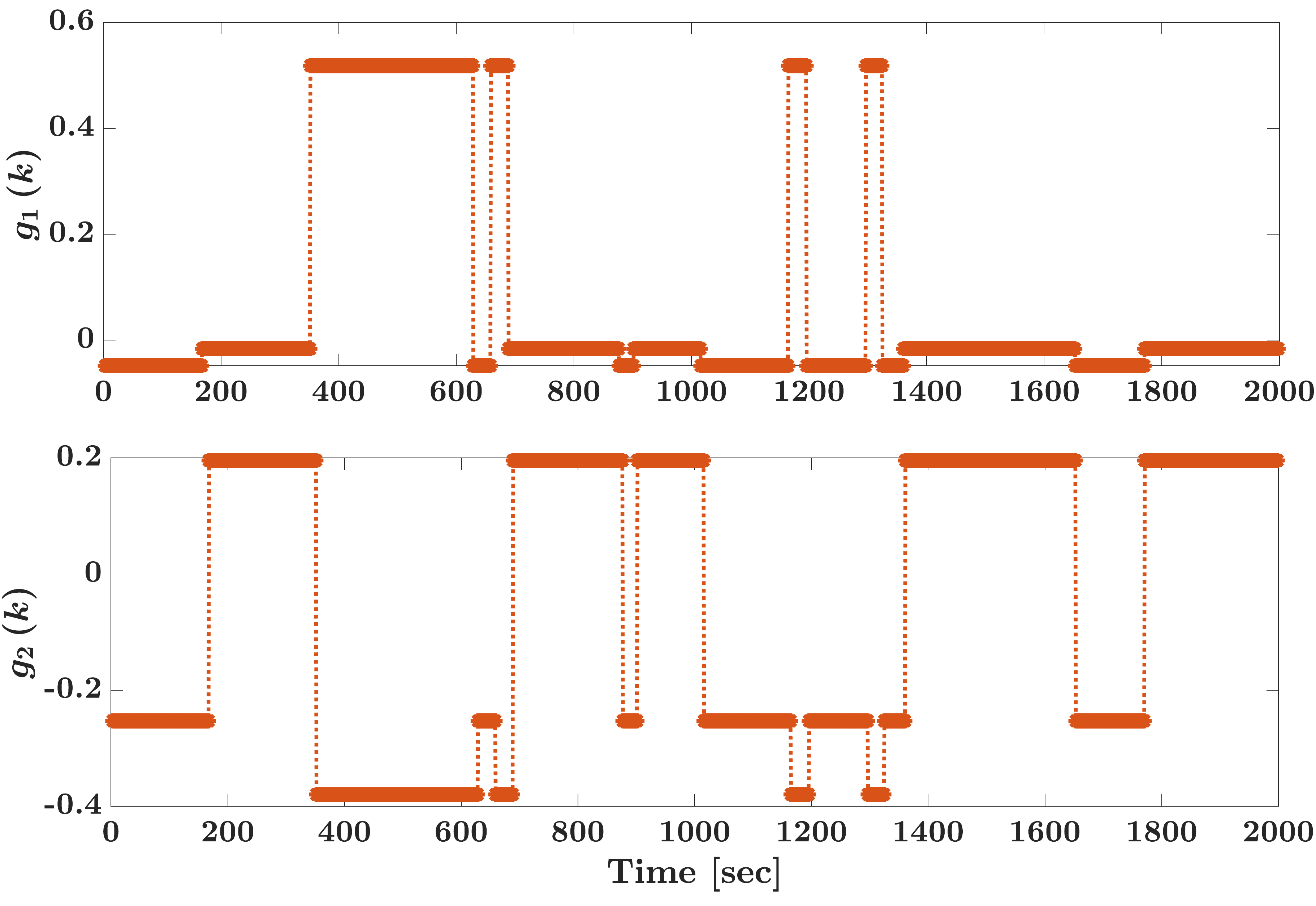}
	\caption{The deadbeat observer gain components $g(k)$.}
	\label{figure10013}
\end{figure}

In Figure~\ref{figure10013}, at 30dB SNR the components of the deadbeat observer gains are plotted for a single run of the algorithm over the interval $(2n,N-2n)$. To show that over-parameterization does not
influence estimation accuracy, with the ordo notation $X=O(\varepsilon)$ meaning $\|X\|_2 \leq \varepsilon$, 
we list $\hat{\theta}(k)$ delivered by Algorithm~5 for the first three long segments:
\begin{eqnarray*}
	\begin{array}{cccccccccc} 
		\theta_1 & = & ( &  0.47  & 1.33 &  0.57 & -0.97 & -0.86 & X_1 & )^T \\
		\theta_2 & = & ( & -1.5  & 0.88 & -0.4 & -4.29 & -0.40 & X_2 & )^T \\
		\theta_3 & = & ( &  2.49  & 0.50 &  0.49 & -8.61 & -0.49 & X_3 & )^T
	\end{array}
\end{eqnarray*}
where $X_k^T=O(10^{-17}) \in \mathbb{R}^4$, $k=1,2,3$. Hence, $X_k \approx 0$. Recall that $\tau=n=2$ 
can be chosen on $[k_i+n,k_{i+1})$ for all $k_i \in \chi$ by Lemma~\ref{deadbeatlem}.

\subsubsection{Average case performance}

We end this numerical example by studying the average case performance of the proposed scheme and comparing it with
another scheme \cite{BakoVanLuongLauerBloch2013} from the literature on hybrid systems. The scheme presented in \cite{BakoVanLuongLauerBloch2013} is similar to the scheme put forward in this paper in that it utilizes 
a sparse optimization stage. But, it assumes availability of the state measurements $x(k)$ which is a rather 
unrealistic assumption unless $C(k)=I_n$ and $D(k)=0$ for all $k$ in (\ref{ssy}). In this work, computational 
complexity of the state estimation was not reported.

\begin{figure}[hbt!]
	\centering
	\includegraphics[width=3in]{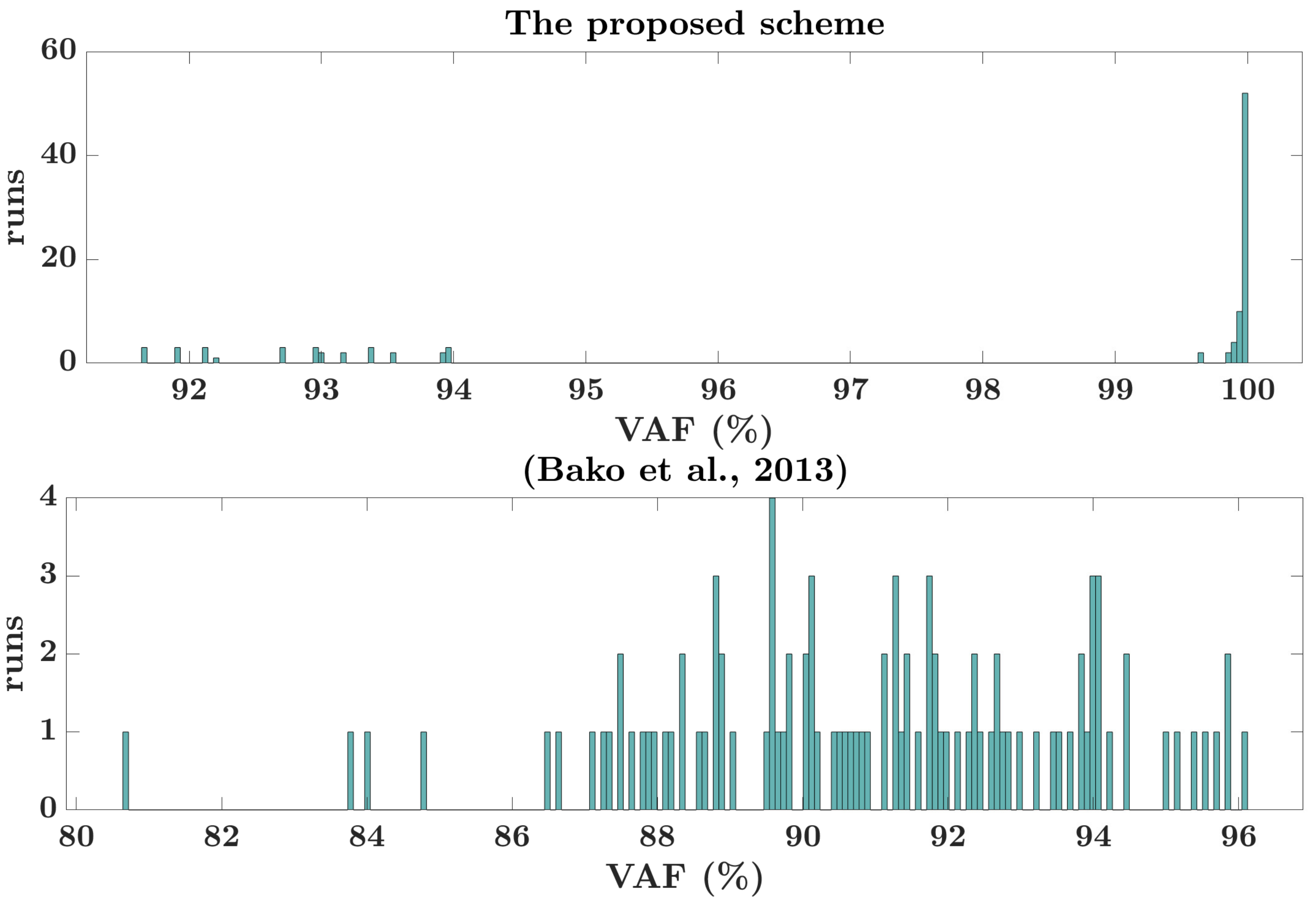}
	\caption{The distribution of VAF  over $100$ noise realizations at 20dB SNR.}
	\label{fig10015}
\end{figure}

We will perform Monte Carlo simulations by producing a new noise realization for the SLS in the beginning of this section. 
The average case performance will first be assessed  by computing the relative error $\|M_\ell-\hat{M}_\ell\|_F/\|M_\ell\|_F$
for the quadruples $\hat{\mathcal P}_l=({\hat A}_\ell,{\hat b}_\ell,\hat{c}_\ell^T,{\hat d}_\ell)$, $\ell \in \mathbb{S}$ 
where
\begin{equation*}
	\hat M_\ell = \left[ {\begin{array}{*{20}{c}}{\hat A}_\ell &{{{\hat b}_\ell}}\\
			{\hat c_\ell^T}&{{{\hat d}_\ell}}\end{array}} \right]
\end{equation*}
is the linear map $(x^T(k)\;\;u^T(k))^T \mapsto (x^T(k+1)\;\;u^T(k))^T$ induced by $\hat{\mathcal P}_l$. This criterion 
was used in \cite{BakoVanLuongLauerBloch2013}. We sum the relative errors over the $\sigma$ discrete states and average 
them over $100$ noise realizations. The result will be denoted by $\delta_r$. In Table~1, the mean value for $\delta_r$ 
is displayed for $20,30$, and $40$dB SNRs for the two schemes. The proposed scheme outperforms the scheme in \cite{BakoVanLuongLauerBloch2013}. 
A second criterion for the performance assessment will be defined on validation data sets as follows. On a fresh data set
for each noise realization, we compute a variance-accounted-for (VAF) value 
\begin{equation}
	{\rm VAF} = \max \left\{ {1 - \frac{{{\mathop{\rm var}} \left( {y\left( k \right) - \hat y\left( k \right)} \right)}}
		{{{\mathop{\rm var}} \left( {y\left( k \right)} \right)}},0} \right\} \times 100\% 
\end{equation}
In Table~1, the mean of VAF for a range of the SNRs are shown. An inspection of the table reveals better performance 
for the presented scheme when subjected to output predictions. The average run time of the proposed scheme per noise 
realization denoted by $t_{\rm CPU}$ equals $55$ secs. For the scheme proposed in \cite{BakoVanLuongLauerBloch2013},
$t_{\rm CPU}=10.5$ secs. The latter scheme is computationally less expensive since it does not include the cost of
state estimation. Figure~\ref{fig10015} reinforces our conclusion drawn earlier. It shows the distribution of VAF 
over $100$ noise realizations. The VAF concentrates at $100$ for the proposed scheme whereas they are
scattered for the scheme in \cite{BakoVanLuongLauerBloch2013}.

\begin{table}
	\begin{center}
		\begin{tabular}{|c|c|c|c|c|c|c|} 
			\hline
			\textbf{Schemes}  & \multicolumn{3}{c|}{$\delta_r$} & \multicolumn{3}{c|}{\textbf{VAF} ($\%$)} \\
			\hline 
			SNR (dB) & 40dB & 30dB & 20dB & 40dB & 30dB & 20dB \\ 
			\hline
			\texttt{Proposed}  & 0.0110 & 0.0234 &  0.1507 & 99.99 & 99.25 & 97.93 \\ 
			\hline
			\cite{BakoVanLuongLauerBloch2013}  & 0.0122 &  0.0270 &  0.1509 & 99.83 & 98.08 & 90.90\\
			\hline
		\end{tabular}
		\caption{$\delta_r$ and the means of VAF ($\%$) computed over $100$ noise realizations.}
		\label{table1}
	\end{center}
\end{table}

\begin{figure}[hbt!]
	\centering
	\includegraphics[width=3in]{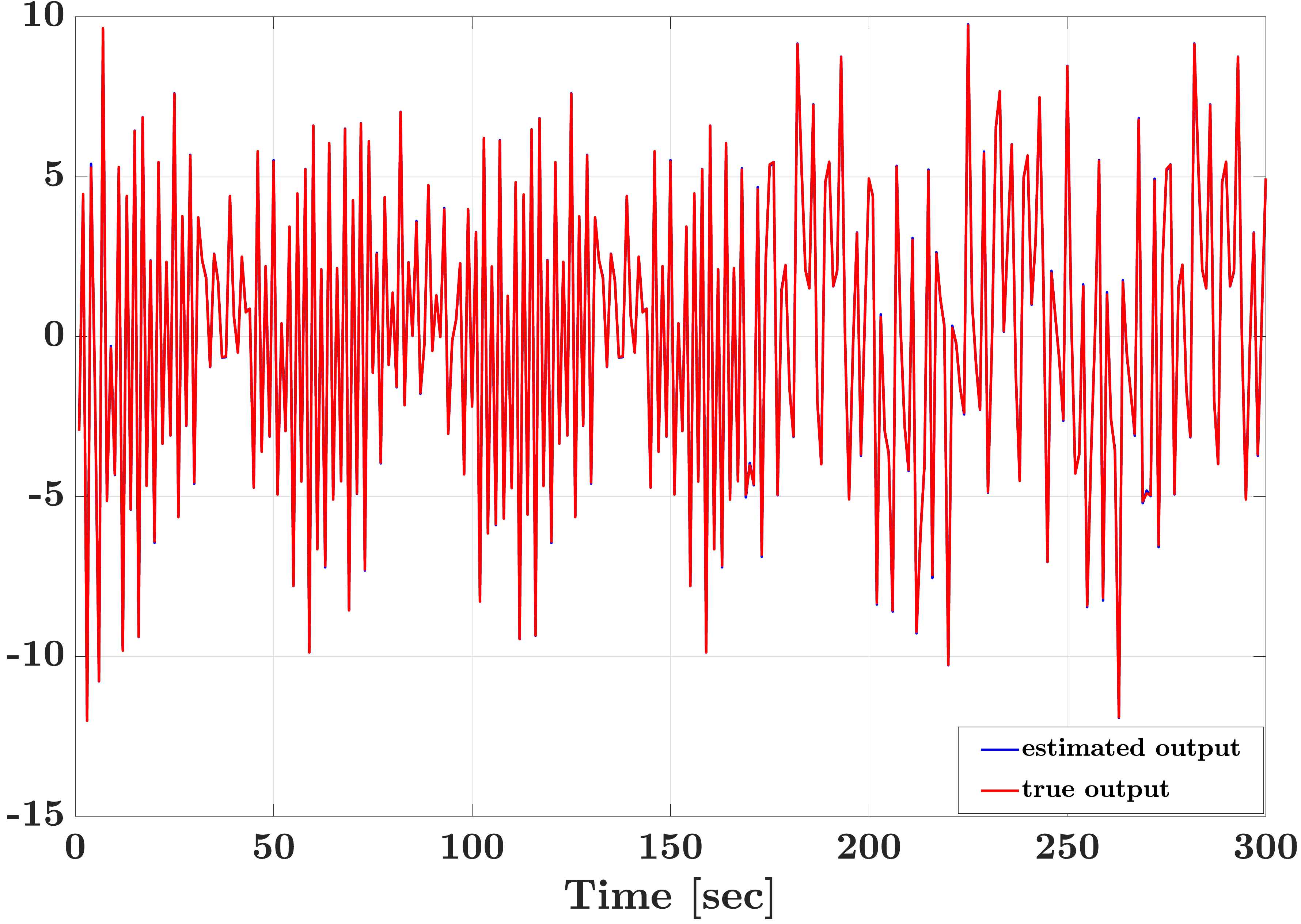}
	\caption{The true and the estimated outputs on the validation data set. The SNR is 30dB and VAF is 99.39$\%$}
	\label{figure10014}
\end{figure}

In Figure~\ref{figure10014}, the true and the predicted outputs are plotted on the validation data set. 
Only the first $300$ data points are displayed for visualization purpose. A high value for VAF indicates 
superior quality of the prediction.	We calculated $\hat{y}(k)$ after applying a basis transformation to
all submodel estimates in the set $\hat{\mathcal P}$. The observability canonical form suggested in 
\cite{Mercere&Bako:2011} was used. A basis transformation is required since the system Markov parameters 
in a transition band are calculated from state-space realizations of two different discrete states.

\section{Conclusions}\label{concsec}

This paper extended the identification problem for SARX systems to SLSs in state-space form via a 
deadbeat observer-based transformation which packs an infinite sequence of system Markov
parameters into a finite sequence of observer Markov parameters. The pay-off for this data compression is more
complicated discrete model sets and switching sequences in the transformation domain. A careful study 
of the switch and submodel identifiability issues laid foundations for an integrated approach to the 
identification problem. A non-convex and sparse optimization problem formulation devised an algorithm
to estimate the discrete states up to arbitrary similarity transformations. This stage was complemented 
by the retrieval of switching sequences via a MOESP subspace algorithm. Convex relaxation of the
non-convex optimization problem by the BBPDN method was demonstrated to be effective both theoretically 
and numerically. 

Identifiability of discrete states of an SLS was exhibited as the PE conditions over segments. When a 
PE condition fails over a segment, the parameters of the discrete state that is active in this segment cannot 
be determined uniquely. In fact, this failure was shown to be an intrinsic feature of all deadbeat observers. 
As a result, recovery and robustness guarantees for the sparse optimization stage cannot be derived 
for the entire observation interval. In the literature on hybrid systems, a system identification procedure 
typically starts with the estimation of switching sequence from input-output data. Then, local models are 
parametrically estimated from input-output data over segments. In the current paper, we carried out system
identification in reverse order. This approach facilitated local analysis, which was resorted very often in 
the paper. 

In this paper, we considered SISO-SLSs to avoid delicate parameterization issues. Extension to MIMO-SLSs is not
difficult by recognizing that the derivations in Sections~\ref{observertrans}--\ref{bpdnsec} except for \ref{switchidsec} 
and \ref{PEsubsec} apply to MIMO-SLSs verbatim. Section~\ref{switchidsec} requires only notational changes. The PE 
condition in \ref{PEsubsec} can be generalized by utilizing the matrix-fraction descriptions and the co-prime factorizations, 
see \cite{Kailath:1980}. Replacement of the BBPDN algorithm in Figure~\ref{figure1001} with a greedy 
algorithm, for example the BOMP algorithm, is an open problem we have not tackled in this paper.

\section*{Acknowledgment} 
The authors would like to thank Professor Laurent Bako for providing them a source code of the algorithm 
in \cite{BakoVanLuongLauerBloch2013}.


\section*{Appendix A}\label{appA}

{\em Proof of Lemma~\ref{deadbeatlem}.} For each $1 \leq k \leq N$, there is a gain $g(k)$ satisfying
\begin{equation}\label{jkl107}
	(A_{\varphi(k)}+g(k) c^T_{\varphi(k)})^n=0.
\end{equation}
Extend $g(k)$ such that  $g(k)=g(l)$ if $\varphi(k)=\varphi(l)$. Thus, 
$g(k)$ is a sequence of at most $\sigma$ vectors $g_{\varphi(k_i)}$, $i=0,\cdots,i^*$. 

Let $\epsilon,\ell \in \mathbb{N}$ be such that $0< k-\epsilon < \ell < k$. Write
\begin{eqnarray}\label{jkl2}
	\Phi_{\rm o}(k,k-\epsilon) = \Phi_{\rm o}(k,\ell)\Phi_{\rm o}(\ell,k-\epsilon).
\end{eqnarray}
First, let us assume that $\epsilon \geq 2n-1$. 
At least one of the inequalities $\ell-(k-\epsilon) \geq n$ or $k-\ell \geq n$ must be true then. If not,  
$\ell-(k-\epsilon) < n$ and $k-\ell < n$ or $\ell-(k-\epsilon) \leq n-1$ and $k-\ell \leq n-1$ and summing the 
last two inequalities, we get $\epsilon \leq  2n-2$. We reach a contradiction. 

Suppose $\ell \in \chi$. Then, $\ell=k_i$ for some $0 < i<i^*$. Assume $\delta_{i-1}(\chi) \geq n$ and 
$l-(k-\epsilon) \geq n$. Then, from (\ref{jkl107}) 
\begin{eqnarray*}
	\Phi_{\rm o}(l,k-\epsilon) &=& A_{\rm o} (l-1) \;\cdots\; A_{\rm o} (l-n) \Phi_{\rm o}(l-n,k-\epsilon) \\
	&=& A_{\rm o}^n(k_{i-1})\Phi_{\rm o}(l-n,k-\epsilon)=0.
\end{eqnarray*}
If $k-\ell \geq n$  and $\delta_i(\chi) \geq n-1$, again from (\ref{jkl107})
\begin{eqnarray*}
	\Phi_{\rm o}(k,\ell) &=& \Phi_{\rm o} (k,\ell+n) A_{\rm o} (l+n-1) \;\cdots\;A_{\rm o}(l) \\
	&=& \Phi_{\rm o}(k,\ell+n) A_{\rm o}^n(k_i)=0.
\end{eqnarray*}
Hence, if $\delta_*(\chi) \geq n$, $\epsilon \geq 2n-1$, and there is a switch in the interval $(k-\epsilon\;\;k)$,
from (\ref{jkl2}) we get $\Phi_{\rm o}(k,k-\epsilon)=0$. If there is no switch in $(k-\epsilon\;\;k)$,
$(k-\epsilon\;\;k) \subset [k_{i-1}\;\;k_i)$ for some $i$ and
\[
\Phi_{\rm o}(k,k-\epsilon) = A_{\rm o} (k-1) \;\cdots\; A_{\rm o} (k-\epsilon)=A_{\rm o}^{\epsilon}(k_{i-1})=0.
\]
Since $k$ was arbitrary, $\Phi_{\rm o}(k,k-\epsilon)=0$ for all $k$ if $\delta_*(\chi) \geq n$ and 
$\epsilon \geq 2n-1$. Then, $\tau^* \leq 2n-1$ if $\delta_*(\chi) \geq n$.  

If $\delta_*(\chi) \geq n$, then $\tau^* < 2n$ is a tight bound. In fact, suppose that there exist
three switches $k_{i-1},k_i,k_{i+1}$ satisfying the equalities $k_i-k_{i-1}=k_{i+1}-k_{i}=n-1$. Set 
$k=k_{i+1}$, $\ell=k_i$, and $\epsilon=2n-2$. Then, $\Phi_{\rm o}(l,k-\epsilon) \neq 0$
if $(A_j+g_jc^T_j,c_j^T)$ with $j=\varphi(k_{i-1})$ is observable since $l-k+\epsilon=n-1$. Next, 
$\Phi_{\rm o}(k,l) \neq 0$ if $(A_j+g_jc^T_j,c^T_j)$ with $j=\varphi(k_i)$ is observable 
since $k-l=n-1$. Thus, $\Phi(k,k-\epsilon) \neq 0$ and $\tau^* >2n-2$. 

It follows that $\Phi_{\rm o}(k,k-2n+1)=0$ if $k \geq 2n$ and $\delta_*(\chi)\geq n$. The requirements
$k \geq 2n$ and $\tau^*=2n-1$ are automatically satisfied by selecting a gain $g_{\varphi(k_0)}$ 
from $\delta_0(\chi) \geq 2n$. If for $i>0$ and $k \in [k_i+n\;\;k_{i+1})$, with $g(k)=g_{\varphi(k_i)}$
we have
\[
\Phi_{\rm o}(k,k-n)=(A_{\varphi(k_i)}+g_{\varphi(k_i)}c^T_{\varphi(k_i)})^n=0.
\]

\section*{Appendix B}\label{appB}
{\em Proof Lemma~\ref{lemreal}}. The proof of this lemma is similar to the proof of Lemma~\ref{deadbeatlem}. 
Consider the set of the inequalities $k-\kappa_{\rm c} < \ell <k-1$, $\ell \in \mathbb{N}$ where $\kappa_{\rm c} \geq 2n$. 
Then, at least one of the inequalities $\ell-k+\kappa_{\rm c} \geq n$ or $k-\ell-1 \geq n$ must be true.

Suppose $k-\ell -1 \geq n$ and $\ell \in \chi$. Then,  for some $0 < i<i^*$, $\ell=k_i$. Assume
$\delta_i(\chi) \geq n$. For $\ell \leq j \leq \ell+n-1$, 
\begin{eqnarray*}
	\Phi(k,j+1) b(j) &=& \Phi(k,\ell+n) A(\ell+n-1)\;\cdots\;A(j+1)b(j) \\
	&=& \Phi(k,k_i+n)A^{k_i+n-j-1}(k_i) b(k_i)
\end{eqnarray*}
since $\delta_i(\chi) \geq n$ and $k-\ell>n$. Thus,
\begin{eqnarray*}
	G_{\rm c}(k,\kappa_{\rm c}) &=&  \sum_{j=k-\kappa_{\rm c}}^{k-1} \Phi(k,j+1) b(j) b^T(j) \Phi^T(k,j+1) \\
	&\geq& \sum_{j=k_i}^{k_i+n-1} \Phi(k,j+1) b(j) b^T(j) \Phi^T(k,j+1) \\
	&=& \Phi(k,k_i+n)\sum_{j=k_i}^{k_i+n-1} A_{\varphi(k_i)}^{k_i+n-j-1}b_{\varphi(k_i)} b^T_{\varphi(k_i)} \\
	&{}& \;\;\; \cdot \, [A_{\varphi(k_i)}^{k_i+n-j-1}]^T \Phi^T(k,k_i+n)\\
	&=&  \Phi(k,k_i+n) \sum_{v=0}^{n-1} A_{\varphi(k_i)}^v b_{\varphi(k_i)} b^T_{\varphi(k_i)} \\
	&{}& \;\;\; \cdot \, [A^v_{\varphi(k_i)}]^T \Phi^T(k,k_i+n).
\end{eqnarray*}
Since ${\mathcal P}_{\varphi(k_i)}$ is stable, $\Phi(k,k_i+n)$ is nonsingular. The middle term
is the controllability Grammian of the minimal discrete state ${\mathcal P}_{\varphi(k_i)}$.
Hence, it is positive definite. 

Now, suppose $\ell-k+\kappa_{\rm c} \geq n$ and $\ell \in \chi$ again. Assume $\delta_{i-1}(\chi) \geq n$.
For $\ell-n \leq j \leq  \ell-2$, 
\begin{eqnarray*}
	\Phi(k,j+1) b(j) &=& \Phi(k,\ell)\;A(\ell-1)\;\cdots\;A(j+1) b(j) \\
	&=& \Phi(k,k_i)A^{k_i-j-1}(k_{i-1}) b(k_{i-1})
\end{eqnarray*}
since $\ell-n \geq k-\kappa_{\rm c}$ and $\delta_{i-1}(\chi) \geq n$. Thus,
\begin{eqnarray*}
	G_{\rm c}(k,\kappa_{\rm c}) &\geq& \Phi(k,k_i)\sum_{j=\ell-n}^{\ell-2} A_{\varphi(k_{i-1})}^{k_i-j-1}
	b_{\varphi(k_{i-1})} b^T_{\varphi(k_{i-1})} \\
	&{}& \;\;\; \cdot \, [A_{\varphi(k_{i-1})}^{k_i-j-1}]^T \Phi^T(k,k_i)\\
	&=&  \Phi(k,k_i) \sum_{v=0}^{n-1} A_{\varphi(k_{i-1})}^v b_{\varphi(k_{i-1})} b^T_{\varphi(k_{i-1})} \\
	&{}& \;\;\; \cdot \, [A^v_{\varphi(k_{i-1})}]^T \Phi^T(k,k_i) > 0
\end{eqnarray*}
since ${\mathcal P}_{\varphi(k_i)}$ is stable and hence $\Phi(k,k_i)$ is nonsingular and the middle term being 
the controllability Grammian of the minimal discrete state ${\mathcal P}_{\varphi(k_{i-1})}$. Thus, if
$\delta_*(\chi) \geq n$, $\kappa_{\rm c} \geq 2n$, and there is a switch in $(k-\kappa_{\rm c}\;\;k-1)$, then 
$G_{\rm c}(k,\kappa_{\rm c}) > 0$. If there is no switch in $(k-\kappa_{\rm c}\;\;k-1)$,
$(k-\kappa_{\rm c}\;\;k-1) \subset [k_{i-1}\;\;k_i)$ for some $i$ and from the $l-k+\kappa_{\rm c} \geq n$ case, we 
again have $G_{\rm c}(k,\kappa_{\rm c}) > 0$. Fix $\kappa_{\rm c}$ as $\kappa_{\rm c}=2n$. Since $k$ lies
in the compact set $[1\;\;N]$ and $\sigma<\infty$, one can easily determine some constants $\alpha_0,\alpha_1,\beta_0,\beta_1$, and $\delta_{\rm c}$ in the definition of uniform observability.
The boundedness of (\ref{ssx})--(\ref{varphit}) is obvious. 

Although duality arguments may be used to show that $G_{\rm o}(k,\kappa_{\rm o})$ is positive definite, 
we will prove it directly. This time, we will consider the inequalities $k<\ell < k+\kappa_{\rm o}-1$ with 
$\delta_*(\chi) \geq n$ and fix $\kappa_{\rm o}=2n$. We first examine the $\ell-k \geq n$ case with 
$\ell=k_i$ for some $0<i<i^*$. Then,
\begin{eqnarray*}
	G_{\rm o}(k,\kappa_{\rm o}) &=& \sum_{j=k}^{k+\kappa_{\rm o}-1} \Phi^T(j,k) c(j) c^T(j) \Phi(j,k) \\
	&\geq& \sum_{j=k}^{k+n-1} \Phi^T(j,k) c(j) c^T(j) \Phi(j,k)\\
	&=& \sum_{v=0}^{n-1} [A_{\varphi(k_{i-1})}^v]^T c_{\varphi(k_{i-1})} c^T_{\varphi(k_{i-1})} A^v_{\varphi(k_{i-1})} > 0
\end{eqnarray*}
since $\delta_{i-1}(\chi) \geq n$. If $k+\kappa_{\rm o}-1-\ell \geq n$, from  $\delta_i(\chi) \geq n$
\begin{eqnarray*}
	G_{\rm o}(k,\kappa_{\rm o}) &\geq& \sum_{j=\ell}^{\ell+n-1} \Phi^T(\ell,k)\Phi^T(j,\ell) c(j) c^T(j) \Phi(j,\ell)\Phi(\ell,k)\\
	&{}& \hspace{-10mm} = \, \Phi^T(\ell,k) \sum_{v=0}^{n-1} [A_{\varphi(k_{i})}^v]^T c_{\varphi(k_{i})} c^T_{\varphi(k_{i})} A^v_{\varphi(k_{i})}	\Phi(\ell,k) > 0.  
\end{eqnarray*}
If there is no switch in $(k\;\;k+\kappa_{\rm o}-1)$, we then have for some $i$, $(k\;\;k+\kappa_{\rm o}-1) 
\subset [k_{i-1}\;\;k_i)$ and from the last case above we get $G_{\rm o}(k,\kappa_{\rm o}) > 0$. Similar 
comments to the uniform controllability case apply for the constants $\alpha_0^\prime,\alpha_1^\prime,\beta_0^\prime,\beta_1^\prime$, and $\delta_{\rm c}$ in the definition 
of uniform observability. From the definitions of $G_{\rm c}(k,\kappa_{\rm o})$ and $G_{\rm c}(k,\kappa_{\rm c})$, we see that
(\ref{ssx})--(\ref{varphit}) may be demanded uniform in $[\kappa_{\rm c}+1\;\;N-\kappa_{\rm o}+1]$.   

\section*{Appendix C}\label{appC}
{\em Proof of Lemma~\ref{ARXdwell}}. Suppose $k_i+n \leq k < k_{i+1}$. The first two components of the observer Markov
parameters are 
\[
h_{\rm o}(k,k-v) =\left\{\begin{array}{lr} \left[d(k_i) \;\;0\right], & v=0; \\
	c^T(k_i) B_{\rm o}(k_i), & v=1
\end{array} \right. 
\]
since $d(k)=d(k_i)$, $c(k)=c(k_i)$, and  $B_{\rm o}(k-v)=B_{\rm o}(k_i)$ for all $0 \leq v \leq n$. Recall that $g(k)=g(k_i)$ 
for all $k \in [k_i\;\;k_{i+1})$ from Lemma~\ref{deadbeatlem}. Hence, for all $k \in [k_i\;\;k_{i+1}]$ 
$$
A_{\rm o}(k)=A(k)+g(k) c^T(k)=A(k_i)+g(k_i)=A_{\rm o}(k_i)
$$ 
and therefore for $v \geq 2$ 
\begin{eqnarray*}
	h_{\rm o}(k,k-v)  &=& c^T(k) A_{\rm o}(k-1) \;\cdots\;A_{\rm o}(k-v+1)  B_{\rm o}(k-v) \\
	&=& c^T(k_i) A_{\rm o}^{v-1}(k_i)B_{\rm o}(k_i-v).
\end{eqnarray*}
Thus, $h_{\rm o}(k,k-v)=h_{\rm o}(k_i+n,k_i+n-v)$, $v \in [0\;\;n]$. Then, 
\[
\theta(k)=\theta(k_i+n), \qquad k_i+n \leq k_{i+1}.
\]
Pick $s_j$ as the greatest lower bound on $k_i+n$ and $s_{j+1}$ as the smallest upper bound on 
$k_{i+1}$ in $\chi_\theta$.

\section*{Appendix D}\label{appD}
{\em Proof of Lemma~\ref{ARXdwell2}.} Let $k,l \in [s_j+2n-1,s_{j+1})$. Then for $0 \leq v <2n$, $h_{\rm o}(k-v,k-v)=h_{\rm o}(l-v,l-v)=d(l-v)$ and $h_{\rm o}(k,k-v)=h_{\rm o}(l,l-v)$. In (\ref{gammaki}) in place of $i$ in $\gamma(k,i)$,
substitute $k-v$  
\begin{eqnarray}
	\gamma(k,k-v) &=& h_{\rm o}^{(1)}(k,k-v)+h_{\rm o}^{(2)}(k,k-v) \, h_{\rm o}^{(1)}(k-v,k-v) \nonumber \\
	&=&  h_{\rm o}^{(1)}(l,l-v)+h_{\rm o}^{(2)}(l,l-v) \, h_{\rm o}^{(1)}(l-v,l-v) \nonumber \\
	&=& \gamma(l,l-v), \qquad 0\leq v <2n. \nonumber
\end{eqnarray}
Since $l\geq s_j+2n-1$, notice that $d(l-v)=d(s_j)$. Hence, as long as $k-v \geq s_j$, we let $v \geq 2n$. 
Then, $h_{\rm o}^{(1)}(k-v,k-v)$ is well-defined and equals to $d(s_j)$. If $v\geq 2n$,
$h_{\rm o}^{(1)}(k,k-v)=0$ and $h_{\rm o}^{(2)}(k,k-v)=0$ . Thus, $\gamma(k,k-v)=0$. Combining both cases if 
$\max\{k,l\}<s_{j+1}$, we derive 
\[
\gamma(k,k-v)=\gamma(l,l-v), \qquad s_j+v \leq \min\{k,l\}.
\]

Now, for $v=1$ we have
$$
h(k,k-1)=\gamma(k,k-1)=\gamma(l,l-1)=h(l,l-1).
$$
For $v>1$, we proceed  by induction. Consider the second term on the right-hand side of (\ref{markopasa}). Denote it 
by $J(v)$. Substitute $k-v$ in place of $i$ in $J(v)$ and change the variable $j$ inside the summand to $\mu$. Then, 
change $\mu$ to $\xi=\mu+v-k$: 
\begin{eqnarray*}
	J(v) &=& \sum_{\mu=k-v+1}^{k-1} h_{\rm o}^{(2)}(k,\mu) \, h(\mu,k-v), \;\;\;v\geq 2 \\
	&=& \sum_{\xi=1}^{v-1} h_{\rm o}^{(2)}(k,k-v+\xi) \, h(k-v+\xi,k-v) \\
	&=& \sum_{\xi=1}^{v-1} h_{\rm o}^{(2)}(l,l-v+\xi) \, h(k-v+\xi,k-v).
\end{eqnarray*}
Thus, $J(v)$ is a linear combination of the system Markov parameters $h(k-1,k-v)$, ... , $h(k-v+1,k-v)$.
For $v=2$, $h(k-1,k-2)=\gamma(k-1,k-2)=\gamma(l,l-1)$ and $h(k,k-2)$ becomes a linear combination of 
$\gamma(l,l-2)$ and $\gamma(l,l-1)$. Put $v=3$. Then, $h(k-1,k-3)$ and $h(k-2,k-3)$ are the only terms in the 
linear combination. Hence, $h(k,k-3)$ is a linear combination of $\gamma(l,l-3)$, $\gamma(l,l-2)$, and
$\gamma(l,l-1)$. Assume $h(k,k-v)$ a linear combination of $\gamma(l,l-v)$, ... , $\gamma(l,l-1)$. Then, 
$J(v+1)$ becomes a linear combination of the terms $h(k-1,k-v-1)$, ... , $h(k-v,k-v-1)$ or 
$\gamma(l,l-v)$, ... , $\gamma(l,l-1)$ and so does $h(k,k-v-1)$ a linear combination of $\gamma(l,l-\mu)$ 
for $\mu=1,\cdots,v+1$. This completes the induction and $h(k,k-v)=h(l,l-v)$. The lower bound
$s_j+v \leq \min\{k,l\}$ is not violated throughout the iterations. The iterations stops when 
$\max\{k,l\}<s_{j+1}$ is reached. 

For the last part, observe that (\ref{markopasam}) is driven by the shift-invariant terms 
$h_{\rm o}^{(2)}(k,\xi),$ $0<k-\xi < 2n$. 

\section*{Appendix E}\label{appE}
{\em Proof of Lemma~\ref{cavit2}.} From Lemma~\ref{cavit},  
$$
[s_j+4n-1\;\;s_{j+1}-2n) \subset [k_i\;\;k_{i+1}).  
$$
Suppose $s_j-k_i \geq 4n$. Recall from Lemma~\ref{ARXdwell} that for some $s_\ell$ and $s_{\ell+1}$ in $\chi_\theta$, $[k_i+n\;\;k_{i+1}) \subseteq [s_\ell\;\;s_{\ell+1})$. Then, from  the inequalities $s_\ell \leq k_i+n \leq  s_j-3n$ 
we must have $j>l$. 
Next, from the chain of inequalities $s_{j+1}-2n \leq k_{i+1} \leq s_{l+1} \leq s_j$, we obtain 
$\delta_j(\chi_\theta) \leq 2n$. A contradiction. Thus, $s_j-k_i<4n$. The case $s_j < k_i$ is not 
possible. If it were possible, $k_i=s_\ell$ for some $\ell>j$ since $k_i \in \chi_\theta$. But, $k_i \leq s_j+4n-1$ 
or $s_\ell-s_j < 4n$. Hence, $\delta_j(\chi_\theta) < 4n$. Contradiction. Thus, $s_j-k_i \geq 0$. 

Suppose $s_{j+1}>k_{i+1}$. From Assumption~\ref{main11}, $k_{i+1} \in \chi_\theta$. Then, $s_t=k_{i+1}$
for some $t \leq j$. Since $s_{j+1}-2n \leq k_{i+1}$, we then have $\delta_j(\chi_\theta) \leq 2n$, 
a contradiction. Hence, $s_{j+1} \leq k_{i+1}$. From $[k_i+n\;\;k_{i+1}) \subseteq [s_\ell\;\;s_{\ell+1})$, we then 
have $s_{j+1} \leq k_{i+1} \leq s_{\ell+1}$. Hence $j\leq l$. The case $l>j$ is not possible
because otherwise the inequalities $s_{j+1} \leq s_{\ell} \leq k_i+n \leq s_j+5n-1$ would yield the contradiction
$\delta_j(\chi_\theta) < 5n$. Thus, $j=\ell$ and $s_{j+1}=k_{i+1}$.

\section*{Appendix F}\label{appF}
{\em Proof of Lemma~\ref{ARXdwell3}.} Similarly to (\ref{gHankel}), we define the triangular Hankel matrices
\[
{\mathcal H}_{\rm o}^{(1)}(k) = \left[ \begin{array}{ccc} h_{\rm o}^{(1)}(k,k-1) & \cdots 
	& h_{\rm o}^{(1)}(k,k-\kappa_{\rm c}) \\  
	& \ddots & \vdots \\  & \cdots & h_{\rm o}^{(1)}(k+\kappa_{\rm o}-1,k-\kappa_{\rm c})   \end{array} \right]
\]
and introduce the extended observability and controllability matrices
\begin{eqnarray*}
	{\mathcal O}_{\kappa_{\rm o}}^{(1)}(k) &=& [c(k) \;\cdots\;  \Phi_{\rm o}^T(k+\kappa_{\rm o},k) 
	c(k+\kappa_{\rm o}-1)]^T, \\
	{\mathcal R}_{\kappa_{\rm c}}^{(1)}(k-1) &=& [b_{\rm o}^{(1)}(k-1)\;\cdots\;\Phi_{\rm o}
	(k,k-\kappa_{\rm c}+1) b_{\rm o}^{(1)}(k-\kappa_{\rm c})]
\end{eqnarray*}
for $\delta_{\rm c} <k\leq \delta_{\rm o}$ and factorize ${\mathcal H}_{\rm o}^{(1)}(k)$ as 
$$
{\mathcal H}_{\rm o}^{(1)}(k) = {\mathcal O}_{\kappa_{\rm o}}^{(1)}(k){\mathcal R}_{\kappa_{\rm c}}^{(1)}(k-1).
$$ 
From Lemma~\ref{lemreal2}, ${\rm rank}({\mathcal H}_{\rm o}^{(1)}(k))=n$ for all $k$. Write 
\begin{eqnarray*}
	\theta^{(1)}(k_i) &=& [d(k_i) \;c^T(k_i) {\mathcal R}_{\kappa_{\rm c}}^{(1)}(k_i-1)], \\
	\theta^{(1)}(k_i-1) &=& [d(k_{i-1}) \;\;c^T(k_{i-1}){\mathcal R}_{\kappa_{\rm c}}^{(1)}(k_i-2)].
\end{eqnarray*} 
Since $\delta_i(\chi)>2n$, ${\mathcal R}_{\kappa_{\rm c}}^{(1)}(k_i-1)={\mathcal R}_{\kappa_{\rm c}}^{(1)}(k_i-2)$ 
and from Lemma~\ref{lemreal2},  ${\mathcal R}_{\kappa_{\rm c}}^{(1)}(k_i-1)$ has full rank. Suppose $k_i \in \chi$. 
Then, $\varphi(k_i-1) \neq \varphi(k_i)$ and $[c^T(k_i) \;d(k_i)] \neq [c^T(k_i-1) \;d(k_i-1)]$ from 
Assumption~\ref{distinctas}. Therefore, $\theta^{(1)}(k_i-1) \neq \theta^{(1)}(k_i)$ if $d(k_i) \neq d(k_i-1)$.
If $c(k_i) \neq c^T(k_i-1)$, the same conclusion is drawn since ${\mathcal R}_{\kappa_{\rm c}}^{(1)}(k_i-1)$ 
and ${\mathcal R}_{\kappa_{\rm c}}^{(1)}(k_i-2)$ are equal and have full rank. It follows that 
$\theta(k_i-1) \neq \theta(k_i)$ and $k_i \in \chi_\theta$. 

Suppose $ii.$ holds in Assumption~\ref{distinctas}, construct 
$\mathcal{H}_{\rm o}^{(2)}(k)$ similarly to $\mathcal{H}_{\rm o}^{(1)}(k)$ from the Markov parameters 
$h_{\rm o}^{(2)}(k,i)$. Factorization of $\mathcal{H}_{\rm o}^{(2)}(k)$ shows that 
$\theta^{(2)}(k_i-1) \neq \theta^{(2)}(k_i)$ if $c^T(k_i) \neq c^T(k_{i-1})$. Then, $\theta(k_i-1) \neq \theta(k_i)$.
and $k_i \in \chi_\theta$.

\section*{Appendix G}\label{appG}
{\em Proof of Proposition~\ref{propPE}}. Fix $i$ first and hence $\varphi(k_i)$. 
Assumptions~\ref{PEsls}--\ref{PEinput} are sufficient for the existence of PE inputs for the transfer function 
\[
G(z;k_i)=(1-H_{\rm o}^{(2)}(z;k_i))^{-1} H_{\rm o}^{(1)}(z;k_i).
\]
Thus, (\ref{rankPE}) is true for some $\gamma_i>0$. Let $\gamma=\sup_i \gamma$ which is finite for there are no more 
than $\sigma$ discrete states as $i$ changes. Thus, (\ref{rankPE}) holds with $\gamma$ and all $k_i \in \chi$.

\bibliography{deadpl}

\begin{thebibliography}{10}

\bibitem{Bako:2011}
Laurent Bako.
\newblock Identification of switched linear systems via sparse optimization.
\newblock {\em Automatica}, 47(4):668--677, 2011.

\bibitem{BakoVanLuongLauerBloch2013}
Laurent Bako, Van~Luong Le, Fabien Lauer, and G{\'e}rard Bloch.
\newblock Identification of {MIMO} switched state-space models.
\newblock In {\em 2013 American Control Conference}, pages 71--76, Washington,
  DC, June 2013.

\bibitem{Bako&Mercere&Guillaume&Vidal&Lecoeuche:2009}
Laurent Bako, Guillaume Merc{\`e}re, Ren{\'e} Vidal, and St{\'e}phane
  Lecoeuche.
\newblock Identification of switched linear state space models without minimum
  dwell time.
\newblock {\em IFAC Proceedings Volumes}, 42(10):569--574, 2009.

\bibitem{Barker&Balas:2000}
Jeffrey~M. Barker and Gary~J Balas.
\newblock Comparing linear parameter-varying gain-scheduled control techniques
  for active flutter suppression.
\newblock {\em Journal of Guidance, Control, and Dynamics}, 23(5):948--955,
  2000.

\bibitem{Basseville&Nikiforov:1993}
Michele Basseville and Igor~V. Nikiforov.
\newblock {\em Detection of {A}brupt {C}hanges, {T}heory and {A}pplications},
  volume 104.
\newblock Prentice-Hall, Englewood Cliffs, 1993.

\bibitem{Bemporad&Garulli&Paoletti&Vicino:2005}
Alberto Bemporad, Andrea Garulli, Simone Paoletti, and Antonio Vicino.
\newblock A bounded-error approach to piecewise affine system identification.
\newblock {\em IEEE Transactions on Automatic Control}, 50(10):1567--1580,
  2005.

\bibitem{Bencherki&Turkay&Akcay:2021}
Fethi Bencherki, Semiha T{\"u}rkay, and H{\"u}seyin Ak{\c{c}}ay.
\newblock Basis transform in switched linear system state-space models from
  input-output data.
\newblock {\em ArXiv Preprint, arXiv:2106.10888}, 2021.

\bibitem{Borges&Verdult&Verhaegen&Botto:2005}
Jos{\'e} Borges, Vincent Verdult, Michel Verhaegen, and Miguel~A. Botto.
\newblock A switching detection method based on projected subspace
  classification.
\newblock In {\em 44th IEEE Conf. Decision and Control and the European Control
  Conference 2005}, pages 344--349, Seville, Spain, December 2005.

\bibitem{Candes&Wakin&Boyd:2008}
Emmanuel~J. Candes, Michael~B. Wakin, and Stephen~P. Boyd.
\newblock Enhancing sparsity by reweighted $\ell_1$ minimization.
\newblock {\em Journal of Fourier Analysis and Applications}, 14(5):877--905,
  2008.

\bibitem{Donoho&Elad&Temlyakov:2005}
David~L Donoho, Michael Elad, and Vladimir~N. Temlyakov.
\newblock Stable recovery of sparse overcomplete representations in the
  presence of noise.
\newblock {\em IEEE Transactions on Information Theory}, 52(1):6--18, 2005.

\bibitem{Eldar&Kuppinger&Bolcskei:2010}
Yonina~C. Eldar, Patrick Kuppinger, and Helmut Bolcskei.
\newblock Block-sparse signals: Uncertainty relations and efficient recovery.
\newblock {\em IEEE Transactions on Signal Processing}, 58(6):3042--3054, 2010.

\bibitem{Eldar&Mishali:2009}
Yonina~C. Eldar and Moshe Mishali.
\newblock Robust recovery of signals from a structured union of subspaces.
\newblock {\em IEEE Transactions on Information Theory}, 55(11):5302--5316,
  2009.

\bibitem{EsterKriegelSanderXu1996}
Martin Ester, Hans-Peter Kriegel, J{\"o}rg Sander, and Xiaowei Xu.
\newblock A density-based algorithm for discovering clusters in large spatial
  databases with noise.
\newblock In {\em Second International Conference on Knowledge Discovery and
  Data Mining}, pages 226--231, Portland, OR, August 1996.

\bibitem{Felici&Wingerden&Verhaegen:2007}
Federico Felici, Jan-Willem Van~Wingerden, and Michel Verhaegen.
\newblock Subspace identification of {MIMO} {LPV} systems using a periodic
  scheduling sequence.
\newblock {\em Automatica}, 43(10):1684--1697, 2007.

\bibitem{Ferrari-Trecate&Muselli&Liberati&Morari:2003}
Giancarlo Ferrari-Trecate, Marco Muselli, Diego Liberati, and Manfred Morari.
\newblock A clustering technique for the identification of piecewise affine
  systems.
\newblock {\em Automatica}, 39(2):205--217, 2003.

\bibitem{Giarre&Bauso&Falugi&Bamieh:2006}
Laura Giarr{\'e}, Dario Bauso, Paola Falugi, and Bassam Bamieh.
\newblock {LPV} model identification for gain scheduling control: An
  application to rotating stall and surge control problem.
\newblock {\em Control Engineering Practice}, 14(4):351--361, 2006.

\bibitem{Grant&Boyd:2014}
Michael Grant and Stephen Boyd.
\newblock {CVX}: Matlab software for disciplined convex programming, version
  2.1, 2014.

\bibitem{Hastie&Tibshirani&Friedman:2001}
Trevor Hastie, Robert Tibshirani, and Jerome Friedman.
\newblock {\em The {E}lements of {S}tatistical {L}earning: {D}ata {M}ining,
  {I}nference, and {P}rediction}.
\newblock Springer-Verlag, New York, NY, 2001.

\bibitem{Heemels&DeSchutter&Bemporad:2001}
Wilhemus P.~M.~H. Heemels, Bart De~Schutter, and Alberto Bemporad.
\newblock Equivalence of hybrid dynamical models.
\newblock {\em Automatica}, 37(7):1085--1091, 2001.

\bibitem{Huang&Wagner&Ma:2004}
Kun Huang, Andrew Wagner, and Yi~Ma.
\newblock Identification of hybrid linear time-invariant systems via subspace
  embedding and segmentation ({SES}).
\newblock In {\em 43rd IEEE Conference on Decision and Control}, pages
  3227--3234, Paradise Island, Bahamas, December 2004.

\bibitem{Jikuya&Verhaegen:2002}
Ichiro Jikuya and Michel Verhaegen.
\newblock Deadbeat observer based detection and estimation of a jump in {LTI}
  systems.
\newblock {\em IFAC Proceedings Volumes}, 35(1):377--382, 2002.

\bibitem{Kailath:1980}
Thomas Kailath.
\newblock {\em {L}inear {S}ystems}, volume 156.
\newblock Prentice-Hall, Englewood Cliffs, NJ, 1980.

\bibitem{Majji&Juang&Junkins:2010}
Manoranjan Majji, Jer-Nan Juang, and John~L Junkins.
\newblock Observer/{K}alman-filter time-varying system identification.
\newblock {\em Journal of Guidance, Control, and Dynamics}, 33(3):887--900,
  2010.

\bibitem{Mercere&Bako:2011}
Guillaume Merc{\`e}re and Laurent Bako.
\newblock Parameterization and identification of multivariable state-space
  systems: A canonical approach.
\newblock {\em Automatica}, 47(8):1547--1555, 2011.

\bibitem{Mohammadpour&Scherer:2012}
Javad Mohammadpour and Carsten~W Scherer.
\newblock {\em {C}ontrol of {L}inear {P}arameter {V}arying {S}ystems with
  {A}pplications}.
\newblock Springer, New York, 2012.

\bibitem{Ohlsson&Ljung:2013}
Henrik Ohlsson and Lennart Ljung.
\newblock Identification of switched linear regression models using
  sum-of-norms regularization.
\newblock {\em Automatica}, 49(4):1045--1050, 2013.

\bibitem{Ohlsson&Ljung&Boyd:2010}
Henrik Ohlsson, Lennart Ljung, and Stephen Boyd.
\newblock Segmentation of {ARX}-models using sum-of-norms regularization.
\newblock {\em Automatica}, 46(6):1107--1111, 2010.

\bibitem{Ozay&Sznaier&Lagoa&Camps:2011}
Necmiye Ozay, Mario Sznaier, Constantino~M. Lagoa, and Octavia~I. Camps.
\newblock A sparsification approach to set membership identification of
  switched affine systems.
\newblock {\em IEEE Transactions on Automatic Control}, 57(3):634--648, 2011.

\bibitem{Paoletti&Juloski&Ferrari-Trecate&Vidal:2007}
Simone Paoletti, Aleksandar~Lj. Juloski, Giancarlo Ferrari-Trecate, and
  Ren{\'e} Vidal.
\newblock Identification of hybrid systems a tutorial.
\newblock {\em European Journal of Control}, 13(2-3):242--260, 2007.

\bibitem{Paoletti&Roll&Garulli&Vicino:2007}
Simone Paoletti, Jacob Roll, Andrea Garulli, and Antonio Vicino.
\newblock Input-output realization of piecewise affine state space models.
\newblock In {\em 46th IEEE Conference on Decision and Control}, pages
  3164--3169, New Orleans, LA, December 2007.

\bibitem{Pekpe&Mourot&Gasso&Ragot:2004}
Komi~M. Pekpe, Gilles Mourot, Komi Gasso, and Jos{\'e} Ragot.
\newblock Identification of switching systems using change detection technique
  in the subspace framework.
\newblock In {\em 43rd IEEE Conference on Decision and Control}, pages
  3720--3725, Paradise Island, Bahamas, December 2004.

\bibitem{Petreczky&Toth&Mercere:2016}
Mih{\'a}ly Petreczky, Roland T{\'o}th, and Guillaume Merc{\`e}re.
\newblock Realization theory for {LPV} state-space representations with affine
  dependence.
\newblock {\em IEEE Transactions on Automatic Control}, 62(9):4667--4674, 2016.

\bibitem{Roll&Bemporad&Ljung:2004}
Jacob Roll, Alberto Bemporad, and Lennart Ljung.
\newblock Identification of piecewise affine systems via mixed-integer
  programming.
\newblock {\em Automatica}, 40(1):37--50, 2004.

\bibitem{Scherer:2001}
Carsten~W Scherer.
\newblock {LPV} control and full block multipliers.
\newblock {\em Automatica}, 37(3):361--375, 2001.

\bibitem{Sefidmazgi&Kordmahalleh&Homaifar&Karimoddini&Tunstel:2016}
Mohammad~G. Sefidmazgi, Mina~M. Kordmahalleh, Abdollah Homaifar, Ali
  Karimoddini, and Edward Tunstel.
\newblock A bounded switching approach for identification of switched {MIMO}
  systems.
\newblock In {\em IEEE International Conference on Systems, Man, and
  Cybernetics (SMC)}, pages 4743--4748, Budapest, Hungary, October 2016.

\bibitem{Shokoohi&Silverman:1987}
Shahriar Shokoohi and Leonard~M Silverman.
\newblock Identification and model reduction of time-varying discrete-time
  systems.
\newblock {\em Automatica}, 23(4):509--521, 1987.

\bibitem{Toth:2010}
Roland T{\'o}th.
\newblock {\em {M}odeling and {I}dentification of {L}inear
  {P}arameter-{V}arying {S}ystems}, volume 403.
\newblock Springer, 2010.

\bibitem{Tropp:2006}
Joel~A Tropp.
\newblock Just relax: {C}onvex programming methods for identifying sparse
  signals in noise.
\newblock {\em IEEE Transactions on Information Theory}, 52(3):1030--1051,
  2006.

\bibitem{Tropp:2007}
Joel~A Tropp and Anna~C Gilbert.
\newblock Signal recovery from random measurements via orthogonal matching
  pursuit.
\newblock {\em IEEE Transactions on Information theory}, 53(12):4655--4666,
  2007.

\bibitem{VanWingerden&Verhaegen:2009}
Jan-Willem Van~Wingerden and Michel Verhaegen.
\newblock Subspace identification of bilinear and {LPV} systems for open-and
  closed-loop data.
\newblock {\em Automatica}, 45(2):372--381, 2009.

\bibitem{Vapnik:1998}
Viladimir~N. Vapnik.
\newblock {\em Statistical {L}earning {T}heory}.
\newblock Wiley-Interscience, New York, NY, 1998.

\bibitem{Vassilvitskii&Arthur:2007}
Sergei Vassilvitskii and David Arthur.
\newblock k-means++: The advantages of careful seeding.
\newblock In {\em 18th annual ACM-SIAM Symposium on Discrete algorithms}, pages
  1027--1035, New Orleans, Louisiana, January 2007.

\bibitem{Verdult&Verhaegen:2002}
Vincent Verdult and Michel Verhaegen.
\newblock Subspace identification of multivariable linear parameter-varying
  systems.
\newblock {\em Automatica}, 38(5):805--814, 2002.

\bibitem{Verdult&Verhaegen:2004}
Vincent Verdult and Michel Verhaegen.
\newblock Subspace identification of piecewise linear systems.
\newblock In {\em 43rd IEEE Conference on Decision and Control}, pages
  3838--3843, Paradise Island, Bahamas, December 2004.

\bibitem{Verdult&Verhaegen:2005}
Vincent Verdult and Michel Verhaegen.
\newblock Kernel methods for subspace identification of multivariable {LPV} and
  bilinear systems.
\newblock {\em Automatica}, 41(9):1557--1565, 2005.

\bibitem{Verhaegen:1994}
Michel Verhaegen.
\newblock Identification of the deterministic part of {MIMO} state space models
  given in innovations form from input-output data.
\newblock {\em Automatica}, 30(1):61--74, 1994.

\bibitem{Verhaegen&Dewilde:1992a}
Michel Verhaegen and Patrick Dewilde.
\newblock Subspace model identification, {P}art 1. {T}he output-error
  state-space model identification class of algorithm.
\newblock {\em International Journal of Control}, 56(5):1187--1210, 1992.

\bibitem{Verhaegen&Dewilde:1992b}
Michel Verhaegen and Patrick Dewilde.
\newblock Subspace model identification, {P}art 2. {A}nalysis of the elementary
  output-error state-space model identification algorithm.
\newblock {\em International Journal of Control}, 56(5):1211--1241, 1992.

\bibitem{Verhaegen&Verdult:2007}
Michel Verhaegen and Vincent Verdult.
\newblock {\em {F}iltering and {S}ystem {I}dentification: {A} {L}east {S}quares
  {A}pproach}.
\newblock Cambridge {U}niversity {P}ress, New York, 2007.

\bibitem{Vidal&Chiuso&Soatto:2002}
Ren{\'e} Vidal, Alessandro Chiuso, and Stefano Soatto.
\newblock Observability and identifiability of jump linear systems.
\newblock In {\em 41st IEEE Conference on Decision and Control}, pages
  3614--3619, Las Vegas, NV, December 2002.

\bibitem{Vidal&Soatto&Ma&Sastry:2003}
Ren{\'e} Vidal, Stefano Soatto, Yi~Ma, and Shankar Sastry.
\newblock An algebraic geometric approach to the identification of a class of
  linear hybrid systems.
\newblock In {\em 42nd IEEE International Conference on Decision and Control},
  pages 167--172, Maui, HI, December 2003.

\bibitem{Wen&Zhou&Liu&Lai&Tang:2019}
Jinming Wen, Zhengchun Zhou, Zilong Liu, Ming-Jun Lai, and Xiaohu Tang.
\newblock Sharp sufficient conditions for stable recovery of block sparse
  signals by block orthogonal matching pursuit.
\newblock {\em Applied and Computational Harmonic Analysis}, 47(3):948--974,
  2019.

\bibitem{Willsky:1976}
Alan~S Willsky.
\newblock A survey of design methods for failure detection in dynamic systems.
\newblock {\em Automatica}, 12(6):601--611, 1976.

\end{thebibliography}

\end{document}